\documentclass[nofootinbib,showpacs, twocolumn,preprintnumbers,amsmath,amssymb,amsfonts,superscriptaddress]{revtex4-1}
\usepackage{amsmath}
\usepackage{graphicx}
\usepackage{dcolumn}
\usepackage{bbm}
\usepackage{subfigure}
\usepackage{amssymb}
\usepackage{gensymb}
\usepackage{latexsym}
\usepackage{makecell}
\usepackage{footnote}
\usepackage{soul}
\usepackage{bm}
\usepackage{tikz-feynman}
\usetikzlibrary{arrows,calc}
\usepackage{mathrsfs}
\usepackage{tikz} 
\usetikzlibrary{arrows.meta,bending,decorations.markings}
\tikzset{>=stealth} 

\usepackage{color}
 \usepackage[colorlinks,urlcolor=blue,linkcolor=blue,anchorcolor=blue,citecolor=blue]{hyperref}

\tikzset{middlearrow/.style={
        decoration={markings,
            mark= at position 0.5 with {\arrow{#1}} ,
        },
        postaction={decorate}
    }
}

\begin{document}
\title{Helical superfluid in a frustrated honeycomb Bose-Hubbard model}

\author{Tzu-Chi~Hsieh}\affiliation{Department of Physics and Center
  for Theory of Quantum Matter, University of Colorado, Boulder, Colorado 80309, USA}
\author{Han~Ma}\affiliation{Perimeter Institute for Theoretical Physics, Waterloo, Ontario N2L 2Y5, Canada}\affiliation{Department of Physics and Center
  for Theory of Quantum Matter, University of Colorado, Boulder, Colorado 80309, USA}
\author{Leo~Radzihovsky}\affiliation{Department of Physics and Center
  for Theory of Quantum Matter, University of Colorado, Boulder, Colorado 80309, USA}

\date{\today}

\begin{abstract}
  We study a ``helical" superfluid, a nonzero-momentum condensate in a
  frustrated bosonic model. At mean-field Bogoliubov level, such a novel
  state exhibits ``smectic" fluctuation that are qualitatively
  stronger than that of a conventional superfluid. We develop a phase
  diagram and compute a variety of its physical properties, including
  the spectrum, structure factor, condensate depletion, momentum
  distribution, all of which are qualitatively distinct from that of a
  conventional superfluid. Interplay of fluctuations, interaction and
  lattice effects gives rise to the phenomenon of order-by-disorder,
  leading to a crossover from the smectic superfluid regime to the
  anisotropic XY superfluid phase. We complement the microscopic
  lattice analysis with a field theoretic description for such a
  helical superfluid, which we derive from microscopics and justify on
  general symmetry grounds, reassuringly finding full
  consistency.  Possible experimental realizations are discussed.

\end{abstract}

\maketitle

\section{Introduction}

Frustrated systems exhibit novel emergent phenomena as they often host
extensive degeneracy that is sensitive to even weak perturbations
\cite{moessner_low-temperature_1998,balents_spin_2010}. This manifests itself in rich
low-temperature phenomenology, as in the frustrated magnetism, which
commonly remains disordered down to temperatures low compared to the
Curie-Weiss temperature set by the exchange interaction
\cite{ramirez_strong_1990,gaulin_spin_1992,ramirez_strongly_1994,uemura_spin_1994,dunsiger_muon_1996}. At
even lower temperature, generically order develops through the
so-called ``order-by-disorder'' process
\cite{villain_order_1980,henley_ordering_1989}, where fluctuations (e.g.,
entropically) select ordered states among many competing nearly
free-energetically degenerate phases. This contrasts with an even more
exotic possibility of a putative quantum liquid, that fails to order
down to zero temperature, as believed to be exemplified by some 2d
Kagome magnetic materials with a flat-band dispersion
\cite{sachdev_kagome_1992,hastings_dirac_2000,ran_projected-wave-function_2007,hermele_properties_2008,evenbly_frustrated_2010,yan_spin-liquid_2011,depenbrock_nature_2012,jiang_identifying_2012,iqbal_gapless_2013,iqbal_vanishing_2014,he_chiral_2014,gong_global_2015,norman_colloquium_2016,mei_gapped_2017,he_signatures_2017,zhu_entanglement_2018}.

Another rich class of ``codimension-one'' frustrated systems
\cite{bergman_order-by-disorder_2007,lee_theory_2008,mulder_spiral_2010,matsuda_disordered_2010,gao_spiral_2017,biswas_semiclassical_2018,niggemann_classical_2019,shimokawa_ripple_2019,gao_fractional_2020,liu_featureless_2020,balla_affine_2019,balla_degenerate_2020,yao_generic_2021,bordelon_frustrated_2021,gao_spiral_2022}, characterized by a bare dispersion that is nearly degenerate along
$d-1$ dimensions ($d$ the spatial dimension), develop even in
the bipartite lattice materials, frustrated by competing
interactions. Examples include
spinel materials such as $\textrm{MnSc}_2\textrm{S}_4$ on 3d diamond
lattice
\cite{bergman_order-by-disorder_2007,lee_theory_2008,gao_spiral_2017,bordelon_frustrated_2021}
and $\textrm{FeCl}_3$ on layered honeycomb lattice
\cite{gao_spiral_2022}. Theoretically, they are well described by
classical spin models with nearest-neighbor and next-nearest-neighbor exchange
interactions. Within certain regime of frustration, the dispersion
minima form a degenerate manifold with codimension-one in the
reciprocal space. However, at nonzero temperature a set of wavevectors
is selected entropically based on their lowest free energy.

The phenomenon of order-by-disorder, however, is elusive in magnetic
systems, often obscured by further-neighbor interactions, which
instead can lead to a low-temperature multi-step ordering sequence of
phases \cite{gao_spiral_2017}. Alternatively, the codimension-one bare
dispersion and the associate phenomenology can be realized in bosonic
systems with frustrated hoppings
\cite{varney_quantum_2012,sedrakyan_absence_2014} or Rashba spin-orbital coupling
\cite{sedrakyan_composite_2012}, engineered in a controlled way in cold
atom experiments \cite{wintersperger_realization_2020,zhai_degenerate_2015}. More exotic
constructions involve ring exchange interactions that induce an
emergent Bose metal \cite{paramekanti_ring_2002,tay_possible_2011}.

It is thus of interest to elucidate the nature of a state that
emerges from a nonzero density of interacting frustrated bosons
condensed at nonzero momentum on a dispersion minimum contour, and in
particular to explore its stability to a putative quantum liquid state
\cite{sur_metallic_2019,lake_bose-luttinger_2021}.  To this end, we take two complementary
approaches, a paradigmatic two-dimensional Bose-Hubbard model on
honeycomb lattice and a continuum field theory that we derive from it.
The model is frustrated by competing nearest-neighbor and next-nearest-neighbor
hoppings, $t_1$ and $t_2$, and thereby (for a broad range of parameters)
exhibits a bare dispersion with a minimum on a closed contour at
nonzero momentum $k_0$ set by the hopping matrix elements. We thus
explore in detail the rich phenomenology of the resulting helical
superfluid state. In particular, as summarized in detail in
Sec.~\ref{sec:sum}, we analyze the helical superfluid within the
lattice Bogoliubov approximation, computing its dispersion, condensate
depletion, momentum distribution, structure factor, and equation of
state, all of which different qualitatively from that of a
conventional superfluid. We complement this honeycomb lattice analysis
by deriving from it (guided by generalized dipolar
symmetry \cite{gromov_towards_2019,radzihovsky_quantum_2020,yuan_fractonic_2020,gorantla_global_2022,lake_dipolar_2022}) a superfluid
smectic field theory \cite{radzihovsky_fluctuations_2011,radzihovsky_quantum_2020} to more generally
explore the properties of the helical condensate. By going beyond the
Bogoliubov approximation, we analyze the quantum and thermal
order-by-disorder phenomenon that sets in at low energies and leads to
a crossover to a more conventional but highly anisotropic XY superfluid.


The outline of this paper is as follows. In Sec~\ref{sec:sum}, we give
a summary of our primary results. In Sec.~\ref{sec:model}, we study a
model of bosons hopping on a frustrated honeycomb lattice and
its superfluidity within the Bogoliubov approximation. In
Sec.~\ref{sec:eff_th}, we construct a smectic field theory for the
helical superfluid, and use it to study its stability to quantum and
thermal fluctuations and its properties in the isotropic continuum,
thereby complementing the microscopic lattice model analysis in
Sec~\ref{sec:model}. Sec.~\ref{sec:order_by_disorder} is devoted to
the quantum and thermal order-by-disorder phenomenon, that reduces the
degeneracy of the dispersion minimum contour down to a discrete set of
six minima, and thereby stabilizes the helical superfluid state. We
conclude in Sec.~\ref{sec:conclusion} with a discussion of results in
the contexts of current experiments and remaining open questions.

\section{Summary of results\label{sec:sum}}

\begin{figure}[h]
\includegraphics[width=.4\textwidth]{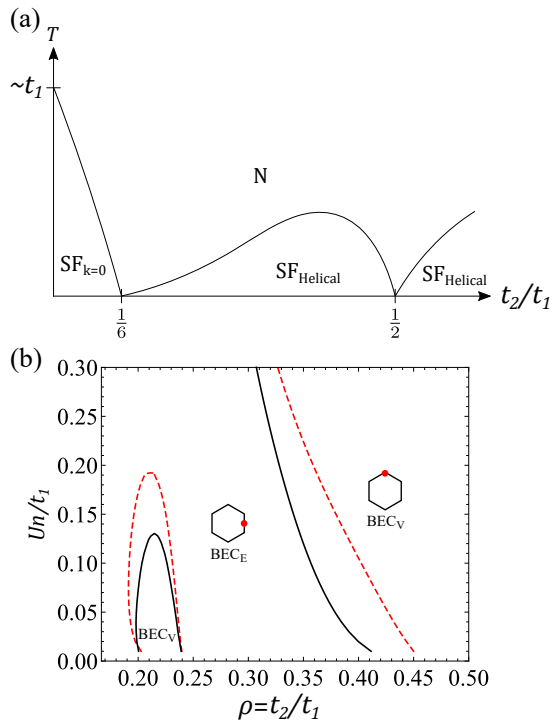}
\caption{(a) Schematic phase diagram of the frustrated honeycomb
  Bose-Hubbard model. With the inclusion of lattice
  order-by-disorder effects, the finite-temperature phase boundaries
  from a conventional ($\text{SF}_{k=0}$) or a helical
  ($\text{SF}_\text{Helical}$) superfluid to a normal state (N) are of
  Berezinskii-Kosterlitz-Thouless universality class. To leading
  order, the phase transition temperature is given by
  $T_{\text{KT}}\sim n_0\sqrt{BB_\perp}$, where $B$ and $B_\perp$ are the superfluid stiffnesses
  defined in the Lagrangian (\ref{eq:L_low_energy}). For $t_2/t_1<1/6$, $B\sim B_\perp\sim t_1$, giving $T_{\text{KT}}\propto t_1$. In contrast, for $t_2/t_1>1/6$,
  $B_\perp$ is induced perturbatively via interaction together with
  lattice effects, giving $T_{\text{KT}}\propto U^{5/8}t_1^{3/8}$ in
  the weakly-interacting limit. In the absence of lattice effects,
  $\text{SF}_\text{Helical}$ is always destabilized to the Normal
  state by divergent thermal fluctuations. (b) Phase diagram of the helical condensate in the regime $1/6<t_2/t_1<1/2$. Bosons can condense at the Vertex ($\text{BEC}_\text{V}$) or the Edge ($\text{BEC}_\text{E}$) of the dispersion minimum contour as a result of the order-by-disorder, obtained by numerically evaluating the one-loop correction to the thermodynamic potential (\ref{eq:Omega^(1)}). The black (red-dashed) curves are zero temperature ($T=0.1Un$) phase boundaries, across which there is a first order transition.\label{fig:phase_diagram}}
\end{figure}

\begin{figure}[h]
\includegraphics[width=.4\textwidth]{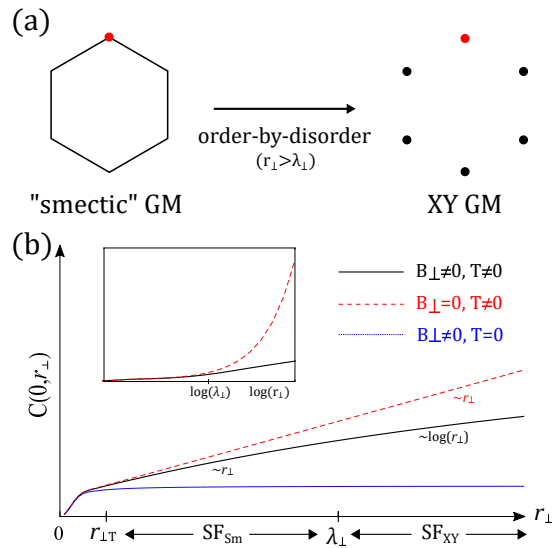}
\caption{(a) Minima of the thermodynamic potential as a function of the condensate momentum ${\bf k}_0$ for intermediate frustration $1/6<t_2/t_1<1/2$. The order-by-disorder effects, which manifest at long scales $r_\perp>\lambda_\perp$, break the degenerate ground state manifold from a (schematic) hexagon down to six points. The helical condensate (red dot) spontaneously breaks the U(1) ($\text{C}_\text{6}$) symmetry in the left (right) panel, resulting in a low-energy smectic-like (conventional, XY) Goldstone mode (GM). (b) Two-point correlation function $C(
{\bf r})=\langle[\phi({\bf r})-\phi(0)]^2\rangle$ of (\ref{eq:L_low_energy}) along $r_\perp$ with $r_\parallel = 0$. For general cases (black curve), the helical superfluid exhibits quantum ($r_\perp < r_{\perp T}$), thermal ($r_\perp > r_{\perp T}$), smectic ($\text{SF}_\text{Sm}$, $r_\perp < \lambda_\perp$), and XY ($\text{SF}_\text{XY}$, $r_\perp > \lambda_\perp$) fluctuations, separated by two crossover scales, $r_{\perp T}$ and $\lambda_\perp$. In the absence of the order-by-disorder ($\lambda_\perp\to\infty$, red-dashed curve), the helical superfluid is characterized by linear in $r_\perp$ smectic fluctuations for $r_\perp > r_{\perp T}$. In the zero temperature limit ($r_{\perp T}\to\infty$, blue-dotted curve), the superfluid exhibits a long-range order with $C(0,r_\perp\to\infty)\to\text{const}$. Inset: same plot with logarithmic scale in the $r_\perp$ axis, where $r_\perp$ is plotted to longer scale for clarity of the crossover. \label{fig:summary}}
\end{figure}

Before turning to a detailed analysis, we summarize the key results of
this work. We study a Bose-Hubbard model with frustrated
nearest-neighbor (NN) and next-nearest-neighbor (NNN) hoppings, $t_1$
and $t_2$. For intermediate frustration $1/6<\rho\equiv t_2/t_1 <1/2$,
its band structure features a closed contour minimum centered around
the $\Gamma$ point (${\bf k}=0$). We then consider a bosonic
condensate with a macroscopic wavefunction
$\Phi=\sqrt{n_0+\pi}e^{i{\bf k}_0\cdot{\bf r}+i\phi}$ at a single
momentum ${\bf k}_0$ on the contour minimum -- a ``helical
condensate", where $n_0$ is the condensate density, $\phi$ and $\pi$
are the phase and density fluctuations, respectively. As illustrated
in Fig.~\ref{fig:phase_diagram}, the helical superfluid is a stable
phase on the high symmetry points of the contour with its transition temperature to the normal state (below that of a
conventional superfluid in the unfrustrated regime $\rho<1/6$) strongly
suppressed by quantum and thermal fluctuations. Quantum
Lifshitz transitions at $\rho=1/6,1/2$\footnote{The quantum Lifshitz
  transition at $\rho=1/2$ only happens at the vertices of the
  hexagonal contour (see Fig.~\ref{fig:contour}), which is
  energetically stable at one-loop order as shown in
  Fig.~\ref{fig:phase_diagram}(b). Higher-order fluctuations may
  shift the positions of the Lifshitz transitions, or make the
  transition at $\rho=1/2$ split into two Lifshitz points
  characterized by smectic-like Goldstone modes with Lagrangian
  (\ref{eq:L_phi_0}), as allowed by the $\text{C}_\text{6}$ lattice
  symmetry.} are characterized by ``soft'' Goldstone modes that
exhibit an anisotropic quartic dispersion, which suppresses the
transition temperature to zero.

As summarized in
Fig.~\ref{fig:summary}, at zero temperature ($T=0$) and on scale
shorter than quantum order-by-disorder length
$\lambda_{\perp Q}$, the highly anisotropic superfluid exhibits an unconventional
smectic-like Goldstone mode (GM) with a low-energy
dispersion [for full dispersion, see Eq.~(\ref{eq:disp})]
\begin{align}\label{eq:E_q_Sm}
    E_{\bf q} \sim \sqrt{Bq_\parallel^2 + Kq_\perp^4},
\end{align}
where the subscript $\parallel$ ($\perp$) denotes the direction that
is parallel (perpendicular) to condensate momentum ${\bf k}_0$,
located at the high symmetry points of the dispersion contour. At
temperatures higher than the interaction $U$, the crossover scale is
modified to $\lambda_{\perp T}$. On longer scales
the so-called order-by-disorder sets in, driven by quantum, thermal
and lattice effects, and the system crossovers to a more conventional
superfluid with highly anisotropic linear dispersion. In the weakly-interacting limit, the order-by-disorder length is long, given by
\begin{align}
\lambda_\perp = \left\lbrace\begin{array}{cc}
         \lambda_{\perp Q}\sim U^{-5/8},\quad T\ll Un_0.\\
         \lambda_{\perp T}\sim U^{-1/8}T^{-1/2},\quad T\gg Un_0.
     \end{array} \right.
\end{align}

To further explain the order-by-disorder phenomenology (discussed in
detail in Sec.~\ref{sec:order_by_disorder}), we first note that the
thermodynamic potential of the helical state near the condensate
momentum ${\bf k}_0$ [see the right panel of
Fig.~\ref{fig:summary}(a)] exhibits a property
\begin{align}\label{eq:Omega_exp}
    \Omega({\bf k}_0+{\bf q}) - \Omega({\bf k}_0) \approx N_0\left[bq_\parallel^2 + b_\perp q_\perp^2 + b_4 q_\perp^4\right],
\end{align}
where the coefficients $b({\bf k}_0)$ and $b_4({\bf k}_0)$ are dominated by the bare dispersion (see Appendix~\ref{app:exp_disp_contour}), while $b_\perp({\bf k}_0)$ is generated perturbatively by the lattice order-by-disorder via interaction $U$. In the above, $N_0$ is the condensate number. The corresponding low-energy Goldstone mode Lagrangian density is given by
\begin{align}\label{eq:L_low_energy}
    \mathcal{L} = n_0\left[ B_\tau(\partial_\tau\phi)^2 + B(\partial_\parallel\phi)^2 + B_\perp(\partial_\perp\phi)^2 + K(\partial_\perp^2\phi)^2\right],
\end{align}
which when treated at a quantum level leads to the dispersion
(\ref{eq:E_q_Sm}) for
$a\ll q_\perp^{-1}\ll\lambda_\perp=\sqrt{K/B_\perp}$ ($a$ the lattice constant). We observe that
a difference choice of the helical condensation,
${\bf k}_0\to {\bf k}_0 + {\bf q}$, equivalently corresponds to a
linear in ${\bf r}$ phase rotation, $\phi\to\phi+{\bf q}\cdot{\bf
  r}$.  By applying the two transformations to the thermodynamic
potential and the Goldstone mode Hamiltonian respectively, and
requiring the shifted energies to be identical, we obtain the
following Ward identities,
\begin{align}\label{eq:B=b}
    b=B,\quad b_\perp=B_\perp.
\end{align}
We expect the relations (\ref{eq:B=b}) to hold at all order of a
perturbation theory (see Sec.~\ref{sec:order_by_disorder}). In
particular, at the zeroth order, the thermodynamic potential [see the
left panel of Fig.~\ref{fig:summary}(a)] is featured by a degenerate
contour minimum, and thus $b_\perp=B_\perp=0$. At one-loop (first)
order, the correction to the thermodynamic potential is given by the
zero-point energy of the Bogoliubov quasiparticles together with
entropic contributions, which in turn gives a $b_\perp\propto U^{5/4}$
($b_\perp\propto U^{1/4}T$) for $T\ll Un_0$ ($T\gg Un_0$) in the
weakly-interacting limit. We also perform a complementary calculation
of $B_\perp$ verifying (\ref{eq:B=b}). In contrast, while $b_4=K$ at
zeroth-order, the above symmetry does not constrain it to hold
generically.


The Goldstone mode theory (\ref{eq:L_low_energy}) predicts the low-energy properties of the helical superfluid. In particular, the off-diagonal order of the superfluid $\langle\Psi^\ast({\bf r})\Psi(0)\rangle\propto e^{-C({\bf r})}$, where $C({\bf r}) = \langle[\phi({\bf r})-\phi(0)]^2\rangle$. As shown in Fig.~\ref{fig:summary}(b), the two-point correlation function $C({\bf r})$ exhibits several qualitatively distinct limits. Here we focus on the behavior along the $\perp$ direction, leaving detailed discussions in Sec.~\ref{sec:stability}. At zero temperature (see the blue-dotted curve), for either $B_\perp=0$ or $B_\perp\neq 0$, $C(r_\parallel=0,r_\perp\to\infty)\to \text{const}$, indicating a long-range order of the helical superfluid. At nonzero temperature, as illustrated by the black (red-dashed) curve, $C(0,r_\perp\to\infty)\sim \log(r_\perp)$ ($\sim r_\perp$) for $B_\perp\neq 0$ ($B_\perp=0$), which signifies a quasi-long-range (short-range) order as a consequence of the XY (smectic) GM fluctuations. For a general case $T\neq 0$ and $B_\perp\neq 0$, there are two important crossover scales, $r_{\perp T}$ and $\lambda_\perp$. The former separates the low-temperature quantum and higher-temperature classical regimes, while the latter, as discussed above, separates the short-distance smectic and long-distance XY regimes. 

\begin{widetext}

\begin{table}[h]
{\setcellgapes{1.5ex}\makegapedcells
    \centering
    \begin{tabular}{c|c|c}
       \hline
       Observables & 2d Helical SF & 2d conventional SF
       \\ \hline
       Static structure factor, $S$ & $\sim n_0\sqrt{\xi^2 q_\parallel^2+\xi^2\lambda^2 q_\perp^4}$ & $\sim n_0\tilde{\xi}q$
       \\ \hline
       Condensate depletion, $n_d/n$ & $\sim U^{3/4}n^{-1/4}$ & $\sim U$ \\ \hline
       Momentum distribution, $n_{\bf q}$ & \thead{$\frac{1}{\xi q_\parallel} f_n \biggl(\lambda\frac{q_\perp^2}{q_\parallel}\biggr)$, when $q_\parallel \xi \ll 1$ \\ \vspace{2mm} $\frac{1}{(\xi q_\parallel)^4} f_n \biggl(\lambda\frac{q_\perp^2}{q_\parallel}\biggr) $, when $q_\parallel \xi \gg 1$} & \thead{$1/q$, when $q \tilde{\xi} \ll 1$ \\ \vspace{2mm} $C/q^4$, when $q  \tilde{\xi} \gg 1$ } \\ \hline 
       Superfluid stiffness, $(\rho_s)_{ij}$ & $8Jn k_{0i} k_{0j} $ & $\delta_{ij} n/m$  \\ \hline
       Chemical potential correction, $\mu^{(1)}$ & $-Un\mathcal{C}\Big(\frac{U^3}{n B^2K}\Big)^{1/4}$ & logarithmic correction \cite{popov_theory_1972,salasnich_zero-point_2016} \\ \hline
    \end{tabular}}
    \caption{Comparison of the helical superfluid (within the order-by-disorder crossover scale) and a conventional superfluid at zero temperature in 2d. Note ${\bf q}={\bf k}-{\bf k}_0$ (${\bf q}={\bf k}$) for the helical (conventional) superfluid. For helical superfluid with low-energy Lagrangian (\ref{eq:L_low_energy}), the coherence length $\xi=\sqrt{B/2Un}$, the anisotropy length scale $\lambda=\sqrt{K/B}$, the condensate momentum ${\bf k_0}$, $J$ and $\mathcal{C}$ are constants, and $f_n$ is a scaling function given in (\ref{eq:fn}). For conventional superfluid, $C$ is the Tan's contact and $\tilde{\xi}=1/\sqrt{4mUn}$ is the coherence length of interacting Bose gas with $m$, $U$ and $n$ being the mass, interaction strength and particle density respectively.}
    \label{tab:SF_helicalSF}
\end{table}

\end{widetext}

Consequently, the helical state is characterized by an anisotropic XY
superfluidity at the longest scale. However, within a large (for small
$U$) order-by-disorder crossover scale,
$(r_\parallel^*,r_\perp^*)=(\lambda_\perp^2/\xi,\lambda_\perp)$ ($\xi$ the coherence length), the
system exhibits soft ($B_\perp\approx 0$) quantum and thermal
fluctuations, with its physical properties qualitatively distinct from
that of a conventional superfluid. Abstracting from the Bose-Hubbard
lattice model, we develop a field theory that captures the key
qualitative features of the helical superfluid within this regime, and
calculate a number of its physical observables, with the
zero-temperature results summarized in
Table~\ref{tab:SF_helicalSF}. At nonzero temperature, thermal
fluctuations set in at scales beyond
$(r_{\parallel T},r_{\perp T}) =
(\frac{Un_0}{T}\xi,\sqrt{\frac{Un_0}{T}}\sqrt{\lambda\xi})$ ($\lambda=\sqrt{K/B}$), as
illustrated in Fig.~\ref{fig:summary}(b). Such quantum-to-classical
crossover is observable in real space- or momentum-resolved
quantities, as exemplified by the structure factor in
Sec.~\ref{sec:Structure_fac}. We now turn to detailed analysis that
leads to the results enumerated above.

\section{Frustrated Bose-Hubbard model on honeycomb lattice\label{sec:model}}

To explore the phenomenon of nonzero momentum helical superfluidity, we study a frustrated Bose-Hubbard model on honeycomb lattice with Hamiltonian $H = H_0 + H_{\text{int}}$, where 
\begin{align}\label{eq:H_int}
    H_{\text{int}}=\frac{U}{2}\sum_{i}\sum_{s=1,2} n_{i,s}(n_{i,s}-1)
\end{align}
is the usual on-site interactions with the subscripts $i$ and $s$
labeling the Bravais lattice sites and the two sublattices
respectively. $n_{i,s}=a^\dag_{i,s}a_{i,s}$ is the boson number
operator with $a^\dag_{i,s}$ and $a_{i,s}$ the corresponding creation
and annihilation operators. The kinetic part of the Hamiltonian is
given by
\begin{equation}
H_0 = -t_1\sum_{\langle ij\rangle}  a^\dag_{i,1} a_{j,2}+t_2\sum_{\langle \langle ij \rangle\rangle}  (a^\dag_{i,1} a_{j,1}+a^\dag_{i,2} a_{j,2} )+h.c.,
\label{eq:H_0}
\end{equation}
where $t_1$ and $t_2$ are the nearest-neighbor (NN) and the
next-nearest-neighbor (NNN) hopping parameters, respectively. We focus
on the case of $t_1>0$, taking the advantage of the symmetry of the
model: $t_1\rightarrow -t_1$, $a_2 \rightarrow -a_2$. In contrast to
the NN hopping, we take the NNN hopping, $-t_{2}$, to be ``frustrated
antiferromagnetic'', $-t_{2}<0$. This has been engineered in atomic
systems in an optical lattice through Floquet techniques
\cite{wintersperger_realization_2020}. Throughout our analysis below, we use
dimensionless parameter $\rho=t_2/t_1>0$ to quantify the frustration
in this model.

\subsection{Band structure of non-interacting Hamiltonian}

\begin{figure}[h]
\includegraphics[width=.45\textwidth]{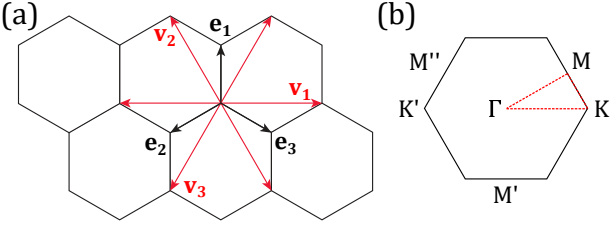}
\caption{(a) Honeycomb lattice with NN (black) and NNN (red)
  vectors. The two sublattices are connected by
  ${\bm \delta}={\bf e}_1$. The lattice spacing is set to ${\rm a}=1$
  throughout this paper.  (b) First Brillouin zone, high
  symmetry points \protect\footnotemark{}, and high symmetry (red-dashed) lines. Equivalent high symmetry
  points are connected by the reciprocal vectors
  (\ref{eq:recip_vec}). \label{fig:lattice} }
\end{figure}

Below we first review the band structure of $H_0$, identifying the
condition when a degenerate minimum along a closed contour develops, and then study a Bose condensate
on the this dispersion minimum contour via Bogoliubov
approximation. We start with specifying the NN and NNN vectors as [see
Fig.~\ref{fig:lattice}(a)]
\begin{align}
{\bf e}_1 &=(0,1) ,\quad {\bf e}_2 = (-\frac{\sqrt{3}}{2},-\frac{1}{2}) ,\quad  {\bf e}_3=(\frac{\sqrt{3}}{2},-\frac{1}{2}),
\nonumber\\
{\bf v}_1 &=(\sqrt{3},0) ,\quad {\bf v}_2 =(-\frac{\sqrt{3}}{2},\frac{3}{2}) ,\quad {\bf v}_3 =(-\frac{\sqrt{3}}{2},-\frac{3}{2})
\end{align}
with the lattice vectors ${\bf v}_i$ spanning the Bravais triangular lattice and the unit vectors ${\bf e}_i$ spanning the honeycomb lattice (the lattice constant $a=1$). The corresponding reciprocal vectors are [see Fig.~\ref{fig:lattice}(b)]
\begin{align}
{\bf G}_1 =(0,\frac{4\pi}{3}) ,\ \ {\bf G}_2 =(-\frac{2\pi}{\sqrt{3}},-\frac{2\pi}{3}) ,\ \ {\bf G}_3 =(\frac{2\pi}{\sqrt{3}},-\frac{2\pi}{3}).\label{eq:recip_vec}
\end{align}
 
\footnotetext{The high symmetry points are given by $\Gamma=(0,0)$, $K=(\frac{4\pi}{3\sqrt{3}},0)$, $K'=(-\frac{4\pi}{3\sqrt{3}},0)$, $M=(\frac{\pi}{\sqrt{3}},\frac{\pi}{3})$, $M'=(0,-\frac{2\pi}{3})$, and $M''=(-\frac{\pi}{\sqrt{3}},\frac{\pi}{3})$.}

\begin{figure}
\includegraphics[width=.4\textwidth]{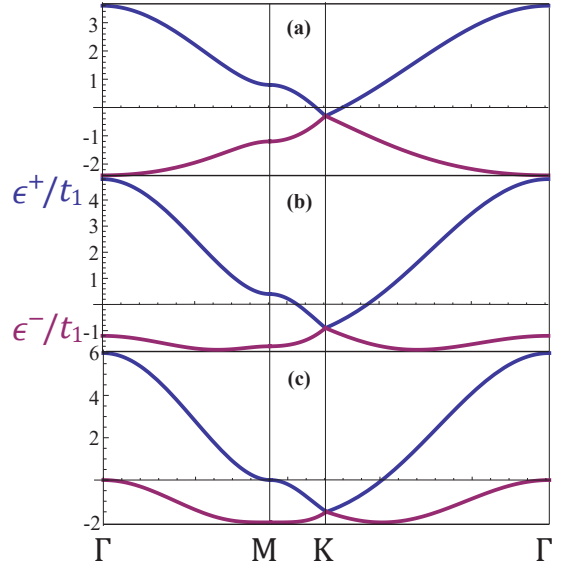}
\caption{Band structure (\ref{eq:epsilon}) of the frustrated bosonic tight-binding model at different $\rho$. (a) $\rho=0.1 < 1/6$. In this case, there is only one energy minimum at ${\bf k}=0$ ($\Gamma$ point). Bosons form a condensate at zero momentum.  (b) $1/6 <\rho=0.3 < 1/2$. Energy minimum shifts away from $\Gamma$ point to nonzero momentum ${\bf k}_0$, forming a closed contour. This contour is enlarged and deformed as $\rho$ increases from $1/6$ to $1/2$  (c) $\rho=0.5$. Now the contour is a perfect hexagon with corner at the center of each edge of first Brillouin zone ($M$ point). \label{fig:band}}
\end{figure}

The Hamiltonian (\ref{eq:H_0}) is straightforwardly diagonalized in momentum space (see Appendix~\ref{app:diago})
\begin{align}
    H_0 =&\ -t_1\sum_{\bf k}(\Gamma_{\bf q}a^\dag_{{\bf k},1}a_{{\bf k},2} + \Gamma_{\bf q}^*a^\dag_{{\bf k},2}a_{{\bf k},1}) \nonumber\\
    &\  + t_2\sum_{\bf k}\epsilon_{\bf k}(a^\dag_{{\bf k},1}a_{{\bf k},1} + a^\dag_{{\bf k},2}a_{{\bf k},2})
    \nonumber\\
    =&\ \sum_{\bf k}\epsilon^-_{\bf k}d_{{\bf k},-}^\dag d_{{\bf k},-} + \sum_{\bf k}\epsilon^+_{\bf k}d_{{\bf k},+}^\dag d_{{\bf k},+},
\end{align}
where
\begin{align}
    d_{{\bf k},\pm} =&\ \frac{1}{\sqrt{2}}(e^{-\frac{\theta_{\bf k}}{2}}a_{{\bf k},1}\mp e^{\frac{\theta_{\bf k}}{2}}a_{{\bf k},2})
\end{align}
create bosonic excitations in the two bands,
\begin{equation}
\epsilon^-_{\bf k}= t_2 \epsilon_{{\bf k}} -t_1 |\Gamma_{{\bf k}}|, \quad \epsilon^+_{{\bf k}} = t_2 \epsilon_{{\bf k}} + t_1 |\Gamma_{{\bf k}}|, \label{eq:epsilon}
\end{equation}
exhibiting Dirac nodes at the $K$ points (the corners of the first Brillouin zone), made famous in graphene but irrelevant here for bosonic system dominated by condensation at the bottom of the band. In the above,
\begin{equation}
\epsilon_{{\bf k}} =2\sum_i \cos {\bf k}\cdot {\bf v}_i, \quad \Gamma_{{\bf k}} = \sum_i \exp\left(-i{\bf k}\cdot{\bf e}_i\right) \label{eq:eplison_and_Gamma}
\end{equation}
with $|\Gamma_{{\bf k}}| =\sqrt{3+\epsilon_{{\bf k}}}$. The dependence of the band structure on $\rho=t_2/t_1$ is depicted in Fig.~\ref{fig:band}. 
Momenta $\bar{\bf k}_0$ is the dispersion minimum that satisfies
\begin{equation}
\nabla_k\epsilon^-_{\bf k}|_{{\bf k}=\bar{\bf k}_0} =0 \rightarrow \rho=\frac{1}{2|\Gamma_{\bar{\bf k}_0}|}.
\label{eq:k0}
\end{equation}

\begin{figure}[h]
\includegraphics[width=.45\textwidth]{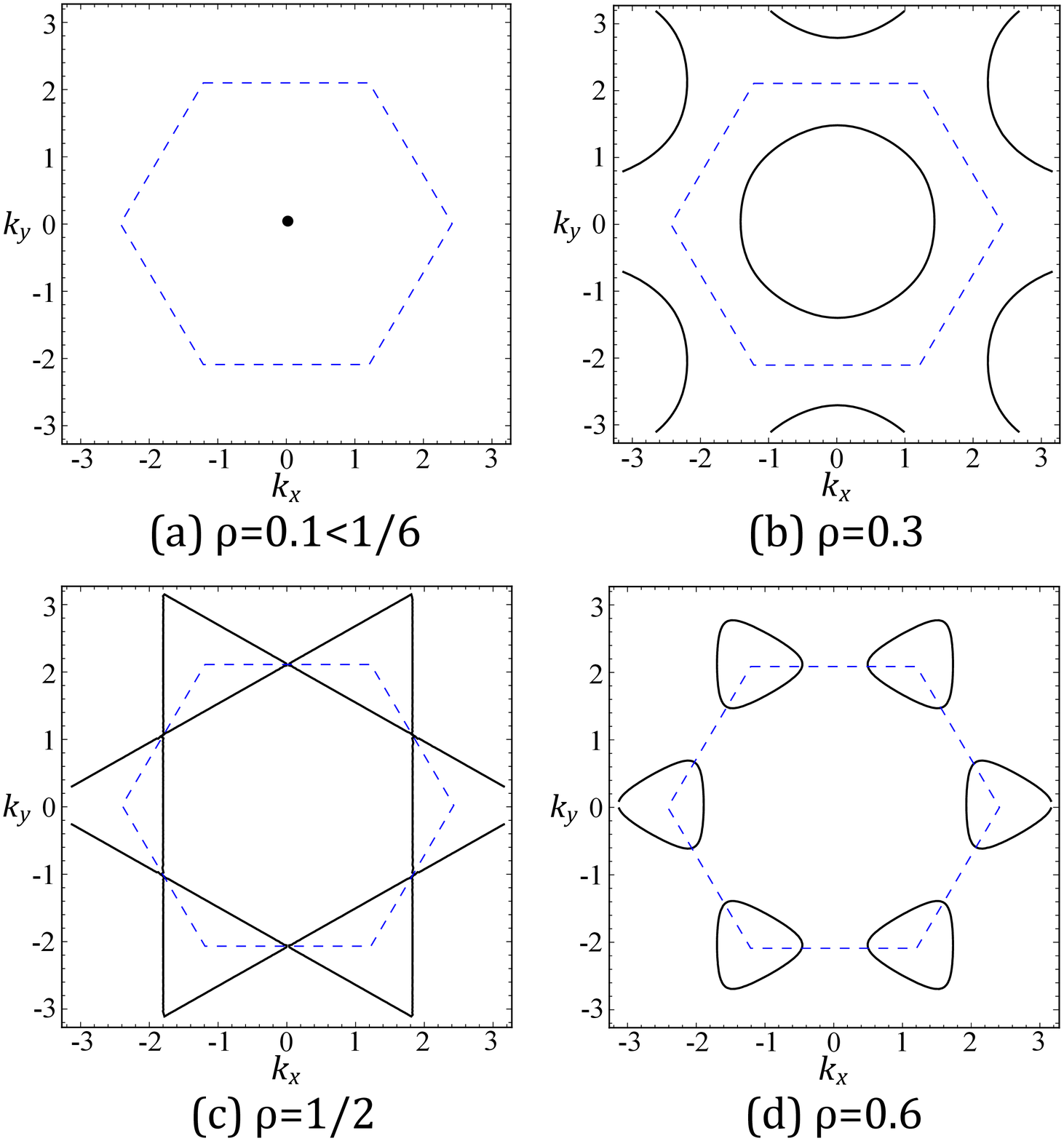}
\caption{Solution of energy minimum, plotted in unit of $a^{-1}$. Blue-dashed lines denote the first Brillouin zone boundaries. (a) This is the case shown in Fig.~\ref{fig:band}(a), where the energy minimum is a point. (b) This case corresponds to Fig.~\ref{fig:band}(b) with a dispersion minimum contour. (c) This is the case denoted in Fig.~\ref{fig:band}(c) with a perfect hexagon contour minimum. (d) When $\rho >1/2$, the energy minima form a contour centered at every corner of the first Brillouin zone. We are not focus on this part in this paper.  \label{fig:contour}}
\end{figure}

As shown in Fig.~\ref{fig:contour}, $\rho$ controls the form of the dispersion minimum in the reciprocal space. In detail, for $\rho<1/6$, the minimum is at the $\Gamma$ point ($\bar{\bf k}_0=(0,0)$), while for $\rho\rightarrow \infty$, the minima are located at the $K$ points. For intermediate frustration $1/6<\rho<1/2$ ($1/2<\rho<\infty$), the dispersion minima are extended to closed contour(s) centered around the $\Gamma$ (K) point(s). This macroscopic degeneracy makes the system sensitive to perturbations. Below we discuss these three cases respectively.

\subsubsection{$\rho<1/6$}

The simplest is the weakly-frustrated case of $\rho<1/6$, for which
the bosons condense at the $\Gamma$ point [see
Fig.~\ref{fig:contour}(a)], exhibiting conventional zero momentum
superfluidity. In the XY spin language, this corresponds to a
ferromagnetic spin state.

\subsubsection{Intermediate frustration, $\rho >\frac{1}{2}$}

In this case, the energy minima form closed contours encircling six corners of the first Brillouin zone, as shown in Fig.~\ref{fig:contour}(d), with contours shrinking as $\rho$ increases to $\infty$. 

\begin{figure}[h]
\includegraphics[width=.3\textwidth]{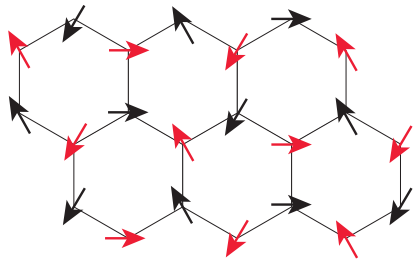}
\caption{Spin representation of the bosonic condensates on the two
  sublattices (black and red), condensed at the first Brillouin zone corner $K$ or
  $K'$. These become independent in the $\rho \rightarrow \infty$
  limit, coupled only by the boson interaction.
  \label{fig:spin120}}
\end{figure}

As $\rho \rightarrow \infty$, two sublattices decouple, with bosons on
each triangular lattice condensing at a K point of the reciprocal
lattice.\footnote{In XY spin language, this corresponds to the
  coplanar $120^\circ$ state, the ground state of XXZ model on the
  triangular lattice.} Generally, such superfluid is described by two
order parameters
$\langle a_{s,i} \rangle = c_s e^{i{\bf K}_s\cdot {\bf r}_{s,i}}$
where $c_{s=1,2}$ are two independent complex condensate amplitudes at
${\bf K}_s \in \lbrace {\bf K}, {\bf K'}\rbrace$. Each of these
site-dependent order parameters can be written as a two component XY
vector
$\vec{S}_{\bf r}=(\textrm{Re} \langle a_{{\bf r}} \rangle, \textrm{Im}
\langle a_{{\bf r}} \rangle)$. For a particular choice of $c_{1,2}$, a
configuration of $\vec{S}_{\bf r}$ is pictorially illustrated in
Fig.~\ref{fig:spin120}.

\subsubsection{$1/6<\rho<1/2$}

The intermediate-frustration regime -- the main focus of this paper --
exhibits a closed dispersion minimum centered around the $\Gamma$
point, as illustrated in Fig.~\ref{fig:contour}(b). Generically, the
contour exhibits a $\text{C}_\text{6}$ lattice symmetry. The contour
degenerates into a circle [hexagon] for $\rho \to \frac{1}{6}^+$, see
Fig.~\ref{fig:contour}(b) [$\rho \to \frac{1}{2}^-$, see
Fig.~\ref{fig:contour}(c)].

\subsection{Helical superfliud in Bogoliubov approximation}

The interplay of interactions and the macroscopic degeneracy discussed
above presents a rich and challenging problem. However, for weak
interactions, a Bose condensate on a finite set of ${\bf k}_{0i}$
points\footnote{This is to be contrasted with an exotic Bose-Luttinger
  liquid state, where the boson amplitude is ordered with ${\bf k}_0$
  fluctuating across the entire contour, as recently studied by Sur
  and Yang \cite{sur_metallic_2019} and Lake et al. \cite{lake_bose-luttinger_2021}} on the
minimum of the dispersion contour is at least a locally stable state
of matter. For a set of non-collinear ${\bf k}_{0i}$ points, the
ground state is a supersolid that generically breaking the underlying
crystal lattice and the U(1) symmetries. A simpler collinear
superfluid states are represented by a helical FF-like condensate at a
${\bf k}_0$ and an LO-like condensate at $\{{\bf k}_0,-{\bf k}_0\}$
\cite{radzihovsky_fluctuations_2011,choi_finite-momentum_2011}. For interacting bosons, finding a
ground state even among these set of supersolids (characterized by a
set of ${\bf k}_{0i}$) is a challenging problem that we do not address
here and is best studied numerically.

Here we focus on a helical superfluid corresponding to a condensation
at a single ${\bf k}_0$ that we find exhibits rich phenomenology that
we explore in detail. Such a state can also be described by a helical
spin density wave
$\vec{S}_{\bf r}=(\textrm{Re}\langle a_{\bf r} \rangle,
\textrm{Im}\langle a_{\bf r} \rangle) = \sqrt{N_0}(\cos {\bf k_0}
\cdot{\bf r}, \sin {\bf k_0} \cdot {\bf r})$ that breaks the lattice
rotational and time-reversal symmetries. However, in this FF helical
state, the physical observables remain (lattice-) translationally
invariant.

Thus, focusing on such a helical condensate at ${\bf k}_0$, we study
interacting bosons on a frustrated honeycomb lattice (\ref{eq:H_0}) within the Bogoliubov approximation in a canonical ensemble. In momentum space, the interaction (\ref{eq:H_int}) is given by
\begin{align}
H_{\text{int}} =&\ \frac{U}{2V} \sum_{s=1,2}\sum_{{\bf k}_1,{\bf k}_2,{\bf p}} a^\dag_{{\bf k}_1+\frac{{\bf p}}{2},s} a^\dag_{{\bf k}_2-\frac{{\bf p}}{2},s} a_{{\bf k}_1-\frac{{\bf p}}{2},s} a_{{\bf k}_2+\frac{{\bf p}}{2},s},  \label{eq:interaction}
\end{align}
where $V$ is the number of the unit cells, which is one-half of the number of the sites, $N_s=2V$. At low energies, neglecting inter-band excitations, we can focus on the lowest band, and express the site boson operators $a_{{\bf k},1}$ and $a_{{\bf k},2}$  in terms of the lower-band operator $d_{{\bf k},-}$,
\begin{equation}
a_{{\bf k},1} \approx \frac{1}{\sqrt{2}} e^{i \frac{\theta_{{\bf k}}}{2}} d_{{\bf k},-},\quad a_{{\bf k},2} \approx \frac{1}{\sqrt{2}} e^{-i \frac{\theta_{{\bf k}}}{2}} d_{{\bf k},-},
\label{eq:lower_band_operator}
\end{equation}
where $\theta_{\bf k} = \textrm{Arg} (\Gamma_{\bf k})$ defined in Eq.~(\ref{eq:eplison_and_Gamma}). Expressing the Hamiltonian in terms of excitations
\begin{equation}\label{eq:d_decompose}
d_{{\bf k}_0+{\bf q},-} \approx - \sqrt{N_0}\delta_{{\bf k},{\bf k}_0} + d_{{\bf k},-},
\end{equation}
out of a helical condensate $N_0$, to quadratic order we have the Bogoliubov Hamiltonian,
\begin{align}
H \approx &\ \epsilon^-_{{\bf k}_0} N+ \frac{nNU}{2} - \frac{1}{2}\sum_{{\bf q}} \varepsilon'_{{\bf q}} \\
&+ \frac{1}{2} \sum_{{\bf q}} \left( \begin{array}{cc}
d^\dag_{{\bf k}_0+{\bf q},-} &d_{{\bf k}_0-{\bf q},-}
\end{array}\right)h_0^- \left( \begin{array}{c}
d_{{\bf k}_0+{\bf q},-} \\ d^\dag_{{\bf k}_0-{\bf q},-}
\end{array}\right), \label{eq:hammatrix}
\end{align}
where
\begin{align}
h_0^- = \left( \begin{array}{cc}
\varepsilon'_{\bf q} & Un\cos(\Theta_{\bf q}) \\
Un\cos(\Theta_{\bf q}) & \varepsilon'_{\bf -q}
\end{array}\right)
\end{align}
and
\begin{align}
\varepsilon'_{\bf q} =&\ \epsilon^-_{{\bf k}_0+{\bf q}}-\epsilon^-_{{\bf k}_0}+Un , \nonumber\\
\Theta_{\bf q} =&\ \frac{\theta_{{\bf k}_0+{\bf q}}+\theta_{{\bf k}_0-{\bf q}}}{2}-\theta_{{\bf k}_0},
\nonumber\\
n =&\ \frac{N}{N_s} = \frac{N}{2V},\nonumber\\
N =&\ N_0 + \sum_{{\bf q}}d_{{\bf k}_0+{\bf q},-}^\dag d_{{\bf k}_0+{\bf q},-}, \label{eq:defOmega}
\end{align}
deferring details to Appendix~\ref{app:hint}.
Bogoliubov-diagonalizing the Hamiltonian gives
\begin{align}
H 
=&\ E_{gs} + \sum_{{\bf q}} E_{\bf q} \alpha_{{\bf k}_0+{\bf q}}^\dag \alpha_{{\bf k}_0+{\bf q}}
,\label{eq:hamsimple}
\end{align}
where the ground state energy
\begin{align}
E_{gs} &= \epsilon^-_{{\bf k}_0} N+ \frac{nNU}{2} + \frac{1}{2} \sum_{{\bf q}} (E_{{\bf q}}-\varepsilon'_{{\bf q}}) \label{eq:E_gs}
\end{align}
and the dispersion
\begin{align}
E_{\bf q}
=&\  \mathcal{E}_{2,{\bf q}} + \mathcal{E}_{1,{\bf q}},
\label{eq:disp}
\end{align}
with
\begin{align}
 \mathcal{E}_{1,{\bf q}} = \sqrt{\mathcal{E}_{{\bf q}}^2 +2Un\mathcal{E}_{{\bf q}} + U^2n^2\sin^2(\Theta_{\bf q})}
\end{align}
and
\begin{align}
\mathcal{E}_{\bf q} =&\ \frac{\epsilon^-_{{\bf k}_0+{\bf q}}+\epsilon^-_{{\bf k}_0-{\bf q}}}{2}-\epsilon^-_{{\bf k}_0},\quad \mathcal{E}_{2,{\bf q}} = \frac{\epsilon^-_{{\bf k}_0+{\bf q}}-\epsilon^-_{{\bf k}_0-{\bf q}}}{2}.\label{eq:disp_Epsilon_1_and_2}
\end{align}

We postpone the discussion of $E_{gs}$ until Sec.~\ref{sec:order_by_disorder}, where we show that fluctuations together with the lattice effects lead to order-by-disorder selection of the condensation at high symmetry points of the contour. In anticipation of this, here we note that there are two classes (and $\text{C}_\text{6}$ equivalents) of high symmetry points represented by ${\bf k}_0=(0,k_{0V})$ ``Vertex" $\text{BEC}_\text{V}$ and $(k_{0E},0)$ ``Edge" $\text{BEC}_\text{E}$, with (\ref{eq:k0}) giving, 
\begin{align}
k_{0V} =&\ \frac{2}{3}\textrm{arccos}\left[\frac{1}{16\rho^2}-\frac{5}{4}\right],
\nonumber\\
k_{0E} =&\ \frac{2}{\sqrt{3}}\textrm{arccos}\left[\frac{1}{4\rho}-\frac{1}{2}\right].
\end{align}

We next examine the details of the Bogoliubov dispersion $E_{\bf q}$ that is positive-semi-definite (gapless at ${\bf q}\to 0$) for all $U$ and ${\bf q}$, which for $U\to 0$ reduces to $\epsilon^-_{{\bf k}_0+{\bf q}}-\epsilon^-_{{\bf k}_0}>0$. We thus find that the helical condensate at the Bogoliubov level is locally stable.

\begin{figure}[h]
\includegraphics[width=0.4\textwidth]{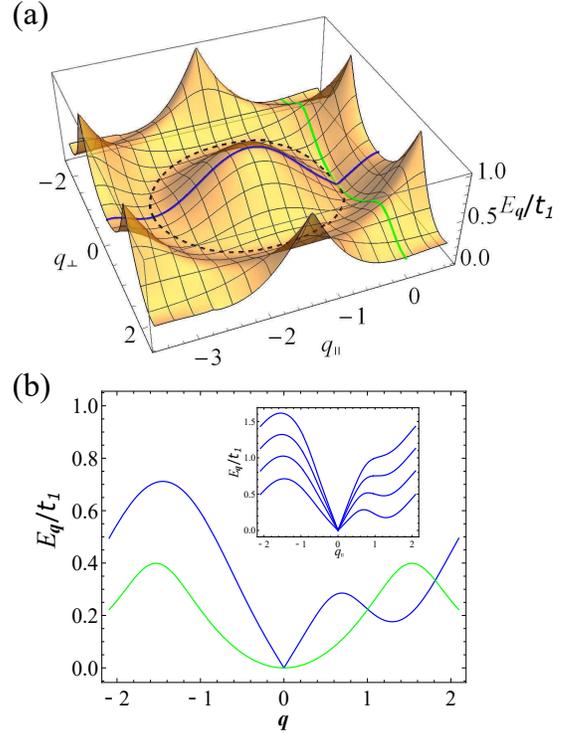}
\caption{(a) The helical superfluid dispersion $E_{\bf q}$ from
  Eq.~(\ref{eq:disp}) with ${\bf k}_0=(0,k_{0V})$, $\rho=0.3$,
  $Un/t_1=0.2$ and ${\bf q}$ in unit of $a^{-1}$. The black-dashed contour indicates the
  degenerate minimum of the noninteracting band $\epsilon^-_{\bf k}$
  in Eq.~(\ref{eq:epsilon}). In the presence of interaction, the minimum
  appears at the origin, ${\bf q}=0$. The blue and green curves are
  dispersion cuts along and perpendicular to ${\bf k}_0$, indicating
  small $q$ linear and quadratic dispersions, respectively. The
  corresponding long wavelength description in (b) illustrates the
  symmetry-expected smectic form (\ref{eq:disp_small_q}). The inset of (b) shows dispersions
  along $q_\parallel$ for a set of interaction strengths, from the
  lowest to the highest given by
  $Un/t_1=0.2,0.6,1,1.4$. \label{fig:dispersion_micro}}
\end{figure}

In Fig.~\ref{fig:dispersion_micro}, we plot the quasiparticle dispersion $E_{\bf q}$ as a function of ${\bf q} = {\bf q}_\parallel + {\bf q}_\perp
$, where
${\bf q}_\parallel=({\bf q}\cdot\hat{{\bf k}}_0)\hat{{\bf k}}_0$ and
${\bf q}_\perp= \hat{{\bf k}}_0\times{\bf q}\times\hat{{\bf k}}_0
$. Without loss of generality, we plot the case
${\bf k}_0=(0,k_{0V})$, which illustrates general properties of the
dispersion (\ref{eq:disp}). We note that it is asymmetric under
$q_\parallel\to -q_\parallel$ due to the $\mathcal{E}_{2,{\bf q}}$
contribution in (\ref{eq:disp_Epsilon_1_and_2}). Physically, this
broken inversion symmetry is a consequence of the spontaneously-chosen ${\bf k}_0$. At
$q_\parallel = \text{mod}(-2k_0,4\pi/3) \sim 1.5$, the disperion also
displays a metastable saddle point inherited from the contour minimum
in the noninteracting dispersion $\epsilon^-_{\bf k}$, as shown by the
blue curve in Fig.~\ref{fig:dispersion_micro}(b). This feature
disappears gradually with increasing interaction [see inset of
Fig.~\ref{fig:dispersion_micro}(b)]. In the small ${\bf q}$ limit, as
expected on rotational symmetry grounds (here neglecting lattice
order-by-disorder effects, but see Sec.~\ref{sec:order_by_disorder})
of a helical superfluid (striped) state, the dispersion is
well-described by an anisotropic smectic form,
\begin{align}
     E_{\bf q} \approx &\ \sqrt{2Un(Bq_\parallel^2+K_\perp q_\perp^4+K_\parallel q_\parallel^4+K_{\parallel\perp} q_\parallel^2q_\perp^2)} \label{eq:disp_small_q}
     \nonumber\\
     \sim &\ \left\lbrace\begin{array}{cc}
         q_\perp^2,\quad q_\parallel\ll\lambda q_\perp^2,\\
         |q_\parallel|,\quad q_\parallel\gg\lambda q_\perp^2,
     \end{array} \right.
\end{align}
with a ``soft" $\perp$ direction [see the green curve in
Fig.~\ref{fig:dispersion_micro}(b)], where
$\lambda=\sqrt{K_\perp/B}$. For $q_\parallel \sim q_\perp^2$, the
$K_\parallel$ and $K_{\parallel\perp}$ can be neglected at small
${\bf q}$, while in the rotational invariant limit $\rho\to 1/6^{+}$,
$K_\parallel\approx K_\perp\approx K_{\parallel\perp}/2$. The
parameters in Eq.~(\ref{eq:disp_small_q}) can be obtained by expanding
the free dispersion $\epsilon^-_{{{\bf k}_0}+{\bf q}}$ in ${\bf q}$
with the results given in Appendix~\ref{app:exp_disp_contour}. We note
that at the Lifshitz transition, $B(\rho\to 1/6)\to 0$. Then, the
(above neglected) higher-order contribution
$K_\parallel q_\parallel^4$ should be taken into account, with the
Goldstone mode exhibiting an anisotropic quartic dispersion in all
directions.

\section{field theory of helical superfluid on a closed contour minimum \label{sec:eff_th}}

We now turn to a complementary continuum field-theoretic description of the helical superfuild state, characterized by Bose condensation on a closed contour minimum in momentum space. In the isotropic case, neglecting lattice effects, this physics can be encoded through a quartic dispersion
\begin{equation}
\varepsilon_{{\bf k}} = J({\bf k}^2-\bar{k}_0^2)^2 + \varepsilon_0 \label{eq:dispersion_isotropic}
\end{equation}
with minima on a contour ${\bf k}^2=\bar{k}_0^2$. As we demonstrated in Sec.~\ref{sec:model}, such dispersion is indeed realized for bosons on a honeycomb lattice for $\rho \approx 1/6^{+}$, with
\begin{align}
J &= t_1\left(-\frac{3}{64}+\frac{27\rho}{32}\right),
\nonumber\\
\bar{k}_0 &= \sqrt{\frac{8-48\rho}{1-18\rho}},
\nonumber\\
\varepsilon_0 &= t_1\left[\frac{3(1-6\rho)^2}{1-18\rho} -3 + 6\rho\right]. \label{eq:parameters_isotropic}
\end{align}
We can encode this physics in a noninteracting Hamiltonian $H_0=\int_{\bf r}\hat{\Phi}_{\bf r}^\dag\hat{\varepsilon}_{{\bf r}}\hat{\Phi}_{\bf r}$ with
\begin{equation}
\hat{\varepsilon}_{{\bf r}} = J(-\nabla^2-\bar{k}_0^2)^2+\varepsilon_0. \label{eq:eff_th_free_dispersion}
\end{equation}
We then extend this to an interacting field theory and study it in a grand canonical ensemble, with the corresponding imaginary-time action $S= \int d^2 r d\tau (\mathcal{L}_0 +\mathcal{L}_1)$,
\begin{eqnarray}
\mathcal{L}_0 &=&  \Phi^\ast \partial_\tau \Phi +J | \nabla^2 \Phi |^2-2J\bar{k}_0^2 |\nabla \Phi |^2 + (\tilde{\varepsilon}_0-\mu)|\Phi |^2
\nonumber\\
\mathcal{L}_1 &=& \frac{U}{2} |\Phi |^4, \label{eq:L_psi}
\end{eqnarray}
$\Phi({\bf r},\tau)$ a complex bosonic field and $\mu$ the chemical potential. We also defined an energy offset $\tilde{\varepsilon}_0=J \bar{k}_0^4 +\varepsilon_0 $ and interaction $U$ parameter inherited from the Hubbard model.

As with the lattice model in Sec.~\ref{sec:model}, here we focus on the helical superfluid state -- a condensate at a single point ${\bf k}_0$ on the dispersion contour minimum -- encoded in $\varepsilon_{\bf k}$, (\ref{eq:dispersion_isotropic}), and $\mathcal{L}_0$ in (\ref{eq:L_psi}). In the density-phase representation, the state is characterized by
\begin{equation}
\Phi ({\bf r}) = \sqrt{n} e^{i {\bf k}_0 \cdot {\bf r} + i\phi} = \sqrt{n_0 +\pi} e^{i {\bf k}_0 \cdot {\bf r} + i\phi} \label{eq:psi_density_phase}
\end{equation}
with a nonzero condensate density (at mean-field level)
$n_0=(\mu-\varepsilon_0)/U$ and momentum $k_0=\bar{k}_0$, for
$\mu>\varepsilon_0$ (vanishing otherwise), and density and phase
fluctuations, $\pi$ and $\phi$, respectively. To quadratic order, the
helical superfluid is then described by a Goldstone mode Lagrangian
density \footnote{ Each term in Eq.~(\ref{eq:L_psi}) can be rewritten
  in terms of $\pi$ and $\phi$ as
\begin{align*}
\partial_\tau \Phi =& \frac{e^{i {\bf k}_0 \cdot {\bf r} + i\phi}}{\sqrt{n}} \biggl( \frac{1}{2}\partial_\tau \pi + i n \partial_\tau \phi  \biggr),
\nonumber\\
\nabla \Phi =& \frac{e^{i {\bf k}_0 \cdot {\bf r} + i\phi}}{\sqrt{n}} \biggl[ \frac{1}{2}\nabla \pi + i n ({\bf k}_0 + \nabla \phi)  \biggr],
\nonumber\\
\nabla^2 \Phi =& \frac{e^{i {\bf k}_0 \cdot {\bf r} + i\phi}}{n^{3/2}} \biggl[\frac{1}{2} n \nabla^2 \pi  -n^2({\bf k}_0 + \nabla \phi)^2 -\frac{1}{4} (\nabla \pi)^2
\nonumber\\
&+i n (\nabla \pi \cdot {\bf k}_0 +\nabla \pi \cdot \nabla \phi + n \nabla^2 \phi)\biggr].
\end{align*}}

\begin{align}
\mathcal{L}'_0 =& i \pi \partial_\tau \phi + J\biggl[ \frac{k_0^2}{n_0}(\partial_\parallel\pi)^2 + \frac{1}{4n_0}(\nabla^2\pi)^2 + 4n_0k_0^2(\partial_\parallel\phi)^2
\nonumber\\
&+ n_0(\nabla^2\phi)^2 + 4k_0(\partial_\parallel\pi)(\nabla^2\phi)\biggr] + \frac{U}{2}\pi^2, \label{eq:eff_th_L}
\end{align}
where the subscripts $\parallel$ and $\perp$ denote axes parallel and
perpendicular to ${\bf k}_0$, respectively. In
Sec.~\ref{sec:order_by_disorder}, we will consider higher-order
fluctuations that renormalize the parameters $n_0$, ${\bf k}_0$ and
reduce the symmetry of the Lagrangian by $C_6$ lattice effects. We
note that the phase $\phi$ is compact with $2\pi$ periodicity, and
thereby allows vortex configurations -- dislocation in the helical
pattern of the state. However, because we are primarily interested in
the ordered helical superfluid at zero temperature, where these
defects are confined, we will neglect them in most of our
analysis. They are however important for the finite temperature
Kosterlitz-Thouless superfluid-to-normal transition in
Fig.~\ref{fig:phase_diagram}(a).

Integrating out $\pi$ field in Eq.~(\ref{eq:eff_th_L}), we obtain the effective theory of $\phi$ described by a quantum smectic Lagrangian (consistent with the honeycomb lattice model analysis in Sec.~\ref{sec:model})
\addtocounter{equation}{0}
\begin{subequations}
\begin{align}
\mathcal{L}_\phi=&~ n_0\phi\left[\frac{-(\partial_\tau + \hat{\mathcal{E}}_2)^2}{\hat{\mathcal{E}}+B_\tau^{-1}} + \hat{\mathcal{E}}\right]\phi\label{eq:L_phi}\\
\approx &~n_0 \left[ B_\tau (\partial_\tau \phi)^2 +  B ( \partial_\parallel \phi )^2 +  K (\partial_\perp^2 \phi)^2 \right], \label{eq:L_phi_0}
\end{align}
\end{subequations}
where Fourier transform of the operators $\hat{\mathcal{E}}$ and $\hat{\mathcal{E}}_2$ are given by
\begin{align}
\mathcal{E}_{\bf q} = B q_\parallel^2 + Kq^4,\quad\mathcal{E}_{2,{\bf q}} = 2\sqrt{BK}q_\parallel q^2, \label{eq:Epsilon_and_Epsilon_2}
\end{align}
with the couplings\footnote{At $T=0$ and weak interaction, we neglect the difference between the boson and condensate densities, $n$ and $n_0$, with $(n-n_0)/n\sim U^{3/4}$, see Eq.~(\ref{eq:eff_th_n_d_T_0}).}
\begin{equation}
B_\tau=\frac{1}{2Un_0},\quad B= 4 J k_0^2,\quad  K = J.  \label{eq:BK}
\end{equation}
The final form in Eq.~(\ref{eq:L_phi_0}) corresponds to the asymptotic long wavelength limit. The superfluid mode $\phi$ exhibits an unusual dispersion given by
\begin{align}
E_{\bf q} =&\ \mathcal{E}_{2,{\bf q}} + \sqrt{\mathcal{E}_{\bf q}^2 + B_\tau^{-1}\mathcal{E}_{\bf q}}
\nonumber\\
\sim &\ \left\lbrace\begin{array}{cc}
         q_\perp^2,\quad q_\parallel\ll\lambda q_\perp^2,\quad q_{\perp}\ll q_{\perp c},\\
         |q_\parallel|,\quad q_\parallel\gg\lambda q_\perp^2,\quad q_{\parallel}\ll q_{\parallel c},
     \end{array} \right.
 \label{eq:dispersioneffective_comp}
\end{align}
consistent with our lattice analysis,  Eq.~(\ref{eq:disp}) and the features illustrated in Fig.~\ref{fig:dispersion_micro}. Two length scales emerge from the zero-sound dispersion (\ref{eq:dispersioneffective_comp}). The crossover scale $\xi=\sqrt{BB_\tau}$ is the coherence length along ${\bf k}_0$, beyond which the dispersion is controlled by interaction. The length $\lambda=\sqrt{K/B}=1/(2k_0)$ characterizes the anisotropy along ($\parallel$) and transverse ($\perp$) to ${\bf k}_0$. The corresponding momentum ${\bf q}_c=(q_{\parallel c},q_{\perp c})\equiv (\xi^{-1}, (\lambda \xi)^{-1/2} )$ separates the noninteracting regime, $E_{\bf q} \approx \Big(\sqrt{B}q_{\parallel}+\sqrt{K}q^2\Big)^2$, for ${\bf q} \gg  {\bf q}_c$ from the long-wavelength interacting form, $E_{\bf q}\approx B_{\tau}^{-1/2}\sqrt{B q_\parallel^2 + Kq^4}$, for ${\bf q} \ll  {\bf q}_c$. The latter agrees with the predictions of the lattice model, Eq.~(\ref{eq:disp_small_q}), if the lattice symmetry breaking effects are neglected.

Importantly, we note that, in the absent of lattice effects,
$(\nabla_\perp \phi)^2$ is absent in the expansion of
Eq.~(\ref{eq:L_phi_0}) at all orders. This is enforced by the
underlying rotational symmetry that is {\em spontaneously} broken by
the helical superfluid state. More explicitly, an infinitesimal shift
of the condensate momentum
${\bf k}_0 \rightarrow {\bf k}_0 + \delta {\bf k}_0$ along the
dispersion minimum contour is a symmetry\footnote{This is closely
  related to the spontaneous breaking of dipole conservation (in
  addition to charge) of Bose condensation, which prohibits linear
  derivative terms, as recently studied in the dipolar Bose-Hubbard
  model by Lake et al. \cite{lake_dipolar_2022} Here, the helical
  superfluid is characterized by an emerged dipole conservation
  perpendicular to ${\bf k}_0$, which however, only holds
  infinitesimally.}. According to (\ref{eq:psi_density_phase}), this
transforms
$\phi_{\bf r}\to\phi_{\bf r}+ \delta {\bf k}_0 \cdot {\bf r}$, which
then enforces the exact vanishing of the coefficient of
$(\nabla_\perp \phi )^2$.

\subsection{Stability\label{sec:stability}}

We now turn to the analysis of the stability of the helical condensate at ${\bf k}_0$ to quantum and thermal fluctuations, characterized by Goldstone mode fluctuations, $\langle \phi^2 \rangle$. The divergence of this fluctuations in the thermodynamic limit is a signature of instability. We generalize our analysis to $d+1$ space-time dimensions, with two ``hard" axes, $\parallel$ and $\tau$, and $d-1$ transverse ``soft" axes. A complementary generalization to ``columnar" Goldstone mode (with $d$ ``hard" and $1$ ``soft" axes) is relegated to Appendix~\ref{app:fluc}. At $d=2$, both generalizations reduce to the physical case of interest, but corresponding to distinct analytical continuation of the physical problem. Below, we employ the low-energy Goldstone mode theory, (\ref{eq:L_phi_0}), valid within a UV cutoff $\Lambda_U$ defined by $\mathcal{E}_{\bf q} < Un_0$.

\subsubsection{Local fluctuations $\langle \phi^2 \rangle$\label{sec:fluc_instability}}

At zero temperature, quantum fluctuations are quantified by
\begin{align}
\langle \phi^2 \rangle_Q &=\frac{1}{2n_0}\int^{\Lambda_U} \frac{d\omega dq_\parallel d^{d-1} q_\perp }{(2\pi)^{d+1}} \frac{1}{ B_\tau \omega^2 +  K q_\perp^4 + B q_\parallel^2 }
\nonumber\\
&= \mathcal{O}(1)\times n_0^{-1}\xi^{-\frac{d+1}{2}}\lambda^{-\frac{d-1}{2}}
,\quad d>1, \label{eq:phi_squ_Q}
\end{align}
and is finite (system size independent) in the thermodynamic limit. This thereby demonstrates the stability of the helical condensate to quantum fluctuations.

At nonzero temperature, thermal fluctuations are dominated by the zeroth Matsubara frequency $\omega_{n=0}$, where $\omega_n=2\pi n /\beta$, with $\beta=1/T$. This gives
\begin{align}
\langle \phi^2 \rangle_T \approx &\ \frac{T}{2n_0}\int_{L^{-1}}^{\Lambda_U}  \frac{dq_{\parallel}d^{d-1}q_{\perp}}{(2\pi)^d} \frac{1}{ K q_\perp^4 + B q_\parallel^2 }
\nonumber\\
\approx&\ \frac{T\langle \phi^2 \rangle_Q}{Un_0}\times\left\lbrace\begin{array}{ccc}
\min\left(\hat{L}_\parallel^{1/2},\hat{L}_\perp\right)^{3-d}, &\quad \ d<3,
\\
\ln\left[\min\left(\hat{L}_\parallel^{1/2},\hat{L}_\perp\right)\right], &\quad \ d=3,
\\
\mathcal{O}(1), &\quad \ d>3,
\end{array}\right.\label{eq:phi_squ_T}
\end{align}
where the dimensionless parameters $\hat{L}_\parallel=L_\parallel/\xi$
and $\hat{L}_\perp=L_\perp/\sqrt{\xi\lambda}$. $L_\parallel$ and
$L_\perp$ are the system size along the $\parallel$ and $\perp$
directions, respectively. We thus observe that for $d\le 3$, (in a
striking contrast to a conventional superfluid) thermal fluctuations
diverge with system size, and thereby destabilize the helical
condensate at any nonzero temperature in the thermodynamic limit. For
the physical case of $d=2$, the length scales for this instability are
set by
$ \xi_\perp^T = \sqrt{\xi_\parallel^T\lambda} \sim
\frac{Un_0}{T\langle \phi^2 \rangle_Q}\sqrt{\xi\lambda}$.

\subsubsection{Two-point correlation function of $\phi$}

Next, we calculate the static two-point correlation function of $\phi({\bf r},\tau)$, in the physical case given by
\begin{align}
C({\bf r})=&\langle[\phi({\bf r},0)-\phi(0,0)]^2\rangle
\nonumber\\
=&\frac{T}{n_0} \sum_{\omega_n} \int^{\Lambda_U}\frac{dq_\parallel dq_\perp}{(2\pi)^2} \frac{1-e^{i{\bf q}\cdot {\bf r}}}{B_\tau\omega_n^2 + B q_\parallel^2 + K q_\perp^4}.
\label{eq:C(r)}
\end{align}

We are interested in its asymptotics in the high-temperature classical
and low-temperature quantum limits.  At $T=0$, the Matsubara converts
into an integral over a continuous frequency $\omega$, giving
\begin{align}
C_Q({\bf r})=&\ U\int^{\Lambda_U}\frac{dq_\parallel dq_\perp}{(2\pi)^2}\frac{1-e^{i{\bf q}\cdot {\bf r}}}{\sqrt{2Un_0(B q_\parallel^2 + Kq_\perp^4)}}
\nonumber\\
\approx &\ n_0^{-1}\xi^{-3/2}\lambda^{-1/2}
, \quad \quad \ r\gg\Lambda^{-1},
\label{eq:C(r)_quantum}
\end{align}
growing quadratically to a finite asymptote $\langle \phi^2\rangle_Q$, (\ref{eq:phi_squ_Q}), at $d=2$.

In the high-temperature limit, only $\omega_n=0$ contributes to the correlation function, giving
\begin{align}
C_T({\bf r})=&\frac{2U}{\beta}\int\frac{dq_\parallel dq_\perp}{(2\pi)^2}\frac{1-e^{i{\bf q}\cdot {\bf r}}}{2Un_0(B q_\parallel^2 + Kq_\perp^4)}
\nonumber\\
=& \frac{T}{2Un_0^2\xi^2}\left[\left(\frac{r_\parallel}{4\pi\lambda}\right)^{1/2}e^{-\frac{r_\perp^2}{4\lambda r_\parallel}}+\frac{r_\perp}{4\lambda}\text{erf}\left(\frac{r_\perp}{\sqrt{4\lambda r_\parallel}}\right)\right]
\nonumber\\
\approx&\left\lbrace\begin{array}{cc}
\frac{T\xi^{-3/2}\lambda^{-1/2}}{4Un_0^2}\left(\frac{r_\parallel}{\pi\xi}\right)^{1/2}, &\quad r_\parallel\gg\frac{r_\perp^2}{\lambda},\ r\gg\Lambda_U^{-1},
\\
\frac{T\xi^{-3/2}\lambda^{-1/2}}{8Un_0^2}\left(\frac{r_\perp^2}{\xi\lambda}\right)^{1/2}, &\quad r_\parallel\ll\frac{r_\perp^2}{\lambda},\ r\gg\Lambda_U^{-1},
\end{array}\right.\label{eq:C(r)_thermal}
\end{align}
where $\text{erf}(x)$ is the error function \cite{toner_smectic_1981,radzihovsky_fluctuations_2011}.

The above analysis indicates a crossover between the low and high temperature limits of $C({\bf r})$. To this end, we perform Matsubara summation in Eq.~(\ref{eq:C(r)}) using
\begin{equation}
\frac{1}{\beta}\sum_{\omega_n}\frac{e^{i\omega_n\tau}}{i\omega_n-\epsilon}=-\frac{e^{\tau\epsilon}}{e^{\beta\epsilon}-1},\quad \textrm{for }\tau\ge 0, \label{eq:matsubara_sum}
\end{equation}
and define a crossover scale
\begin{eqnarray}
{\bf q}_T = (q_{\parallel T},q_{\perp T}) = \left(\frac{T}{Un_0}q_{\parallel c},\sqrt{\frac{T}{Un_0}}q_{\perp c}\right),
\label{eq:crossover_q}
\end{eqnarray}
separating the quantum (${\bf q}>{\bf q}_T$) and classical (${\bf q}<{\bf q}_T$) regions of momenta.
This gives
\begin{align}
C({\bf r})=&\ U\left(\int^{q_{T}}+\int_{q_{T}}^{\Lambda_U}\right)\frac{1-e^{i{\bf q}\cdot {\bf r}}}{E_{\bf q}}\coth(\beta E_{\bf q}/2)
\nonumber\\
\approx &\ \frac{2U}{\beta}\int^{q_{T}}\frac{1-e^{i{\bf q}\cdot {\bf r}}}{E_{\bf q}^2} + U\int_{q_{T}}^{\Lambda_U}\frac{1-e^{i{\bf q}\cdot {\bf r}}}{E_{\bf q}},
\end{align}
where $E_{\bf q} = \sqrt{B_\tau^{-1}(B q_\parallel^2 + Kq_\perp^4)}$. 
At low temperature, $\beta \rightarrow \infty$ and $q_T \rightarrow 0$, the quantum, second term dominates, giving $C({\bf r}) \approx C_Q({\bf r})$. While at high temperature, $\beta \rightarrow 0$ and $q_T \rightarrow \Lambda_U$, $C({\bf r})$ is dominated by the classical, first term, giving $C({\bf r}) \approx C_T({\bf r})$.

\begin{figure*}[t]
\includegraphics[width=.65\textwidth]{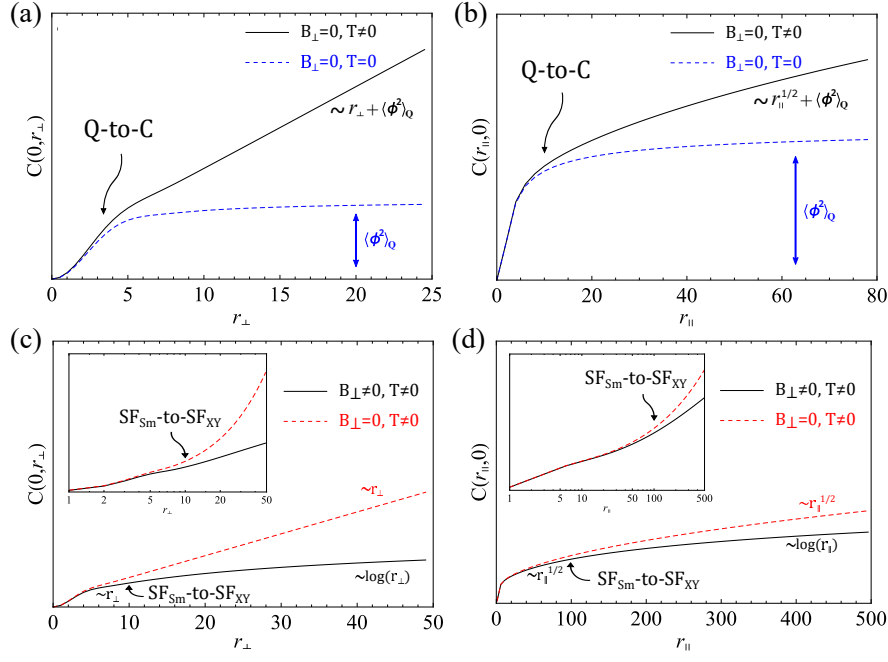}
\caption{Numerical plots ($r_\parallel$ and $r_\perp$ in units of $\xi$ and $\sqrt{\xi\lambda}$, respectively) of the correlation function (\ref{eq:C(r)}) (a) along $r_\perp$ with $r_\parallel=0$ and (b) along $r_\parallel$ with $r_\perp=0$. The helical superfluid exhibits a quantum-to-classical (Q-to-C) crossover around $r_\perp=\sqrt{\xi\lambda}\sqrt{Un_0/T}$ and $r_\parallel=\xi Un_0/T$ with $Un_0/T=10$, where the finite-temperature (black-solid) curves deviate from the zero-temperature (blue-dashed) curves. For $B_\perp\neq 0$, the correlation function (c) along $r_\perp$ with $r_\parallel=0$ and (d) along $r_\parallel$ with $r_\perp=0$ (black curves) shows a crossover from smectic superfluidity ($\text{SF}_\text{Sm}$) to conventional XY superfluidity ($\text{SF}_\text{XY}$) around $r_\perp=\lambda_\perp$ and $r_\parallel=\lambda_\perp^2/\xi$ with $\lambda_\perp=10\xi$. The red-dashed curves show the case $B_\perp=0$ for comparison. Inset of (c)(d): logarithmic scale in ${\bf r}$ axes. \label{fig:QC_crossover}}
\end{figure*}

In Fig.~\ref{fig:QC_crossover}, we plot $C({\bf r})$ in (\ref{eq:C(r)}), computed numerically along $r_\perp$ and $r_\parallel$, showing its asymptotic behaviors in the low-temperature (\ref{eq:C(r)_quantum}) and high-temperature (\ref{eq:C(r)_thermal}) regimes. The crossover scale in real space is given by the inverse of the thermal wavevectors,
\begin{align}
r_{\parallel T}=q_{\parallel T}^{-1}=\frac{Un_0}{T}\xi,\quad r_{\perp T}=q_{\perp T}^{-1}=\sqrt{\frac{Un_0}{T}}\sqrt{\lambda\xi}. \label{eq:r_T}
\end{align}

\subsubsection{Order-by-disorder}
Although the effective theory introduced in this section is unstable
to thermal fluctuations for $d \le 3$, we anticipate that the presence
of the underlying honeycomb lattice explicitly breaks the degenerate
contour in Eq.~(\ref{eq:eff_th_free_dispersion}) down to six-fold
minima through quantum and thermal fluctuations, that probe all
momentum scales (detailed in Sec.~\ref{sec:order_by_disorder}). This
in turn introduces perpendicular stiffness
$B_\perp(\nabla_\perp\phi)^2$, modifying the low-energy Lagrangian to
\begin{align}
    \frac{\mathcal{L}_\phi}{n_0} = B_\tau (\partial_\tau \phi)^2 +  B ( \partial_\parallel \phi )^2 + B_\perp ( \partial_\perp \phi )^2 +  K (\partial_\perp^2 \phi)^2,
\end{align}
with suppressed fluctuations. As discussed in
Sec.~\ref{sec:order_by_disorder}, $B_\perp \propto U^{5/4}$
($B_\perp \propto U^{1/4}T$) for $T\ll Un_0$ ($T\gg Un_0$) in the
weakly-interacting limit $Un_0\ll t_1$. Importantly, this
leads to a crossover length scale
$\lambda_\perp=\sqrt{K/B_\perp}$. Consequently, as illustrated in
Fig.~\ref{fig:QC_crossover}(c)(d), the helical superfluid exhibits an
extended crossover from a smectic regime to an anisotropic XY regime
(but with the condensate at finite momentum), separated by
$(r_\parallel^*,r_\perp^*)=
(\lambda_\perp^2/\xi,\lambda_\perp)$. Since $B_\perp$ is
perturbatively small in $U$, $\lambda_\perp$ can be very large,
exhibiting long range of anomalous smectic superfluidity, that we
study in detail below.

\subsection{Physical observables in smectic superfluid regime}\label{sec:phy_obs}

\subsubsection{Structure factor}\label{sec:Structure_fac}

To calculate the structure factor, we employ the Lagrangian (\ref{eq:eff_th_L}). In Fourier space, it is given by
\begin{equation}
\mathcal{L}'_0 = \frac{1}{2} \biggl( \begin{array}{cc}
\phi_{-{\bf q},-\omega_n} & \pi_{-{\bf q},-\omega_n} 
\end{array} \biggr) D^{-1}({\bf q},\omega_n) \biggl(\begin{array}{c}
\phi_{{\bf q},\omega_n} \\ \pi_{{\bf q},\omega_n}
\end{array}\biggr), \label{eq:eff_th_S}
\end{equation}
where
\begin{align}
D({\bf q},\omega_n) =&\ \frac{1}{(\omega_n+i\mathcal{E}_{2,{\bf q}})^2 + \mathcal{E}_{\bf q}^2 +2Un_0\mathcal{E}_{\bf q}}
\nonumber\\
&\times \biggl( \begin{array}{cc}
 \frac{1}{2n_0}\mathcal{E}_{\bf q} + U & \omega_n + i\mathcal{E}_{2,{\bf q}} \\
-\omega_n - i\mathcal{E}_{2,{\bf q}} & 2n_0\mathcal{E}_{\bf q}
\end{array}\biggr) \label{eq:eff_th_D}
\end{align}
with $\mathcal{E}_{\bf q}$ and $\mathcal{E}_{2,{\bf q}}$ defined in Eq.~(\ref{eq:Epsilon_and_Epsilon_2}).

The structure factor is the density-density correlation function, in the Fourier space given by
\begin{equation}
S({\bf q},\omega_n) = D_{22} ({\bf q},\omega_n)=\frac{2n_0\mathcal{E}_{\bf q}}{(\omega_n+i\mathcal{E}_{2,{\bf q}})^2 + \mathcal{E}_{\bf q}^2 +2Un_0\mathcal{E}_{\bf q}}.
\end{equation}
This static structure factor is then given by
\begin{align}
S({\bf q}) &= \frac{1}{\beta} \sum_{\omega_n} S({\bf q},\omega_n) 
\nonumber\\
&= \frac{n_0\mathcal{E}_{\bf q}}{E_{\bf q}-\mathcal{E}_{2,{\bf q}}}\frac{\sinh[\beta(E_{\bf q}-\mathcal{E}_{2,{\bf q}})]}{\cosh[\beta(E_{\bf q}-\mathcal{E}_{2,{\bf q}})]-\cosh(\beta\mathcal{E}_{2,{\bf q}})},
\label{eq:static_S}
\end{align}
where $E_{\bf q}$ is defined in Eq.~(\ref{eq:dispersioneffective_comp}) and we used Eq.~(\ref{eq:matsubara_sum}) to carry out the Matsubara sum.

\begin{figure*}[t]
\centering
\includegraphics[width=\textwidth]{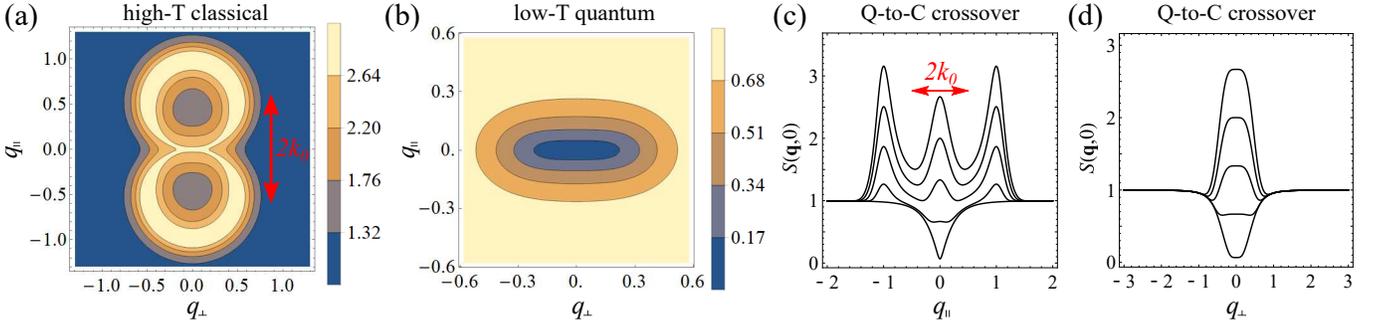}
\caption{Static structure factor (\ref{eq:static_S}) (in unit of $n_0$) as a function of ${\bf q}$ (in the unit of $\lambda^{-1}$) with $\lambda/\xi=0.3$. (a) High-temperature classical limit with $q_{\parallel T}\lambda=0.4$. (b) Low-temperature quantum limit $q_{\parallel T}\lambda=0.01$. Quantum-to-classical crossover (c) along $q_\parallel$ with $q_\perp=0$ and (d) along $q_\perp$ with $q_\parallel=0$. From top to bottom the temperature decreases with $q_{\parallel T}\lambda=0.4,0.3,0.2,0.1,0.01$. The red double-headed arrows in (a) and (c) indicate the scale of condensate momentum $2k_0=\lambda^{-1}$.}\label{fig:static_S}
\end{figure*}

The structure factor (\ref{eq:static_S}) depends on the a number of scales: coherence length $\xi=\sqrt{BB_\tau}$, anisotropic length factor $\lambda=\sqrt{K/B}$ and parallel thermal wavevector $q_{\parallel T}=(2T^2/Un_0 B)^{1/2}$. Below we first present the results in the low- and high-temperature limits, and then discuss the crossover between them.

In the high-temperature limit, $\beta\to 0$, the structure factor becomes
\begin{align}\label{eq:S_T}
S_T({\bf q}) =&\ \frac{2n_0 T\mathcal{E}_{\bf q}}{2Un_0\mathcal{E}_{\bf q} + \mathcal{E}_{\bf q}^2-\mathcal{E}_{2,{\bf q}}^2}
\nonumber\\
\approx&\ \biggl\{ \begin{array}{cc}
T/U,&\quad q_\parallel\xi,\ q_\perp\sqrt{\lambda\xi}\ll 1,
\\
0,&\quad q_\parallel\xi,\ q_\perp\sqrt{\lambda\xi}\gg 1,
\end{array}
\end{align}
exhibiting a maximum on a closed contour (inherited from the noninteracting dispersion minimum) with a height $T/U$ and a width increasing with $U$, illustrated at high temperature in Fig.~\ref{fig:static_S}(a).

In the complementary low-temperature limit, $\beta\to\infty$, the structure factor becomes
\begin{align}
S_Q({\bf q}) =&\ n_0\sqrt{\frac{\mathcal{E}_{\bf q}}{\mathcal{E}_{\bf q}+2Un_0}}
\nonumber\\
\approx&\ \biggl\{ \begin{array}{cc}
\sqrt{n_0\mathcal{E}_{\bf q}/2U},&\quad q_\parallel\xi,\ q_\perp\sqrt{\lambda\xi}\ll 1,
\\
n_0,&\quad q_\parallel\xi,\ q_\perp\sqrt{\lambda\xi}\gg 1,
\end{array}
\end{align}
in contrast exhibiting a single minimum at ${\bf q}=0$, which becomes more shallow with increases $U$, as shown in Fig.~\ref{fig:static_S}(b). It can be written as a scaling form
\begin{equation}
S_Q({\bf q},0) \sim n_0(\xi q_\parallel)^\alpha \hat{S} \biggl(\lambda\frac{q^2}{q_\parallel}\biggr)
\end{equation}
in the limits, $\xi q_\parallel \ll 1$ and $\xi q_\parallel \gg 1$, with
\begin{align}
    \hat{S} (x) &= \sqrt{1+x^2},\quad \alpha = 1,\quad &\text{for }\xi q_\parallel \ll 1,
    \nonumber\\
    &\hat{S} (x) = 1,\quad \alpha=0,\quad &\text{for }\xi q_\parallel \gg 1.
\end{align}

The structure factor exhibits a quantum-to-classical crossover for ${\bf q}$ separated by the thermal wavevectors ${\bf q}_T$ in Eq.~(\ref{eq:crossover_q}), as illustrated in Fig.~\ref{fig:static_S}(c)(d). 

\subsubsection{Condensate depletion}\label{sec:depletion}
The condensate depletion is given by
\begin{align}
n_d =&\ n - n_0 
= \langle\Phi({\bf r},0){\Phi}^\ast({\bf r},\eta)\rangle - n_0 \nonumber\\
=&\ \frac{1}{\beta}\int\frac{d^2q}{(2\pi)^2}\sum_{\omega_n}e^{-i\omega_n\eta}G({\bf q},\omega_n), \label{eq:eff_th_n_d_def}
\end{align}
where $\eta=0^{+}$ is an infinitesimal positive number to encode operator time ordering and with $\Phi$ in Eq.~(\ref{eq:psi_density_phase}) expanded up to quadratic order in $\pi$ and $\phi$,
\begin{align}
G({\bf q},\omega_n)
\approx &\ \frac{1}{4n_0}D_{22}({\bf q},\omega_n)+n_0 D_{11}({\bf q},\omega_n)
\nonumber\\
&+\frac{i}{2}D_{12}({\bf q},\omega_n)-\frac{i}{2}D_{21}({\bf q},\omega_n)
\nonumber\\
=&\ \frac{u_{\bf q}^2}{-i\omega_n+E_{\bf q}} + \frac{v_{\bf q}^2}{i\omega_n+E_{-\bf q}} \label{eq:G_q}
\end{align}
with matrix $D$ given in Eq.~(\ref{eq:eff_th_D}), $E_{\bf q}$ given in Eq.~(\ref{eq:dispersioneffective_comp}) and
\begin{align}
u_{\bf q}^2 =&\frac{1}{2}\biggl(\frac{\mathcal{E}_{\bf q} + Un_0}{\sqrt{\mathcal{E}_{\bf q}^2+2Un_0\mathcal{E}_{\bf q}}}+1\biggr),
\nonumber\\
v_{\bf q}^2 =&\frac{1}{2}\biggl(\frac{\mathcal{E}_{\bf q} + Un_0}{\sqrt{\mathcal{E}_{\bf q}^2+2Un_0\mathcal{E}_{\bf q}}}-1\biggr). \label{eq:eff_th_basis_coef_u_v}
\end{align}
After performing the Matsubara summation in Eq.~(\ref{eq:eff_th_n_d_def}) and using the identity (\ref{eq:matsubara_sum}),
\begin{align}
n_d = & \int\frac{d^2q}{(2\pi)^2}\biggl(v_{\bf q}^2 + \frac{u_{\bf q}^2 + v_{\bf q}^2}{e^{\beta E_{\bf q}} - 1} \biggr), \label{eq:eff_th_n_d}
\end{align}
where the first term gives the zero-temperature interaction driven depletion while the second term describes additional depletion due to thermal excitations.

At zero temperature, the depletion (with details relegated to Appendix~\ref{app:depletion}) is given by
\begin{align}
n_d(T=0) =&\ \int\frac{d^2q}{(2\pi)^2}v_{\bf q}^2 = \mathcal{C}n\biggl(\frac{U^3}{n_0 B^2 K}\biggr)^{1/4}
\nonumber\\
\sim &\ nU^{3/4}n_0^{-1/4}\sim (Un_0)^{3/4},\footnotemark{} \label{eq:eff_th_n_d_T_0}
\end{align}
where $\mathcal{C}=2^{3/4}I_\theta |f_e(0)|/(96\pi^2) \approx 0.035$ with $I_\theta\approx 10.4882$ and $f_e(0) \approx -3.70815$.

\addtocounter{footnote}{-1}
\footnotetext{In the weakly interacting limit, we neglect the difference between $n$ and $n_0$ as it is of higher order in $U$.}

At nonzero temperature, the depletion in Eq.~(\ref{eq:eff_th_n_d})
diverges in the thermodynamic limit for $d\leq 3$, as analyzed in
Appendix~\ref{app:depletion}, indicating helical superfluid thermal
instability as found in Sec.~\ref{sec:fluc_instability}.

For the microscopic lattice model discussed in Sec.~\ref{sec:model}, the depleting is still given by (\ref{eq:eff_th_n_d}), but with $u_{\bf q}$ and $v_{\bf q}$ given by
\begin{equation}
\begin{aligned}
u_{{\bf q}}^2 &= \frac{1}{2} \left( \frac{\mathcal{E}_{{\bf q}}+Un}{\sqrt{\mathcal{E}^2_{{\bf q}} +2Un \mathcal{E}_{{\bf q}} + U^2n^2\sin^2(\Theta_{\bf q})}}+1\right), \\ 
v_{{\bf q}}^2 &= \frac{1}{2} \left( \frac{\mathcal{E}_{{\bf q}}+Un }{\sqrt{\mathcal{E}^2_{{\bf q}} +2Un \mathcal{E}_{{\bf q}} + U^2n^2\sin^2(\Theta_{\bf q})}}-1\right),
\end{aligned} \label{eq:basis_coef_u_v}
\end{equation}
where the lattice effects are encoded in the periodic (in momentum space) forms of $\Theta_{\bf q}$ and $\mathcal{E}_{\bf q}$ defined in Eqs.~(\ref{eq:defOmega}) and~(\ref{eq:disp_Epsilon_1_and_2}).

\begin{figure}
\includegraphics[width=.4\textwidth]{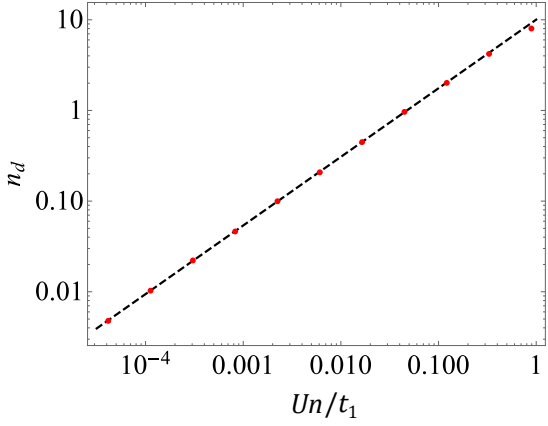}
\caption{The log-log plot of the zero temperature depletion $n_d$ (in unit of $a^{-2}$) in Eq.~(\ref{eq:eff_th_n_d_T_0}) as a function of $Un/t_1$ for $\rho=0.2$, and ${\bf k}_0 = (0,k_{0V})$. The red dots are calculated with the lattice model expression (\ref{eq:basis_coef_u_v}). The dashed line gives a fitting in the small $Un$ region with the predicted slope $=3/4$. \label{fig:depletion}}
\end{figure}

At zero temperature, the depletion as a function of $Un$ is computed numerically and presented in Fig.~\ref{fig:depletion}. At small $Un$, it is a power-law as found in Eq.~(\ref{eq:eff_th_n_d_T_0}). As $Un$ increases, $n_d$ deviates from this perfect power-law given by a weakly interacting field theory. At nonzero temperature, thermal fluctuations still destroy the condensate, as discussed above.

\subsubsection{Momentum distribution}

Another important characterization of the helical superfluid is the momentum distribution, $n_{\bf q} = \langle d^\dag_{{\bf k}_0+{\bf q},-}d_{{\bf k}_0+{\bf q},-} \rangle$. At $T=0$, it is given by
\begin{equation}
n_{{\bf q}} = |v_{{\bf q}}|^2 = \frac{1}{2} \biggl( \frac{1+\frac{U n_0}{\mathcal{E}_{{\bf q}}}}{\sqrt{1+\frac{2Un_0}{\mathcal{E}_{{\bf q}}}}} -1 \biggr) .
\end{equation}
The behavior of $n_{\bf q}$ below ($\xi q_\parallel\gg 1$) and beyond ($\xi q_\parallel\ll 1$) the coherence length $\xi =\sqrt{BB_\tau}$ is described by the following scaling form,
\begin{equation}
n_{{\bf q}} \sim (\xi q_\parallel)^\alpha f_n\biggl(\lambda\frac{q_\perp^2}{q_\parallel}\biggr), \label{eq:nqscalingform}
\end{equation}
with the anisotropy length scale given by $\lambda =\sqrt{K/B}$ and
\begin{align}\label{eq:fn}
    f_n(x)=\frac{1}{\sqrt{1+x^2}},\quad \alpha =&\ -1,\quad \text{for }\xi q_\parallel \ll 1,
    \nonumber\\
    f_n(x)=\frac{1}{(1+x^2)^2},\quad \alpha =&\ -4,\quad \text{for }\xi q_\parallel \gg 1.
\end{align}

We note that for $\lambda q_\perp^2/q_\parallel\ll 1$, $n_{\bf q} \sim (\xi q_{\parallel})^\alpha$. In the opposite limit, $\lambda q_\perp^2/q_\parallel \gg 1$, the momentum distribution is asymptotically given by
\begin{align}
n_{{\bf q}}\sim \biggl\{\begin{array}{cc} \vspace{1mm}
1/q_\perp^2, & \textrm{ for } q_\parallel\xi \ll 1.\\ 
q_\parallel^3/q_\perp^8, & \textrm{ for } q_\parallel\xi \gg 1.
\end{array}
\end{align}
This contrasts qualitatively with that of a conventional superfluid (see Table~\ref{tab:SF_helicalSF}).

\begin{figure}[h]
\includegraphics[width=.4\textwidth]{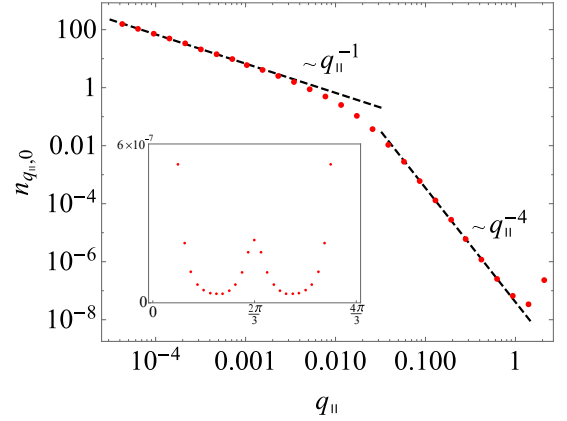}
\caption{Momentum distribution $n_{q_\parallel,q_\perp=0}$ vs. $q_\parallel$ (in unit of $a^{-1}$) for $\rho=0.2$, $Un/t_1=10^{-4}$, and ${\bf k}_0 = (0,k_{0V})$. The red dots are calculated from lattice model expression (\ref{eq:basis_coef_u_v}). The dashed line gives the asymptotic behavior for $q_\parallel\xi\ll 1$ and $q_\parallel\xi\gg 1$. Inset: linear scale of the same plot. \label{fig:n_q}}
\end{figure}

Lattice effects beyond this field theoretical treatment are straightforwardly incorporated with $n_{\bf q}=v_{\bf q}^2$ at $T=0$, with the lattice form of $v_{\bf q}$ given by Eq.~(\ref{eq:basis_coef_u_v}). In Fig.~\ref{fig:n_q}, we plot $n_{q_\parallel,q_\perp=0}$ as a function of $q_{\parallel}$ which verifies the scaling behavior $n_{q_\parallel,q_\perp=0}\sim 1/q_\parallel$ ($1/q_\parallel^4$) for $q_\parallel\xi\ll 1$ ($q_\parallel\xi\gg 1$). Note, somewhat surprisingly the ($T=0$) momentum distribution is symmetric under the inversion of momentum $n_{\bf q}=n_{-{\bf q}}$, and its profile has the reflection symmetry, e.g. with respect to $q_\parallel=2\pi/3$, due to the underlying lattice structure, as shown in the inset of Fig.~\ref{fig:n_q}. As a result, the $1/q_\parallel^4$ tail is only visible for sufficiently small $Un$.

\subsubsection{Superflow}
As in a conventional superfluid a spatial gradient of $\phi$ gives rise to a supercurrent,
\begin{align}
{\bf j}_s = n {\bf v}_s
\end{align}
characterized by the anisotropic superfluid velocity with its form obtained from the Noether's theorem or, equivalently by gauging (see Appendix~\ref{app:superflow})
\begin{align}
{\bf v}_s = 4J\left[(\nabla\phi)^2+2{\bf k}_0\cdot\nabla\phi\right]({\bf k}_0+\nabla\phi)-2J\nabla^3\phi.\label{eq:v_s}
\end{align}
We emphasize that in ${\bf v}_s$ of (\ref{eq:v_s}) we included higher-order terms, previously left out in the quadratic Lagrangian, (\ref{eq:eff_th_L}). For the simplest uniform gradient configuration $\phi={\bf q}\cdot {\bf r}$, we have
\begin{align}
{\bf v}_s(\phi={\bf q}\cdot {\bf r}) =& 4J\left[q^2+2{\bf k}_0\cdot {\bf q}\right]({\bf k}_0+{\bf q})
\nonumber\\
=& \nabla_k \varepsilon_{\bf k}|_{{\bf k}={\bf k}_0+{\bf q}}
\end{align}
with
\begin{equation}
\varepsilon_{\bf k} = J(k^2-k_0^2)^2 + \varepsilon_0,\quad {\bf k}={\bf k}_0+{\bf q}.
\end{equation}

In the long-wavelength limit to linear order, the supercurrent is given by
\begin{align}
(j_s)_i \underset{\nabla\phi\to 0}{=} \sum_{j}(\rho_s)_{ij} \nabla_j\phi
\end{align}
with the superfluid stiffness $(\rho_s)_{ij}=8Jn(k_0)_i (k_0)_j$, which vanishes transversely to ${\bf k}_0$. This helical condensate thus (in the absent of stabilizing lattice effects) cannot maintain a superflow perpendicular to ${\bf k}_0$, and thus by this criterion is not a superfluid.

\subsubsection{Chemical potential}

\begin{figure}
\includegraphics[width=0.4\textwidth]{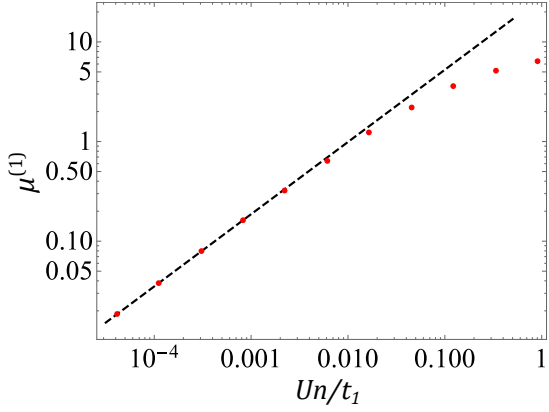}
\caption{$\mu^{(1)}$ (in unit of $Ua^{-2}$) vs. $Un/t_1$ for $\rho=0.2$, and ${\bf k}_0 = (0,k_{0V})$. The red dots are calculated from (\ref{eq:mu}). The dashed line gives a fitting to (\ref{eq:mu_approx}) in the small $Un$ region with a slope $=3/4$. \label{fig:mu}}
\end{figure}

We next calculate the equation of state, given by the chemical
potential as a function of the boson density $n$,
\begin{align}
    \mu=\frac{\partial E_{gs}}{\partial N} = \mu^{(0)} + \mu^{(1)},
\end{align}
where ground state energy $E_{gs}$ is given by
Eq.~(\ref{eq:E_gs}). The lowest-order mean-field contribution and the
one-loop correction are respectively given by
\begin{align}\label{eq:mu}
\mu^{(0)} =&\ \epsilon^{-}_{{\bf k}_0} + Un,
\nonumber\\
\mu^{(1)} =&\ - \frac{U}{4} \int_{\bf q} \left( 1-\frac{\mathcal{E}_{{\bf q}} + Un(\sin \Phi_{\bf q})^2}{\sqrt{\mathcal{E}^2_{{\bf q}} + 2Un \mathcal{E}_{\bf q} + (Un\sin \Phi_{\bf q})^2}}\right)
\nonumber\\
\approx &\ - \frac{U}{4} \int_{\bf q} \left( 1-\frac{\mathcal{E}_{{\bf q}}}{\sqrt{\mathcal{E}^2_{{\bf q}} + 2Un \mathcal{E}_{\bf q}}}\right),
\end{align}
where in the last line we neglected lattice effects and took $\mathcal{E}_{{\bf q}} \approx B q_\parallel^2 + K q_\perp^4$.

Performing the integration (see Appendix~\ref{app:chemical potential}), we obtain
\begin{equation}
\mu^{(1)} \approx - Un\mathcal{C}\biggl( \frac{U^3 }{n B^{2} K} \biggr)^{1/4}, \label{eq:mu_approx}
\end{equation}
where constants $\mathcal{C} = 2^{3/4}I_\theta I_\sigma/(32\pi^2) \approx 0.069$, $I_\sigma \approx 1.23605$ and $I_\theta \approx 10.4882$. The dimensionless factor in $\mu^{(1)}$ expressed in terms of microscopic parameters is given by
\begin{align}
\biggl( \frac{U^3}{n B^{2} K} \biggr)^{1/4}\propto n^{-1/4}U^{3/4}t_1^{-3/4}\bar{k}_0^{-1},\label{eq:dimensionless_microscopic}
\end{align}
where $\bar{k}_0(\rho)$ is given by Eq.~(\ref{eq:parameters_isotropic}) in the limit of $\rho\to 1/6$.
For a general $d$ dimensional helical superfluid, this dimensionless factor is given by
\begin{equation}
\mathcal{Q}_s= \biggl(\frac{U^{d+1} n^{d-3} }{B^{2}K^{d-1}}\biggr)^{1/4}.
\label{eq:dimensionless}
\end{equation}

In Fig.~\ref{fig:mu}, we plot an exact numerical evaluation of the chemical potential in Eq.~(\ref{eq:mu}) with lattice effects as a function of $Un$. At small $Un$, Eq.~(\ref{eq:mu_approx}) fits the data well. We note, as indicated in Eq.~(\ref{eq:mu_approx}), $\mu/U$ decreases (increases) if $U$($n$) increases, in qualitative distinction from a conventional superfluid (see Table~\ref{tab:SF_helicalSF}).

\section{Order-by-disorder \label{sec:order_by_disorder}}
So far, we studied the helical condensation at a single momentum ${\bf k}_0$ on the dispersion minimum contour at the Bogoliubov level. At this order, it exhibits smectic fluctuations that are qualitatively larger than that of conventional XY superfluids due to the macroscopic degenerate ground states. In this section, we go beyond the quadratic action, and study the order-by-disorder phenomenon that appears due to the interplay of quantum and thermal fluctuations, and the $\text{C}_\text{6}$ lattice effects. This generates a nonzero $B_\perp$ that stabilizes the helical superfluid even at nonzero temperature.

Formally, we work with the coherent state path integral of the lattice model introduced in Sec.~\ref{sec:model}, where the partition function is given by $Z=\int\mathcal{D}\Phi e^{-S\left[\Phi\right]}$, with the imaginary-time action $S=\int d\tau L$ and the corresponding Lagrangian
\begin{eqnarray}
L = \sum_{i,a=1,2} \Phi_{i,a}^* \partial_\tau \Phi_{i,a} + H_0+H_{\text{int}}.
\end{eqnarray}
In the above, $H_{0}$ and $H_{\text{int}}$ are functionals that respectively have the same expressions as the kinetic and the interacting parts of the Hamiltonian introduced in Sec.~\ref{sec:model} with the operators $a_{i,a}$ and $a_{i,a}^\dag$ replaced by coherent state fields $\Phi_{i,a}$ and $\Phi_{i,a}^*$. To study the helical superfluid, we consider the density-phase fluctuations around a mean-field of a helical condensate
with density $n_0$ and momentum ${\bf k}_0$:
\begin{align}\label{eq:pi_phi_rep_lattice}
    \Phi_{i,1} =&\ e^{i\theta_{{\bf k}_0}/2}~e^{i{\bf k}_0\cdot {\bf r}_{i,1} + i \phi_{i,1}}~\sqrt{n_0+\pi_{i,1}},
    \nonumber\\
    \Phi_{i,2} =&\ e^{-i\theta_{{\bf k}_0}/2}~e^{i{\bf k}_0\cdot {\bf r}_{i,2} + i\phi_{i,2}}~\sqrt{n_0+\pi_{i,2}},
\end{align}
where ${\bf r}_{i,a}$ is the position of the lattice site $i$ in $a$ sublattice. Below, we drop the subscripts $i$ and $a$ for notation simplicity. Then, the action can be written as $S[\Phi]=\beta\Omega^{(0)}+\int_{\tau}L^{(0)}$ with the constant mean-field part $\Omega^{(0)}(n_0,{\bf k}_0)$ (zeroth-order thermodynamic potential) and the bare Lagrangian, $L^{(0)}(\pi,\phi)$. Such a procedure parallels the field theoretical approach in Sec.~\ref{sec:eff_th}, but now encodes the underlying $\text{C}_\text{6}$ lattice effects. Analysis of the quadratic part of the action reproduces the Bogoliubov results, predicted in Sec.~\ref{sec:model}. To study the effects of the higher-order nonlinearalities, various approximation schemes can be employed. Here we safely employ an analysis perturbative in $Un_0$, self-consistently verifying the corrections to $B_\perp$ are finite. We thereby formally obtain
\begin{align}
    Z =&\ \int\mathcal{D}\pi\mathcal{D}\phi e^{-\beta\Omega^{(0)}-\int_{\tau}L^{(0)}} = \int\mathcal{D}\pi\mathcal{D}\phi e^{-\beta\Omega-\int_{\tau}L},
\end{align}
where the renormalized thermodynamic potential and Lagrangian are given by perturbation series $\Omega(n_0,{\bf k}_0)=\sum_{n=0}^{\infty}\Omega^{(n)}$ and $L=\sum_{n=0}^{\infty}L^{(n)}$ [``$(n)$" stands for the nth loop correction], respectively. Similar to a conventional superfluid, global U(1) symmetry $\phi\to\phi + \alpha$ ensures that $\phi$ is massless, i.e., in the continuum, only gradients of $\phi$ appears in the Lagrangian. For a helical condensate, another constraint is the equivalence of the transformations ${\bf k}_0\to {\bf k}_0+{\bf q}$ and $\phi\to\phi+{\bf q}\cdot {\bf r}$ in the long wavelength limit. The former transforms the free energy,
\begin{align}
    \Omega({\bf k}_0+{\bf q})-\Omega({\bf k}_0) = N_0\sum_{lm}b_{lm} q_\parallel^l q_\perp^m.
\end{align}
The latter, when applied to the continuum Lagrangian density (discussed in detail in Appendix~\ref{app:C6_effects}),
\begin{align}
    \mathcal{L} = n_0\sum_{lm}B_{lm}(\partial_\parallel\phi)^l(\partial_\perp\phi)^m + ...,
\end{align}
gives an energy shift $\sum_{lm}B_{lm} q_\parallel^l q_\perp^m$. In the above, the ellipsis denotes $\pi$ and higher derivative $\phi$ terms. Since ${\bf q}$ can be chosen arbitrarily, we conclude that
\begin{align}
    b_{lm}=B_{lm},
\end{align}
for any integers $l,m\ge 0$, holding in every nth order of the perturbation theory, $b_{lm}^{(n)}=B_{lm}^{(n)}$. Below, we study the order-by-disorder phenomenon by performing one-loop calculations of $b_\perp^{(1)}=b_{02}^{(1)}$ and $B_\perp^{(1)}=B_{02}^{(1)}$ for both zero and nonzero temperatures. 

\subsection{Thermodynamic potential}

\begin{figure}[h]
\includegraphics[width=.4\textwidth]{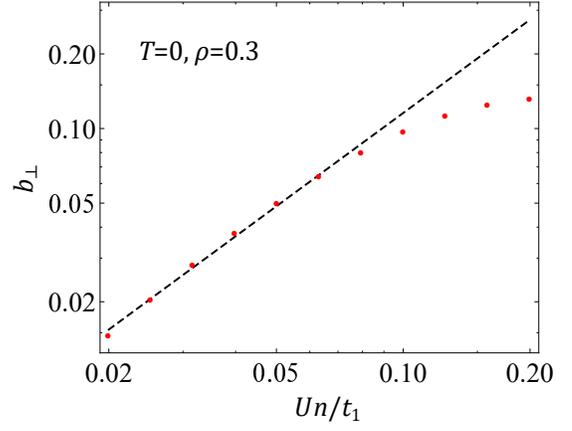}
\caption{$b_\perp$ ($\approx b_\perp^{(1)}$, in unit of $t_1 a^2$) vs. $Un/t_1$ for $\rho=0.3$, $T=0$, $N_0=1$. The black-dashed line gives a fitting to (\ref{eq:b_perp^(1)}) in the small $Un$ region with a slope $=5/4$. \label{fig:order by disorder}}
\end{figure}

Before working in the more convenient functional integral formalism, we note the ground state energy $E_{gs}({\bf k}_0)=E_{gs}^{(0)}({\bf k}_0)+E_{gs}^{(1)}({\bf k}_0)$ in Eq.~(\ref{eq:E_gs}) consists of two contributions, the mean-field and the zero-point energies, respectively, given by
\begin{align}\label{eq:E_gs^(0+1)}
    E_{gs}^{(0)} = \epsilon^-_{{\bf k}_0} N+ \frac{nNU}{2},\quad E_{gs}^{(1)} = \frac{1}{2} \sum_{{\bf q}} (E_{{\bf q}}-\varepsilon'_{{\bf q}}).
\end{align}
The former is the dominate contribution that exhibits a sub-extensive degeneracy on the contour minimum, while the latter is the lowest-order in $Un_0$ correction, corresponding to the one-loop correction in the field theory treatment below. 

In the coherent-state path integral, the zeroth-order thermodynamic potential is equal to the mean-field ground state energy above (with $N\to N_0$)
\begin{align}
\Omega^{(0)}=E^{(0)}_{gs},
\end{align} 
and the one-loop correction (neglecting the upper band, valid for $Un_0/t_1 \ll 1$) is given by
\begin{align}\label{eq:Omega^(1)}
\Omega^{(1)}(N_0,{\bf k}_0) 
=&\ \frac{1}{2\beta}\sum_{{\bf q},\omega_n}\text{Tr}\ln G_0^{-1}({\bf q},\omega_n)
\nonumber\\
=&\ \frac{1}{2\beta}\sum_{{\bf q},\omega_n}\ln \left[(\omega_n+i\mathcal{E}_{2,{\bf q}})^2 + \mathcal{E}_{1,{\bf q}}^2\right]
\nonumber\\
=&\ \frac{1}{\beta}\sum_{{\bf q}}\ln \left[2\sinh(\beta E_{\bf q}/2)\right],
\end{align}
where $G_0({\bf q},\omega_n)$ is defined in Eq.~\eqref{eq:G0}.
In the low-temperature limit $\beta\to \infty$, reduces to the zero-point energy of the Bogoliubov quasiparticles $E^{(1)}_{gs} = \frac{1}{2}\sum_{{\bf q}}(E_{\bf q}-\varepsilon'_{{\bf q}})$ (where we measured related to the zero-point energy of the normal state).
In the above, the Matsubara sum was carried out by using
\begin{align}
    \sum_{\omega_n}\ln \left[(\omega_n+i\mathcal{E}_2)^2 + \mathcal{E}_1^2\right] = \sum_{p=\pm}\ln\left[2\sinh\left(\frac{\mathcal{E}_1+p\mathcal{E}_2}{2T}\right)\right],
\end{align}
obtained by integrating (\ref{eq:matsubara_sum}) over $\epsilon$ or by Poisson summation formula followed by a contour integral \cite{altland_condensed_2010}.

The perturbation theory also gives order-by-order corrections to the condensate momentum, ${\bf k}_0=\sum_{n}{\bf k}_0^{(n)}$. This allows us to study the corrections of the thermodynamic potential and its curvature as an expansion at the mean-field minimum: $\Omega({\bf k}_0)\approx \Omega^{(0)}({\bf k}_0^{(0)})+\Omega^{(1)}({\bf k}_0^{(0)})$ and $2N_0 b_\perp({\bf k}_0)\approx \partial_{\parallel}\partial_{\perp}^2\Omega^{(0)}({\bf k}_0^{(0)})k_0^{(1)} + \partial_{\perp}^2\Omega^{(1)}({\bf k}_0^{(0)})$.\footnote{However, the full expression $\Omega=\Omega^{(0)}+\Omega^{(1)}$ is not real for any ${\bf k}_0$ since the quasiparticle is only guaranteed to be stable when expanding the condensate around the contour minimum ${\bf k}_0^{(0)}$. This is an artifact of the perturbation theory arising from $b_\perp^{(0)}=0$, and can be avoided in a self-consistent or renormalization group analysis.} Numerical evaluation of the integral (\ref{eq:Omega^(1)}) lifts the degeneracy of the bare dispersion contour, replacing it by six minima with the $\text{C}_\text{6}$ lattice symmetry. We refer to this as order-by-disorder phenomenon (demonstrated in the context of frustrated magnetism in \cite{bergman_order-by-disorder_2007,mulder_spiral_2010}). As shown in Fig.~\ref{fig:phase_diagram}(b), the helical condensate is stable on the high symmetry points of the contour, the Vertex ($\text{BEC}_\text{V}$) and the Edge ($\text{BEC}_\text{E}$) condensates, depending on the microscopic parameters $\rho$, $Un_0$, and temperature $T$. The phase boundaries are all first-order. We note that there is a reentrant $\text{BEC}_\text{V}$ around $\rho\approx 0.22$, related to the detailed quasiparticle dispersion at small ${\bf q}$ characterized by $B(\rho)$ and $K(\rho)$. As detailed in Appendix~\ref{app:b_perp}, an expansion at the minimum of the thermodynamic potential gives
\begin{align}\label{eq:b_perp^(1)}
    b_\perp^{(1)} = b_{0\perp}f_b(Un_0/T)
\end{align}
for small $Un_0$, which gives an effective $UV$ cutoff such that $\mathcal{E}_{\bf q}\approx Bq_\parallel^2+Kq_\perp^4 < Un_0$. In the above, 
\begin{align}
b_{0\perp}=(Un_0)^{5/4}t_1^{-1/4}g_b(\rho)
\end{align}
with $g_b(\rho)$ a dimensionless $\mathcal{O}(1)$ constant and the scaling function
\begin{align}\label{eq:scaling_fn_f}
    f_b(x)\sim\left\lbrace\begin{array}{cc}
    1,\quad \text{for } x\gg 1. \\
    1/x,\quad \text{for } x\ll 1. 
    \end{array}\right.
\end{align}
With increasing $Un_0$, higher powers terms in $\mathcal{E}_{\bf q}$ are important, leading to a deviation from the scaling form (\ref{eq:b_perp^(1)}), see Fig.~\ref{fig:order by disorder}.

As a consequence of the order-by-disorder phenomenon, $B_\perp$ is induced given by $B_\perp^{(1)}=b_\perp^{(1)}$ as discussed above. As a complementary analysis, below we perform a direct calculation of $B_\perp^{(1)}$, and show that it reduces to zero in the isotropic, continuum limit.

\subsection{$B_\perp$}\label{sec:aniso}
To obtain $B_\perp$, one first needs to calculate the zeroth-order action $S^{(0)}[\pi,\phi]$ based on the lattice model in Sec.~\ref{sec:model}. A shortcut to this is to expand the mean-field ground state energy $E_{gs}^{(0)}({\bf k}_0+{\bf q})$ in ${\bf q}$, and use the relation $B_{lm}^{(0)}=b_{lm}^{(0)}$. However, this has a shortcoming as it only gives the leading order in ${\bf q}$ couplings. Alternatively, one can start with the lattice model, do an expansion in $\pi$ and $\phi$, and then take the continuum limit. The latter approach is more laborious, but gives the full expression that is valid even at large momenta. We relegate the details of this analysis to Appendix~\ref{app:C6_effects}. For simplicity, we consider a long-wavelength form in analogy to (\ref{eq:L_phi_0}) that is valid within the coherence length ${\bf q}<{\bf q}_c$ in the weakly-interacting limit. By including up to quartic in $\phi$ terms required for one-loop calculations, and choosing ${\bf k}_0$ at one of the $\text{C}_\text{6}$ symmetric points, the zeroth-order Lagrangian takes the following form (with the superscripts $(0)$ dropped for the simplicity of notation)
\begin{widetext}
\begin{align}
\label{eq:L_phi_low_k}
    \frac{\mathcal{L}_\phi}{n_0} =&\ \Delta\left[c(\partial_\parallel\phi)+c^\perp(\partial_\perp\phi)^2\right] + B_\tau(\partial_\tau\phi)^2 + B(\partial_\parallel\phi)^2 + K_{20}(\partial^2_\parallel\phi)^2 + K_{02}(\partial^2_\perp\phi)^2 + K_{11}(\partial_\parallel^2\phi)(\partial^2_\perp\phi)
\nonumber\\
&\ + B_{30}(\partial_\parallel\phi)^3 + B_{12}(\partial_\parallel\phi)(\partial_\perp\phi)^2 + B_{40}(\partial_\parallel\phi)^4 + B_{04}(\partial_\perp\phi)^4 + B_{22}(\partial_\parallel\phi)^2(\partial_\perp\phi)^2,
\end{align}
\end{widetext}
where $B_\tau = 1/2Un_0$ and $\Delta_{{\bf k}_0}=1-|\Gamma_{\bar{{\bf k}}_0}|/|\Gamma_{{\bf k}_0}|$, vanishing at $k_0=\bar{k}_0$. All the other parameters are of the form $t_1g(\rho)$ with $g(\rho)$ a dimensionless $\mathcal{O}(1)$ constant while the explicit expressions can be obtained via $B_{lm}^{(0)} = b_{lm}^{(0)}$ along with $K_{20}^{(0)}=B_{40}^{(0)}$, $K_{02}^{(0)}=B_{04}^{(0)}$ and $K_{11}^{(0)}=B_{22}^{(0)}$, where the latter relations are found by inspecting the explicit form of the quadratic action, (\ref{eq:quadratic_action_pi_phi}), and $b_{lm}^{(0)}$ are given in Eqs.~(\ref{eq:b_lm_Vertex}) and (\ref{eq:b_lm_Edge}) for $\text{BEC}_{\text{V}}$ and $\text{BEC}_{\text{E}}$, respectively. The linear derivative term (tadpole diagram) $\partial_\parallel\phi$ can be eliminated by choosing the condensate momentum such that $\nabla_{k_0}\Omega({\bf k}_0)=0$, which at mean-field level, gives $k_0^{(0)}=\bar{k}_0$. We note $B_\perp=0$ in the Lagrangian (\ref{eq:L_phi_low_k}) even in the presence of $\text{C}_\text{6}$ lattice (but excluding the zero-point energy) effects. This is consistent with the mean-field ground state energy in Eq.~(\ref{eq:E_gs^(0+1)}) that exhibits a degenerate contour minimum. However, higher-order fluctuations generally generate a nonzero $B_\perp$. At one-loop order and the weakly-interacting limit, it is given by (see details in Appendix~\ref{app:B_perp})
\begin{widetext}
\begin{align}
    B_\perp^{(1)} = &\ -\frac{c_\perp B_{12}}{2c}\int_{q}\frac{3q_\parallel^2+q_\perp^2}{B_\tau\omega_n^2 + \mathcal{E}_{\bf q}} + \frac{1}{2}\int_{q}\frac{B_{22}q_\parallel^2+6B_{40} q_\perp^2}{B_\tau\omega_n^2 + \mathcal{E}_{\bf q}} - B_{12}^2\int_{q}\frac{q_\parallel^2q_\perp^2}{(B_\tau\omega_n^2 + \mathcal{E}_{\bf q})^2}
    \nonumber\\
    \approx &\ B_{0\perp}f_B(Un_0/T),
\end{align}
\end{widetext}
where the dimensionless scaling function $f_B$ has the same limits as in Eq.~(\ref{eq:scaling_fn_f}) and $B_{0\perp} = (Un_0)^{5/4}t_1^{-1/4}g_B(\rho)$ with $g_B(\rho)$ a dimensionless $\mathcal{O}(1)$ constant. In the isotropic limit $\rho\to 1/6$, $K_{20}=K_{02}=K_{11}/2$ and $B/k_0^2=B_{30}/k_0=B_{12}/k_0=4B_{40}=4B_{04}=2B_{22}$,
resulting in $B_\perp^{(1)}=0$. This suggests the robustness of the Lifshitz transition point at $\rho=1/6$ at one-loop order and associated relation between harmonic and nonlinear terms enforced by the rotational symmetry of ${\bf k}_0$, also corresponding to the dipole conservation symmetry.

\section{Conclusion\label{sec:conclusion}}

In summary, we studied a nonzero-momentum superfluid state as a
condensate in a frustrated honeycomb Bose-Hubbard model that features
a dispersion minimum contour with a nonzero momentum scale $k_0$, set
by frustrated hopping. We supplemented the lattice model by a
continuum smectic field theory and explored in detail the rich
phenomenology that emerges from the helical (single momentum)
Bose-condensation on this dispersion minimum contour.

Consistent with generalized Hohenberg-Mermin-Wagner theorem \cite{mermin_absence_1966,hohenberg_existence_1967,halperin_hohenbergmerminwagner_2019}, we found
that the helical condensate is stable at zero temperature despite
quantum fluctuations that are qualitatively stronger than that of a
conventional superfluid. The helical state is spontaneously highly
anisotropic in its spectrum of low-energy excitations. We explored in
detail the physical properties of such state, such as the equation of
state (chemical potential), condensate depletion, momentum
distribution, structure factor, of all which are qualitatively
distinct from a conventional XY zero-momentum superfluid.  Because of
the ``soft'' smectic dispersion, at nonzero temperature,
thermal fluctuations that diverge in the absence of lattice effects lead to a vanishing of the helical condensate.

However, somewhat paradoxically, a subtle interplay of lattice effects with quantum and
thermal fluctuations leads to stabilization of
the helical smectic state, through the phenomenon of order-by-disorder
via a crossover to a more conventional stable XY superfluid.

Our study thus predicts a stable helical superfluid state and its rich
phenomenology that can emerge in frustrated bosonic systems,
characterized by a bare dispersion minimum contour. Such sub-extensive
contour degeneracy of the condensate momentum, ${\bf k}_0$, can be
realized in cold atom experiments through Floquet engineering of the
optical lattice with frustrated hopping \cite{struck_quantum_2011,wintersperger_realization_2020} or synthetic
Rashba spin-orbit coupling \cite{zhai_degenerate_2015}. Because the latter
approach is not based on frustration, its closed contour minimum
dispersion would be spoiled by an optical lattice, reduced to a
discrete set of dispersion minima \cite{zhang_superfluid_2013,toniolo_superfluidity_2014,maeland_plane-_2020}. It
would thereby exhibit a bare $B_\perp$ modulus controlled by the depth
of the optical lattice, rather than Feshbach tunable interactions
\cite{chin_feshbach_2010}, as considered in our study.  Yet, in a shallow
optical lattice with a weak $B_\perp$, within a long crossover scale
we expect a nonzero momentum condensate \cite{wu_unconventional_2011,barnett_order_2012,cui_enhancement_2013}
in spin-orbit coupled Bose gases, exhibiting phenomenology
predicted in Table~\ref{tab:SF_helicalSF}.

The frustrated bosonic model that we studied is isomorphic to a
quantum O(2) easy-plane magnet with frustrated exchange couplings,
which is also characterized by a spiral contour as manifold of its
classical ground states. The helical superfluid we explored thus maps
directly on such a coplanar spin spiral state with the U(1) superfluid
phase corresponding to the O(2) direction of the XY spin. Theoretical studies suggest the stability of such spiral states in a quantum easy-plane magnet \cite{liu_featureless_2020} and in XY models (for spin-1/2, equivalent to the Bose-Hubbard model at half-filling in the $U\to\infty$ hard-core limit, in contrast to the weak $U$ limit considered here) with ${\bf k}_0$ at high symmetry points selected by quantum fluctuations \citep{varney_kaleidoscope_2011,di_ciolo_spiral_2014,sedrakyan_spontaneous_2015}. This
contrasts with the corresponding Heisenberg O(3) model
\cite{bergman_order-by-disorder_2007,mulder_spiral_2010}, which at nonzero temperature in 2d is
always unstable (even with the usual linear dispersion) to a
disordered state due to strongly coupled nonlinear spin wave
fluctuations.

Our analysis, however, does not provide the sufficient conditions to
realize the helical superfluid discussed here. Other interesting
supersolid phases, e.g., Bose condensate at multiple momenta, are also
competing ground states that can emerge from the interplay of
interactions and the macroscopic degeneracy of the dispersion
minimum. Selecting between these is a challenging problem that we have
not addressed and likely requires extensive numerical
analysis. Cold-atoms systems where our model is realized using Floquet
engineering will likely be challenged by heating problem as well as
further neighbor hopping and interactions.

Here we neglected the vortices (allowed by the periodicity of the superfluid phase) that can play an important role for the quantum and thermal phase transitions out of the helical superfluid state, sketched at the mean-field level in Fig.~\ref{fig:phase_diagram}(a). In 2+1d it is
isomorphic to the melting of a quantum smectic (at low energies
perturbed by stabilizing modulus $B_\perp$ generated by
order-by-disorder) that has been formulated through a dual gauge
theory in Ref.~\cite{radzihovsky_quantum_2020,zhai_fractonic_2021}, but with critical RG
analysis remaining as an open problem.  We note that, as discussed in
Ref.~\cite{yan_low_2021} in addition to conventional vortices of the
superfluid phase $\phi$, thermal excitations of orientational
(``momentum'') vortices in ${\bf k}_0$ (based on numerical analysis)
appear to play an important role. We leave these and other questions
discussed above for future theoretical research and hope that the rich
phenomenology presented here will stimulate further experimental and
theoretical works on frustrated bosonic and magnetic systems.

\section*{Acknowledgement}
L.~R. thanks Arun Paramekanti for discussions. This work is supported by the Simons Investigator Award to L.~R. from the James Simons Foundation. L.~R. also thanks the KITP for its hospitality as part of the Spring 2022 Pair-Density-Wave Order Rapid Response Workshop, during which part of this work was completed and in part supported by the National Science Foundation under Grant No. NSF PHY-1748958. Research at Perimeter Institute is supported in part by the Government of Canada through the Department of Innovation, Science and Economic Development Canada and by the Province of Ontario through the Ministry of Colleges and Universities.

\appendix

\section{Diagonalization of frustrated tight-binding model on honeycomb lattice \label{app:diago}}

The tight-binding Hamiltonian in Eq.~(\ref{eq:H_0}) can be written in terms of Pauli matrices $H_0= \sum_{\bf k} (a^\dag_{{\bf k},1}, a^\dag_{{\bf k},2}) h_0 (a_{{\bf k},1}, a_{{\bf k},2})^T$, where
\begin{equation}
h_0 =t_2 \epsilon_{\bf k} \mathbbm{1} -t_1|\Gamma_{\bf k}| \cos \theta_{\bf k}~\sigma_x + t_1 |\Gamma_{\bf k}| \sin \theta_{\bf k}~ \sigma_y,
\end{equation}
with
\begin{align}
\epsilon_{{\bf k}} &= 2\sum_i (\cos {\bf k}\cdot {\bf v}_i),
\nonumber\\
\Gamma_{{\bf k}} &= \sum_i \exp(-i{\bf k}\cdot {\bf e}_i),
\nonumber\\
\theta_{\bf k} &= \text{Arg}(\Gamma_{\bf k}).
\end{align}
The Hamiltonian $h_0$ can be diagonalized by the transformation 
\begin{equation}
U =\frac{1}{\sqrt{2}} \biggl( \begin{array}{cc}
\exp(i\frac{\theta_{{\bf k}}}{2}) & \exp(i\frac{\theta_{{\bf k}}}{2})  \\
-\exp(-i\frac{\theta_{{\bf k}}}{2})  & \exp(-i\frac{\theta_{{\bf k}}}{2}) 
\end{array}\biggr),
\end{equation}
which gives
\begin{equation}
U^\dag h_0 U 
= t_2 \epsilon_{{\bf k}} \mathbbm{1} -t_1 \biggl(\begin{array}{cc}
- |\Gamma_{{\bf k}}| & 0 \\
0 & |\Gamma_{{\bf k}}| 
\end{array}\biggr),
\end{equation}
with two bands
\begin{equation}
\epsilon^-_{{\bf k}} = t_2 \epsilon_{{\bf k}} -t_1 |\Gamma_{{\bf k}}| ,\quad \quad \epsilon^+_{{\bf k}} = t_2 \epsilon_{{\bf k}} + t_1 |\Gamma_{{\bf k}}|.
\end{equation}
The lower band exhibits a contour minimum for $1/6<\rho=t_2/t_1<1/2$.

In general, the dispersion minimum contour can appear in a class of Hamiltonians written as
\begin{equation}
h_0=t_1 T + t_2 T^2,
\end{equation}
where $t_1$ and $t_2$ are the nearest-neighbor and next-nearest-neighbor hopping amplitudes and $T$ is a $2\times 2$ hopping matrix, with the form
\begin{equation}
T = \left( \begin{array}{cc}
0 & G_{{\bf k}} \\
G_{{\bf k}}^\dag & 0 
\end{array}\right)
\end{equation}
in the sublattice basis. Then the two band are given by  
\begin{equation}
\epsilon_{{\bf k}}^{\pm} = \pm |t_1| |G_{{\bf k}}| +t_2 |G_{{\bf k}}|^2
\end{equation}
with the lower energy band $\epsilon_{{\bf k}}^-$ exhibiting macroscopic ground state degeneracy along the contour defined by $|G_{{\bf k}}| =|t_1|/2t_2$.

\begin{widetext}

\section{Expansion of $\epsilon^-_{{\bf k}}$ around the dispersion minimum contour \label{app:exp_disp_contour}}

In the weakly-interacting regime, the band dispersion around the bose condensate plays an important role for the low-energy properties of the helical superfluid. This is established through the facts: (i) the zeroth-order thermodynamic potential $\Omega^{(0)}(N_0,{\bf k}_0)=N_0\epsilon^-_{{\bf k}_0}+N_0^2U/4V$ depending on the free dispersion $\epsilon^-_{{\bf k}}$ and (ii) The relation between $\Omega^{(0)}$ and the zeroth-order Goldstone mode theory $\mathcal{L}^{(0)}$ as discussed in Sec.~\ref{sec:order_by_disorder}. Expansion on the lower band dispersion (\ref{eq:epsilon}) up to quartic order gives
\begin{align}
    \epsilon^-_{{\bf k}_0+{\bf q}} - \epsilon^-_{{\bf k}_0} \approx \Delta (cq_\parallel+c_\perp q_\perp^2) + b^{(0)}q_{\parallel}^2 + b_{30}^{(0)} q_{\parallel}^3 + b_{12}^{(0)} q_{\parallel}q_{\perp}^2 + b_{40}^{(0)} q_{\parallel}^4 + b_{04}^{(0)}q_{\perp}^4 + b_{22}^{(0)} q_{\parallel}^2q_{\perp}^2,
\end{align}
where for the Vertex condensate ($\text{BEC}_\text{V}$), ${\bf k}_0 = (0,k_{0V})$, $|\Gamma_{{\bf k}_0}|=\sqrt{5+4\cos(3k_{0V}/2)}$, and the coefficients
\begin{align}\label{eq:b_lm_Vertex}
    c =&\ 6t_2\sin(3k_{0V}/2),\quad c_\perp = \frac{3t_2[2+\cos(3k_{0V}/2)]}{2},\quad b^{(0)} = \frac{9t_2\sin^2(3k_{0V}/2)}{|\Gamma_{{\bf k}_{0V}}|^2},
    \nonumber\\
    b_{30}^{(0)} =&\ \frac{27t_2[3+5\cos(3k_{0V}/2)+\cos(3k_{0V})]\sin(3k_{0V}/2)}{2|\Gamma_{{\bf k}_0}|^4},\quad b_{12}^{(0)} = \frac{9t_2[4\sin(3k_{0V}/2)+\sin(3k_{0V})]}{4|\Gamma_{{\bf k}_0}|^2},
    \nonumber\\
    b_{40}^{(0)} =&\ \frac{27t_2[76+190\cos(3k_{0V}/2)+163\cos(3k_{0V})+50\cos(9k_{0V}/2)+7\cos(6k_{0V})]}{32|\Gamma_{{\bf k}_0}|^6},\quad b_{04}^{(0)} = \frac{9t_2[2+\cos(3k_{0V}/2)]^2}{16|\Gamma_{{\bf k}_0}|^2},
    \nonumber\\
    b_{22}^{(0)} =&\ \frac{27t_2[15+25\cos(3k_{0V}/2)+11\cos(3k_{0V})+3\cos(9k_{0V}/2)]}{16|\Gamma_{{\bf k}_0}|^4}
\end{align}
and for the Edge condensate ($\text{BEC}_\text{E}$), ${\bf k}_0 = (k_{0E},0)$, $|\Gamma_{{\bf k}_0}|=|1+2\cos(\sqrt{3}k_{0E}/2)|$, and the coefficients
\begin{align}\label{eq:b_lm_Edge}
    c =&\ 2\sqrt{3}t_2[\sin(\sqrt{3}k_{0E}/2)+\sin(\sqrt{3}k_{0E})],\quad c_\perp = \frac{9t_2\cos(\sqrt{3}k_{0E}/2)}{2},\quad b^{(0)} = 3t_2\sin^2(\sqrt{3}k_{0E}/2),
    \nonumber\\
    b_{30}^{(0)} =&\ \frac{3\sqrt{3}t_2\sin(\sqrt{3}k_{0E})}{4},\quad b_{12}^{(0)} = \frac{9\sqrt{3}t_2\sin(\sqrt{3}k_{0E})}{4|\Gamma_{{\bf k}_0}|},\quad b_{40}^{(0)} = \frac{3t_2[-1+7\cos(\sqrt{3}k_{0E})]}{32},
    \nonumber\\
    b_{04}^{(0)} =&\ \frac{81t_2[1+\cos(\sqrt{3}k_{0E})]}{32|\Gamma_{{\bf k}_0}|^2},\quad b_{22}^{(0)} = \frac{27t_2[-1+3\cos(\sqrt{3}k_{0E}/2)+3\cos(\sqrt{3}k_{0E})+\cos(3\sqrt{3}k_{0E}/2)]}{16|\Gamma_{{\bf k}_0}|^2}.
\end{align}
In the above, $\Delta = 1-|\Gamma_{\bar{{\bf k}}_0}|/|\Gamma_{{\bf k}_0}|$, which vanishes at ${\bf k}_0=\bar{{\bf k}}_0$. For ${\bf k}_0\neq \bar{{\bf k}}_0$, there are corrections to all the coefficients above, but here we only give the ones for $q_\parallel$ and $q_\perp^2$ that are important for the one-loop calculations in Sec.~\ref{sec:order_by_disorder}. 

\section{Bogoliubov approximation to the frustrated Bose-Hubbard model \label{app:hint}}

For a condensate at ${\bf k}_0$, the bosonic operators are given by
\begin{equation}
a_{{\bf k},1} \approx A_{{\bf k}_0,1}\delta_{{\bf k},{\bf k}_0} + a_{{\bf k}_0 + {\bf q},1},\ \ \ a_{{\bf k},2} \approx A_{{\bf k}_0,2}\delta_{{\bf k},{\bf k}_0} + a_{{\bf k}_0 + {\bf q},2},
\end{equation}
where $A_{{\bf k}_0,1} = \frac{1}{\sqrt{2}} e^{i\frac{\theta_{{\bf k}_0}}{2}} \sqrt{N_0}$ and $ A_{{\bf k}_0,2} =\frac{1}{\sqrt{2}} e^{-i\frac{\theta_{{\bf k}_0}}{2}} \sqrt{N_0}$. We then expand the interacting part of the Hamiltonian, (\ref{eq:interaction}), which up to quadratic order of $a_{{\bf k}_0 + {\bf q}}$ and $a^\dag_{{\bf k}_0 + {\bf q}}$ is given by
\begin{align}
H_{\text{int}} \approx &\ \frac{U}{2V}\sum_{s=1,2}\left( |A_{{\bf k}_0,s}|^4 + 4 |A_{{\bf k}_0,s}|^2 \sum_{{\bf q}} a^\dag_{{\bf k}_0 + {\bf q},s} a_{{\bf k}_0 + {\bf q},s} + (A_{{\bf k}_0,s}^{*})^2 \sum_{{\bf q}} a_{{\bf k}_0-{\bf q},s} a_{{\bf k}_0+{\bf q},s} +  A_{{\bf k}_0,s}^2 \sum_{{\bf q}} a^\dag_{{\bf k}_0-{\bf q},s} a^\dag_{{\bf k}_0+{\bf q},s}\right)
\nonumber\\
= &\ \frac{UN_0^2}{4V}+\frac{UN_0}{4V}\biggl(4\sum_{{\bf q}}a^\dag_{{\bf k}_0+{\bf q},1} a_{{\bf k}_0+{\bf q},1} + e^{-i \theta_{{\bf k}_0}} \sum_{{\bf q}} a_{{\bf k}_0-{\bf q},1} a_{{\bf k}_0+{\bf q},1} + e^{i \theta_{{\bf k}_0}} \sum_{{\bf q}} a^\dag_{{\bf k}_0-{\bf q},1} a^\dag_{{\bf k}_0+{\bf q},1}  \biggr) 
\nonumber\\ & + \frac{UN_0}{4V}\biggl(4\sum_{{\bf q}}a^\dag_{{\bf k}_0+{\bf q},2} a_{{\bf k}_0+{\bf q},2} + e^{i \theta_{{\bf k}_0}} \sum_{{\bf q}} a_{{\bf k}_0-{\bf q},2} a_{{\bf k}_0+{\bf q},2} + e^{-i \theta_{{\bf k}_0}} \sum_{{\bf q}} a^\dag_{{\bf k}_0-{\bf q},2} a^\dag_{{\bf k}_0+{\bf q},2} \biggr).
\end{align}
With a transformation to the band basis
\begin{equation}
a_{{\bf k},1} = \frac{1}{\sqrt{2}} e^{i\frac{\theta_{{\bf k}}}{2}} (d_{{\bf k},+} +d_{{\bf k},-}) ,\quad
a_{{\bf k},2} = \frac{1}{\sqrt{2}} e^{-i\frac{\theta_{{\bf k}}}{2}} (-d_{{\bf k},+} +d_{{\bf k},-}), \label{eq:change_basis}
\end{equation}
followed by neglecting the upper band contributions, $d_{{\bf k},+}$, that only give $O(Un_0/t_1)$ corrections to the ground state, we obtain the following total Hamiltonian
\begin{align}
H =&\ \epsilon^-_{{\bf k}_0} N_0 + \frac{U N_0^2}{4V} + \sum_{{\bf q}}  \epsilon^-_{{\bf k}_0+{\bf q}}  d_{{\bf k}_0+{\bf q},-}^\dag  d_{{\bf k}_0+{\bf q},-} + \frac{U N_0}{2V}\sum_{{\bf q}}\left(  d_{{\bf k}_0+{\bf q},-}^\dag  d_{{\bf k}_0+{\bf q},-} + d_{{\bf k}_0-{\bf q},-}^\dag d_{{\bf k}_0-{\bf q},-} \right)
\nonumber\\
&+ \frac{UN_0}{4V}\sum_{{\bf q}} \cos\left(\theta_{{\bf k}_0}-\frac{\theta_{{\bf k}_0+{\bf q}}+\theta_{{\bf k}_0-{\bf q}}}{2}\right) \left(d_{{\bf k}_0+{\bf q},-}  d_{{\bf k}_0-{\bf q},-} +d_{{\bf k}_0+{\bf q},-}^\dag  d_{{\bf k}_0-{\bf q},-}^\dag \right),
\end{align}
\end{widetext}
which gives the Hamiltonian (\ref{eq:hammatrix}) after using the canonical ensemble relation
\begin{equation}
N =  N_0 + \sum_{{\bf q}}  d_{{\bf k}_0+{\bf q},-}^\dag  d_{{\bf k}_0+{\bf q},-}.
\end{equation}

The Hamiltonian (\ref{eq:hammatrix}) then can be diagonalized by the bosonic Bogoliubov transformation
\begin{equation}
V=\biggl( \begin{array}{cc}
u_{\bf q} & v_{\bf q} \\
v_{-{\bf q}}^* & u_{-{\bf q}}^* 
\end{array}\biggr),
\end{equation}
where $u_{\bf q}$ and $v_{\bf q}$ (chosen to be real) are given in Eq.~(\ref{eq:basis_coef_u_v}) with $u_{\bf q}v_{\bf q} = - \sqrt{u_{\bf q}^2v_{\bf q}^2}$, and $V$ is a nonunitary matrix that preserves the bosonic commutation relation after the basis transformation
\begin{equation}
\biggl( \begin{array}{c}
d_{{\bf k}_0+{\bf q},-} \\ d^\dag_{{\bf k}_0-{\bf q},-}
\end{array}\biggr)=V\biggl( \begin{array}{c}
\alpha_{{\bf k}_0+{\bf q}} \\ \alpha^\dag_{{\bf k}_0-{\bf q}}
\end{array}\biggr).
\end{equation}
The dispersions $E^{\pm}_{\bf q} = E_{\pm\bf q}$ of $\alpha_{{\bf k}_0\pm {\bf q}}$ in Eq.~(\ref{eq:disp}) can be obtained by solving the determinant equation
\begin{equation}
|H - \lambda\sigma_z|=0
\end{equation}
with the solutions $\lambda =  E^{+}_{\bf q}$, $-E^{-}_{\bf q}$, where $E^{\pm}_{\bf q}>0$.

\section{Stability of quantum ``smectic" and ``columnar" phases in d-dimensions: generalized Hohenberg-Mermin-Wagner theorems \label{app:fluc}}

In this appendix, we perform a simple dimensional analysis of the stability of a smectic (columnar) phase in $d$ spatial dimensions, which exhibits a Goldstone mode with $1$ ($d-1$) hard direction(s), $\parallel$, and the other $d-1$ ($1$) soft direction(s), $\perp$. We set $U=B=K=2n_0=1$ for simplicity.

For the smectic phase, the quantum fluctuations at $T=0$ is characterized by
\begin{equation}
\langle \phi^2 \rangle_Q = \int \frac{d\omega}{2\pi} \frac{d q_{\parallel} d^{d-1} q_\perp }{(2\pi)^d}\frac{1}{\omega^2 + q_\parallel^2 +(q_\perp^2)^2}.
\end{equation}
With the change of variables $q_\perp^2=y_\perp$, the above equation becomes
\begin{equation}
\begin{aligned}
\langle \phi^2 \rangle_Q &= \int \frac{d\omega d q_\parallel dq_\perp }{(2\pi)^{d+1}} \frac{q_\perp^{d-2}}{\omega^2 +q_\parallel^2 +q_\perp^4} \\
&\propto \int \frac{d\omega  d q_\parallel dy_\perp }{(2\pi)^{d+1}} \frac{y_\perp^{\frac{d-3}{2}}}{\omega^2 +q_\parallel^2 +y_\perp^2}.
\end{aligned}
\end{equation}
The stability of this state requires the convergence in the IR, which requires
\begin{equation}
3+ \frac{d-3}{2} >2 \Rightarrow d>1.
\end{equation}
For the columnar phase, the quantum fluctuations are
\begin{align}
\langle \phi^2 \rangle_Q&= \int \frac{d\omega  d^{d-1}q_\parallel dq_\perp }{(2\pi)^{d+1}} \frac{1}{\omega^2 + q_\parallel^2 +q_\perp^4}
\nonumber\\
&\propto \int \frac{d \omega dq_\parallel dy_\perp }{(2\pi)^{d+1}} \frac{q_\parallel^{d-2} y^{-1/2}_\perp}{\omega^2 + q_\parallel^2 +y_\perp^2},
\end{align}
which requires  
\begin{equation}
3+(d-2)-\frac{1}{2} >2 \Rightarrow d>\frac{3}{2}
\end{equation}
for the stability. Therefore, for the physical dimension of our interest, $d=2$, both requirements are satisfied, which suggests the helical superfluid stable under quantum fluctuations.

At nonzero temperature, we consider the dominant classical contributions at $\omega_n=0$, which for the smectic are given by
\begin{align}
\langle \phi^2 \rangle_T &= \int \frac{d q_{\parallel} d^{d-1} q_\perp }{(2\pi)^d}\frac{1}{q_\parallel^2 + (q_\perp^2)^2 }
\nonumber\\
&\propto \int \frac{ d q_\parallel dy_\perp }{(2\pi)^{d+1}} \frac{y_\perp^{\frac{d-2}{2}} y^{-1/2}_\perp}{q_\parallel^2 +y_\perp^2}.
\end{align}
Its stability requires 
\begin{equation}
2+\frac{d-3}{2} >2 \Rightarrow d >3.
\end{equation}
While in the columnar phase, 
\begin{align}
\langle \phi^2 \rangle_T &= \int \frac{ d^{d-1} q_\parallel dq_\perp}{(2\pi)^{d}} \frac{1}{q_\parallel^2 +q_\perp^4}
\nonumber\\
&\propto\int \frac{dq_\parallel dy_\perp }{(2\pi)^{d+1}} \frac{q_\parallel^{d-2}y^{-1/2}_\perp}{q_\parallel^2 +y_\perp^2},
\end{align}
which requires
\begin{equation}
2+d-2-\frac{1}{2} >2\Rightarrow d> \frac{5}{2}
\end{equation}
Thus, thermal fluctuations make the 2d smectic/columnar phase unstable.

\section{Calculation details of depletion \label{app:depletion}}
The condensate depletion is given by Eq.~(\ref{eq:eff_th_n_d}) at weak interactions as discussed in the main text. At zero temperature, $\beta =\infty$ leads to
\begin{equation}
n_d= \int_{-\infty}^{\infty} \frac{d q_\parallel dq_\perp}{(2\pi)^2} \frac{1}{2}\biggl( \frac{\mathcal{E}_{{\bf q}} + Un}{\sqrt{\mathcal{E}^2_{{\bf q}} + 2Un \mathcal{E}_{\bf q}} } -1\biggr),
\end{equation}
where $\mathcal{E}_{{\bf q}} \approx B q_\parallel^2 + K q_\perp^4$ as defined in Eq.~(\ref{eq:Epsilon_and_Epsilon_2}). In this section, we perform the integral exactly. With the change of variables: 
\begin{eqnarray}
\sigma_\parallel= \sqrt{\frac{B}{2U n}} q_\parallel , \hspace{0.8mm} \sigma_\perp= \sqrt{\frac{K}{2U n}} q_\perp^2,\hspace{0.8mm} 
\sigma^2=\sigma_\parallel^2+\sigma_\perp^2,
\label{eq:sigma}
\end{eqnarray}
the integral can be rewritten as
\begin{align}
I &= \int_{-\infty}^{\infty} \frac{d q_\parallel dq_\perp}{(2\pi)^2} \frac{1}{2}\biggl( \frac{\mathcal{E}_{{\bf q}}+Un  }{\sqrt{\mathcal{E}^2_{{\bf q}} + 2Un \mathcal{E}_{{\bf q}}} } -1\biggr)
\nonumber\\ 
&=\frac{(2U n)^{3/4}}{B^{1/2} K^{1/4}}  \int_{-\infty}^{\infty} \frac{d\sigma_\parallel d\sigma_\perp}{16\pi^2}  \frac{1}{\sqrt{|\sigma_\perp|}} \biggl( \frac{2+\frac{1}{\sigma_\parallel^2+\sigma_\perp^2}  }{2\sqrt{1 + \frac{1}{ \sigma_\parallel^2+\sigma_\perp^2}} } -1\biggr)
\nonumber\\ 
&=\frac{(2U n)^{\frac{3}{4}}}{B^{\frac{1}{2}} K^{\frac{1}{4}}} \int_{0}^{\infty} \frac{\sqrt{\sigma} d\sigma}{16\pi^2} \int_{0}^{2\pi}   \frac{ d\theta}{\sqrt{|\sin\theta|}} \biggl( \frac{2\sigma^2+1  }{2\sigma \sqrt{\sigma^2 + 1} } -1\biggr)
\nonumber\\ 
&= \frac{(2U n)^{3/4}}{16\pi^2 B^{1/2} K^{1/4}} I_\theta I_\sigma 
,
\label{eq:depletion_integral}
\end{align}
where $I_{\theta}=\int_0^{2\pi} \frac{d\theta }{\sqrt{|\sin \theta|}} \approx 10.4882$ and
\begin{align}
I_\sigma &= \int_0^\infty d\sigma  \biggl( \frac{2\sigma^2+1 }{2\sqrt{\sigma^3 + \sigma} } -\sqrt{\sigma}\biggr)
\nonumber\\
&= \left.-\frac{2}{3}\sigma^{3/2} + \frac{2}{3}\sqrt{\sigma+\sigma^3} +\frac{f_e(\sigma)}{6}\right|_0^\infty
\nonumber\\
&= - \frac{f_e(0)}{6} = \frac{|f_e(0)|}{6}
\end{align}
with
\begin{equation}
f_e(\sigma) = 2 (-1)^{1/4}  \textrm{EllipticF}\left[ i \textrm{ArcSinh} \frac{(-1)^{1/4}}{\sqrt{\sigma}},-1\right]. \label{eq:f_e}
\end{equation}
Consequently, the integral is
\begin{equation}
I = \frac{(2U n)^{3/4}}{16\pi^2 B^{1/2} K^{1/4}} I_\theta \frac{|f_e(0)|}{6},
\end{equation}
which gives the depletion 
\begin{equation}
\frac{n_d}{n} =\biggl(\frac{U^3}{n B^2 K}\biggr)^{1/4} \frac{2^{3/4}I_\theta |f_e(0)|}{96\pi^2}.
\end{equation}

Similarly, we generalize the study of depletion to be in $d=l+m$ dimensions. The system possesses dispersion hard along $l$ directions and soft along $m$ directions. With $q_{\parallel (\perp); i}$ being the momentum along the $i$-th hard (soft) direction, the generalized dispersion is 
\begin{eqnarray}
\mathcal{E}^{(l,m)}_{\bf q} = \sum_{i=1}^m B_{i}q_{\parallel;i}^2 + \sum_{j=1}^m K_{j}q_{\perp;j}^4.
\label{eq:E_lm}
\end{eqnarray}
In terms of variables
\begin{eqnarray}
\sigma_{\parallel;i}&=& \sqrt{\frac{B_i}{2U n}} q_{\parallel;i},\hspace{2mm} \sigma_{\perp;i}= \sqrt{\frac{K_i}{2U n}} q_{\perp;i}^2 ,\nonumber\\
\sigma^2&=& \sum_i^l \sigma_{\parallel;i}^2+\sum_j^m \sigma_{\perp;j}^2,
\label{eq:sigma_i}
\end{eqnarray}
the depletion is given by
\begin{equation}
\begin{aligned}
n_d^{(l,m)} &= \int_{-\infty}^{\infty} \frac{d^{l} q_\parallel d^m q_\perp}{(2\pi)^2} \frac{1}{2}\biggl( \frac{\mathcal{E}_{{\bf q}}^{(l,m)} + Un}{\sqrt{(\mathcal{E}^{(l,m)}_{{\bf q}} )^2+ 2Un \mathcal{E}_{\bf q}^{(l,m)}} } -1\biggr).
\end{aligned}
\end{equation}
The integral can be proceeded as
\begin{eqnarray}
I^{(l,m)} &= \frac{(2U n)^{(2l+m)/4}}{2(\prod_i^l B_i \prod_j^m K_j)^{1/2}}
\int_{-\infty}^{\infty} \frac{d^l \sigma_{\parallel}  d^m\sigma_{\perp} }{(2\pi)^{m+l}} \\&\times \prod_j^m \frac{K_j^{1/4}}{2|\sigma_{\perp,j}|^{1/2}} \biggl( \frac{2\sigma^2+1}{2\sqrt{\sigma^4+\sigma^2}} - 1 \biggr).
\end{eqnarray}
Particularly, when $B_i=B$ and $K_j=K$ for any $i$ and $j$, 
\begin{equation}
I^{(l,m)} \propto \frac{(U n)^{(2l+m)/4}}{B^{l/2}K^{m/4}}.
\end{equation}
The depletion is then given by 
\begin{equation}
\frac{n_d^{(l,m)}}{n} \propto \frac{(U n^{1-\frac{4}{2l+m}})^{(2l+m)/4}}{B^{l/2}K^{m/4}}.
\end{equation}

Now we discuss the thermal corrections to the depletion, which is given by
\begin{eqnarray}
\delta n_d(T) =& n_d(T)-n_d(0) = \int_{-\infty}^{\infty}\frac{d^dq}{(2\pi)^d}\frac{u_{\bf q}^2 + v_{\bf q}^2}{e^{\beta E_{\bf q}} - 1}
\end{eqnarray}
with $E_{\bf q}$ and $\mathcal{E}_{1,\bf q}$ defined in Eq.~(\ref{eq:disp}), and $u_{\bf q}^2$ and $v_{\bf q}^2$ defined in Eq.~(\ref{eq:basis_coef_u_v}). In the long-wavelength limit, $E_{\bf q}\approx \mathcal{E}_{1,\bf q}\approx \sqrt{2Un\mathcal{E}_{\bf q}^{(l,m)}}$.
Then, with the change of variables $\sigma_{\parallel,i}=\beta\sqrt{2BUn_0}q_{\parallel,i}$ and $\sigma_{\perp,j}=\beta\sqrt{2KUn_0}q_{\perp,j}^2$ (set $B_i=B$ and $K_j=K$ for simplicity),
\begin{align}
\delta n_d(T) 
=&\ \int_{-\infty}^{\infty}\frac{d^dq}{(2\pi)^d} \frac{\mathcal{E}_{\bf q}^{(l,m)} + Un}{\mathcal{E}_{1,\bf q}}\frac{1}{e^{\beta E_{\bf q}} - 1}
\nonumber\\
\approx &\ \int_{-\infty}^{\infty}\frac{d^l q_\parallel d^m q_\perp}{(2\pi)^{l+m}} \frac{Un}{\sqrt{2Un\mathcal{E}_{\bf q}^{(l,m)}}}\frac{1}{e^{\beta \sqrt{2Un\mathcal{E}_{\bf q}^{(l,m)}}} - 1}
\nonumber\\
\propto &\ \frac{1}{(2\pi)^{l+m}2^m T}\biggl(\frac{T^2}{2B U n_0}\biggr)^{l/2}\biggl(\frac{T^2}{2K U n_0}\biggr)^{m/4}
\nonumber\\
&\times\int_0^\infty d\sigma \frac{\sigma^{l+m/2-2}}{e^{\sigma}-1},
\end{align}
where the radial integral over $\sigma$ is finite if $l+m/2-2>0$. This suggests the superfluid is stable when
\begin{equation}
2l+m>4.
\end{equation}
For a smectic (columnar) phase with $l=1$ ($l=d-1$) and $m=d-1$ ($m=1$), we reproduce the stability condition $d>3$ ($d>5/2$) as obtained in Appendix~\ref{app:fluc}.

\section{Superflow \label{app:superflow}}

In this section, we derive the expression of the supercurrent. Notice that there involves higher-order derivatives in the Lagrangian $\mathcal{L}_0$ in Eq.~(\ref{eq:L_psi}), its equation of motion is thus modified as
\begin{eqnarray}
&&\frac{\partial \mathcal{L}_0}{\partial \Phi}-\partial_\mu \Big(\frac{\partial \mathcal{L}_0}{\partial (\partial_\mu\Phi)}\Big) + \partial_i \partial_j \Big(\frac{\partial \mathcal{L}_0}{\partial (\partial_i \partial_j \Phi)}\Big)
=0;
\label{eq:EOM}
\end{eqnarray}
where $\mu=(\tau,i,j,k,\dots)$ and $i,j,k,\dots$ are spatial indices. With the detailed form of $\mathcal{L}_0$, We obtain
\begin{eqnarray}
-\partial_\tau {\Phi}^\ast +J ( \nabla^4 {\Phi}^\ast +2k_0^2  \nabla^2 {\Phi}^\ast)=0
\end{eqnarray}
along with a similar equation for ${\Phi}$. In addition, the Lagrangian $\mathcal{L}_0$ respects the global U(1) symmetry. Therefore,
an infinitesimal transformation $\Phi \rightarrow e^{i\varepsilon}\Phi \approx \Phi +i\varepsilon \Phi $ (and ${\Phi}^\ast\rightarrow e^{-i\varepsilon}{\Phi}^\ast \approx {\Phi}^\ast-i\varepsilon {\Phi}^\ast$, where $\varepsilon\ll 1$) results in $\mathcal{L}_0 \rightarrow \mathcal{L}_0+ \varepsilon \Delta \mathcal{L}_0$ with $\Delta \mathcal{L}_0$ being a total derivative, given by
\begin{align}
\Delta \mathcal{L}_0 =&\ \frac{\partial \mathcal{L}_0}{\partial \Phi} (i\Phi) +\Big(\frac{\partial \mathcal{L}_0}{\partial (\partial_\mu\Phi)}\Big)(i\partial_\mu \Phi)
\nonumber\\
&+\Big(\frac{\partial \mathcal{L}_0}{\partial (\partial_i \partial_j \Phi)}\Big)(i\partial_i\partial_j \Phi)
+\Phi \leftrightarrow {\Phi}^\ast
\nonumber\\
=&\ \partial_\mu \Big(\frac{\partial \mathcal{L}_0}{\partial (\partial_\mu\Phi)}(i\Phi)\Big)
+ \partial_i \Big(\frac{\partial \mathcal{L}_0}{\partial (\partial_i \partial_j \Phi)}(i\partial_j \Phi)\Big)
\nonumber\\
&+ \Phi \leftrightarrow {\Phi}^\ast \nonumber,
\end{align}
where we applied the equation of motion in Eq.~(\ref{eq:EOM}) to get the second equality. Then, the Noether current is 
\begin{align}
j^0 =&\ \frac{\partial \mathcal{L}_0}{\partial (\partial_\tau \Phi)} (i \Phi)
+\frac{\partial \mathcal{L}_0}{\partial (\partial_\tau {\Phi}^\ast)} (-i {\Phi}^\ast)
=2i {\Phi}^\ast \Phi
\nonumber\\
j_{i} =&\ \frac{\partial \mathcal{L}_0}{\partial (\partial_i \Phi)} (i \Phi)
+\frac{\partial \mathcal{L}_0}{\partial (\partial_i {\Phi}^\ast)} (-i {\Phi}^\ast)
\nonumber\\
&+ \frac{\partial \mathcal{L}_0}{\partial (\partial_i\partial_j \Phi)} (i \partial_j\Phi)
+\frac{\partial \mathcal{L}_0}{\partial (\partial_i \partial_j {\Phi}^\ast)} (-i \partial_j {\Phi}^\ast)
\nonumber\\
=&\ -i J \Phi (\partial_i \nabla^2 {\Phi}^\ast+ 2k_0^2 \partial_i {\Phi}^\ast )
+i J {\Phi}^\ast (\partial_i \nabla^2 \Phi+2\bar{k}_0^2 \partial_i \Phi )
\nonumber\\
&+ iJ \nabla^2  {\Phi}^\ast  \partial_i \Phi -i J\nabla^2 \Phi \partial_i {\Phi}^\ast,
\end{align}
which satisfy the continuity equation, $\partial_\mu j^\mu =0$.

With $\Phi=\sqrt{n}e^{i{\bf k}_0\cdot{\bf r} + i\phi}$ (and at mean-field level $k_0=\bar{k}_0$), the supercurrent is then a current in space,
\begin{align}
{\bf j}_s=&\ 4Jn\left[(\nabla\phi)^2 + 2{\bf k}_0\cdot\nabla\phi - \frac{1}{n}\nabla^2n + \frac{3}{4n^2}(\nabla n)^2\right]
\nonumber\\
&\times ({\bf k}_0+\nabla\phi)-2J\nabla n\nabla^2\phi-2Jn\nabla^3\phi
\nonumber\\
\approx &\ 4Jn\left[(\nabla\phi)^2+2{\bf k}_0\cdot\nabla\phi\right]({\bf k}_0+\nabla\phi)-2Jn\nabla^3\phi,\label{eq:app_j_s}
\end{align}
generated through the twisting of the superfluid phase $\phi$. In the last line, we consider the long-wavelength limit, where the fluctuation of density is small such that $\nabla n\approx 0$.

The supercurrent density can be written as ${\bf j}_s=n {\bf v}_s$, where ${\bf v}_s$ the superfluid velocity. For a linear in ${\bf r}$ classical phase variation $\phi ={\bf q}\cdot {\bf r}$, the superfluid velocity reduces to the derivative of the bare dispersion
\begin{align}
{\bf v}_s =&\ 4J\left[q^2+2{\bf k}_0\cdot {\bf q}\right]({\bf k}_0+{\bf q})
\nonumber\\
=&\ \nabla_k \varepsilon_{\bf k}|_{{\bf k}={\bf k}_0+{\bf q}}
,
\end{align}
where
\begin{equation}
\varepsilon_{\bf k} = J(k^2-k_0^2)^2 + \varepsilon_0,\quad {\bf k}={\bf k}_0+{\bf q}.
\end{equation}
We emphasize that the existence of the nonlinear $\phi$ terms in Eq.~(\ref{eq:app_j_s}) makes this relation hold.

Alternatively, the supercurrent can be obtained as a response to a background probe gauge field. We first generalize the long-wavelength harmonic Goldstone mode Hamiltonian density [with the corresponding Lagrangian density (\ref{eq:L_phi_0})],
\begin{align}
\mathcal{H}_{0\phi}=4Jnk_0^2(\nabla_\parallel\phi)^2 + Jn(\nabla^2\phi)^2,
\end{align}
to the quartic order in $\phi$ by requiring the rotational symmetry of ${\bf k}_0$, giving
\begin{align}
\mathcal{H}_\phi= 4Jn\left[k_0\nabla_\parallel\phi+\frac{1}{2}(\nabla\phi)^2\right]^2 + Jn(\nabla^2\phi)^2.
\end{align}
As the system coupled with a background U(1) gauge field, the Hamiltonian density is modified as $\mathcal{H}_\phi [\nabla \phi] \to \mathcal{H}_\phi [\nabla \phi +{\bf A}]$. The supercurrent density is then obtained by taking derivative with respect to the gauge field
\begin{align}
{\bf j}_s=&\ \left.\frac{\partial H_\phi}{\partial {\bf A}}\right|_{{\bf A}=0}
\nonumber\\
=&\ 4Jn\left[(\nabla\phi)^2+2k_0\nabla_\parallel\phi\right]({\bf k}_0+\nabla\phi)-2Jn\nabla^3\phi,
\end{align}
which reproduces the long-wavelength expression in Eq.~(\ref{eq:app_j_s}).

\section{Calculation details of the chemical potential \label{app:chemical potential}}

In this appendix, we evaluate the integral in Eq.~(\ref{eq:mu}). As in Appendix~\ref{app:depletion}, we use the dimensionless variables defined in Eq.~(\ref{eq:sigma}), and written integral as
\begin{align}
I&= \int_{-\infty}^{\infty} \frac{dq_\parallel dq_\perp}{(2\pi)^2} \biggl( 1-\frac{\mathcal{E}_{{\bf q}}}{\sqrt{\mathcal{E}_{{\bf q}} ^2+2U n \mathcal{E}_{{\bf q}}}}\biggr)
\nonumber\\
&=\int_{-\infty}^{\infty} \frac{dq_\parallel dq_\perp}{(2\pi)^2} \biggl(1-\frac{1}{\sqrt{1+\frac{2U n}{\mathcal{E}_{{\bf q}}}}} \biggr)
\nonumber\\
&=\int_{-\infty}^{\infty}\frac{d\sigma_\parallel d\sigma_\perp}{8\pi^2} \frac{(2U n)^{3/4}}{B^{1/2} K^{1/4}}  \frac{1}{\sqrt{|\sigma_\perp|}}\biggl( 1-\frac{\sigma}{\sqrt{\sigma^2+1 }}\biggr)
\nonumber\\
&=\frac{(2U n)^{3/4}}{8\pi^2 B^{1/2} K^{1/4}} I_\theta 
I'_\sigma,
\label{eq:chemical_potential_integral}
\end{align}
where $\sigma^2 = \sigma_\parallel^2 + \sigma_\perp^2$, $I_\theta$ is defined and computed in Eq.~(\ref{eq:depletion_integral}), and
\begin{eqnarray}
I'_\sigma &=& \int_{0}^{\infty} \sqrt{\sigma} d\sigma \biggl( 1-\frac{\sigma}{\sqrt{\sigma^2 + 1}}\biggr) \nonumber\\
&=&\left. \biggl(  \frac{2\sqrt{\sigma}}{3}(\sigma -\sqrt{1+\sigma^2}) + \frac{1}{3} f_e(\sigma) \biggr) \right|_0^\infty \nonumber\\
&=&\frac{1}{3} ( f_e(\infty) - f_e(0) ) \approx 1.23605,
\label{eq:integral_chemical_pot_r}
\end{eqnarray}
with $f_e(\sigma)$ is in Eq.~(\ref{eq:f_e}). We note that $f_e(\infty)\equiv \lim_{\sigma \rightarrow \infty} f_e(\sigma)=0$ and $f_e(0) \equiv \lim_{\sigma \rightarrow 0} f_e(\sigma)\approx -3.70815$, which lead to the last approximation in Eq.~(\ref{eq:integral_chemical_pot_r}).

As a result, the chemical potential for the helical superfluid in $2d$ is
\begin{equation}
\frac{\mu}{U} = n \biggl[ 1-\frac{2^{3/4}I_\theta I'_\sigma}{32\pi^2 }\biggl( \frac{U^3}{B^{2} K n} \biggr)^{1/4} \biggr].
\end{equation}

In general, for a $d$ dimensional system with a dispersion $\mathcal{E}_{\bf q}^{(l,m)}$ defined in Eq.~(\ref{eq:E_lm}), the integral in Eq.~(\ref{eq:chemical_potential_integral}) can be generalized as 
\begin{equation}
I^{(l,m)} = \int_{-\infty}^{\infty} \frac{d^l q_{\parallel} d^m q_{\perp} }{(2\pi)^{m+l}}  \biggl( 1-\frac{\mathcal{E}^{(l,m)}_{{\bf q}}}{\sqrt{(\mathcal{E}^{(l,m)}_{{\bf q}})^2+2U n \mathcal{E}^{(l,m)}_{{\bf q}}}} \biggr).
\end{equation}
In terms of the dimensionless variables defined in Eq.~(\ref{eq:sigma_i}), the integral becomes
\begin{align}
I^{(l,m)} =& \int_{-\infty}^{\infty} \frac{ d^l\sigma_{\parallel}  d^m \sigma_{\perp} }{(2\pi)^{m+l}} \frac{(2U n)^{(2l+m)/4}}{(\prod_i^l B_i \prod_j^m K_j)^{1/2}}
\nonumber\\
&\times \prod_j^m \frac{K_j^{1/4}}{2|\sigma_{\perp,j}|^{1/2}} \biggl(1-\frac{\sigma}{\sqrt{\sigma^2+1}} \biggr).
\end{align}
For the special case $B_i=B$ and $K_j=K$ for any $i$ and $j$, we have
\begin{equation}
I^{(l,m)} \propto \frac{(2U n)^{(2l+m)/4}}{B^{l/2}K^{m/4}}.
\end{equation}
Accordingly, in terms of the dimensionless quantity
\begin{equation}
\mathcal{Q}^{l,m}= \frac{(U n^{1-\frac{4}{2l+m}})^{(2l+m)/4}}{B^{l/2}K^{m/4}},
\end{equation}
the chemical potential is 
\begin{equation}
\frac{\mu^{(l,m)}}{U} =n\left[1- \mathcal{I}^{l,m}   \mathcal{Q}^{l,m}\right],
\end{equation}
where $\mathcal{I}^{l,m}$ is a dimensionless $\mathcal{O}(1)$ constant fully determined by $l$ and $m$. For a $d$-dimensional smectic phase with $l=1$ and $m=d-1$, the dimensionless quantity
\begin{equation}
\mathcal{Q}_{\rm smectic}^{1,d-1}= \biggl(\frac{U^{d+1} n^{d-3} }{B^{2}K^{d-1}}\biggr)^{1/4}.
\end{equation}
We can check that with $\textrm{dim}\left[ U n \right] = E$, $\textrm{dim}[n] = L^{-d}$, $\textrm{dim}\left[ B \right] = E L^2$, and $\textrm{dim}\left[ K \right] = E L^4$ ($E$, $L$ with dimensions of energy and length, respectively), $\mathcal{Q}_{\rm smectic}^{1,d-1}$ is in fact dimensionless. On the other hand, in a columnar phase with $l=d-1$ and $m=1$, the dimensionless quantity is given by
\begin{equation}
\mathcal{Q}_{\rm columnar}^{d-1,1}= \biggl(\frac{U^{2d-1} n^{2d-5} }{B^{2d-2}K}\biggr)^{1/4}.
\end{equation}

\widetext

\section{$\phi$ theory with $\text{C}_\text{6}$ lattice effects}\label{app:C6_effects}

We start with the coherent state path integral formalism of the frustrated Bose-Hubbard model introduced in the beginning of Sec.~\ref{sec:order_by_disorder}. In the density-phase representation (\ref{eq:pi_phi_rep_lattice}), the Lagrangian (including the constant part) can be written as
\begin{eqnarray}
L = i\sum_{i,a=1,2}\pi_{i,a}\partial_\tau \phi_{i,a} + H_0+H_{\text{int}},
\end{eqnarray}
where the kinetic part
\begin{align}\label{eq:H_0_pi_phi}
    H_0 =&\ - 2t_1 n_0\sum_{\langle i,j \rangle}\sqrt{1+\frac{\pi_{i,1}}{n_0}+\frac{\pi_{j,2}}{n_0}+\frac{\pi_{i,1}\pi_{j,2}}{n_0^2}}\cos\left[\theta_{{\bf k}_0}+{\bf k}_0 \cdot({\bf r}_{i,1}-{\bf r}_{j,2})+\phi_{i,1}-\phi_{j,2}\right]
    \nonumber\\
    &\ + 2t_2n_0\sum_{a=1,2}\sum_{\langle\langle i,j \rangle\rangle}\sqrt{1+\frac{\pi_{i,a}}{n_0}+\frac{\pi_{j,a}}{n_0}+\frac{\pi_{i,a}\pi_{j,a}}{n_0^2}}\cos\left[{\bf k}_0 \cdot({\bf r}_{i,a}-{\bf r}_{j,a})+\phi_{i,a}-\phi_{j,a}\right]
\end{align}
and the interaction
\begin{align}
    H_{\text{int}} = \frac{U}{2}\sum_i [(n_0+\pi_{i,1})^2 + (n_0+\pi_{i,2})^2].
\end{align}

For a bose condensate, we assume the fluctuations of $\pi$ and $\phi$ fields are small, and thereby expand in a power series, which up to quadratic order in $\pi$ and $\phi$ is given by
\begin{align}
H'_0 = &\ -t_1 n_0\sum_{\langle i,j \rangle}e^{i\theta_{{\bf k}_0}+i{\bf k}_0 \cdot({\bf r}_{i,1}-{\bf r}_{j,2})}\left[ 1 + \frac{\pi_{i,1}}{2n_0}+\frac{\pi_{j,2}}{2n_0} + i(\phi_{i,1}-\phi_{j,2}) -\frac{1}{8n_0^2}(\pi_{i,1}-\pi_{j,2})^2 - \frac{1}{2}(\phi_{i,1}-\phi_{j,2})^2 \right.
\nonumber\\
&\ \left. - i \frac{(\pi_{i,1}+\pi_{j,2})}{2n_0}  (\phi_{i,1}-\phi_{j,2})\right] + t_2n_0\sum_a\sum_{\langle\langle i,j \rangle\rangle}e^{i{\bf k}_0 \cdot({\bf r}_{i,a}-{\bf r}_{j,a})}\left[ 1 + \frac{\pi_{i,a}}{2n_0}+\frac{\pi_{j,a}}{2n_0} + i(\phi_{i,a}-\phi_{j,a}) \right.
\nonumber\\
&\ \left. - \frac{1}{8n_0^2}(\pi_{i,a}-\pi_{j,a})^2 - \frac{1}{2}(\phi_{i,a}-\phi_{j,a})^2 - i \frac{(\pi_{i,a}+\pi_{j,a})}{2n_0}(\phi_{i,a}-\phi_{j,a}) \right] + \frac{U}{4}\sum_{i,a}(n_0+\pi_{i,a})^2 + \text{h.c.}.
\end{align}
To perform one-loop calculations, we include the cubic and quartic terms of $\phi$, given by
\begin{align}
    H'_1 =&\ -\frac{t_1n_0}{6}\sum_{\langle i,j \rangle}e^{i\theta_{{\bf k}_0}+i{\bf k}_0 \cdot({\bf r}_{i,1}-{\bf r}_{j,2})}\left[-i(\phi_{i,1}-\phi_{j,2})^3 + \frac{1}{4}(\phi_{i,1}-\phi_{j,2})^4\right]
    \nonumber\\
    &+ \frac{t_2n_0}{6}\sum_a\sum_{\langle\langle i,j \rangle\rangle}e^{i{\bf k}_0 \cdot({\bf r}_{i,a}-{\bf r}_{j,a})}\left[-i(\phi_{i,a}-\phi_{j,a})^3 + \frac{1}{4}(\phi_{i,a}-\phi_{j,a})^4\right] + ... + \text{h.c.} 
\end{align}

At this point, one can either directly take the continuum limit in the real space or go to momentum space to get an effective action. The calculation for the former is simpler but misses some contributions even at long length scale, while the later is tedious but valid for all momentum. We discuss both methods below.

\subsection{Taking continuum limit in the real space \label{app:c6_real_space}}

To take the continuum limit, we consider the fields ${\phi}_{i,\alpha}$ and ${\pi}_{i,\alpha}$ on lattice sites as slowly-varying functions of continuous spacetime coordinates $({\bf r},\tau)$ compared to the lattice constant. The time coordinate is already continuous. Therefore, the difference of two operators at distinct spatial locations can be expanded as
\begin{align}\label{eq:continuum_limit}
     {\phi}({\bf r}_{i,a})- {\phi}({\bf r}_{j,a'}) &= \sum_{n=1}^{\infty}\frac{1}{n!}(\Delta{\bf r}\cdot\nabla )^n {\phi}|_{{\bf r}={\bf r}_{j,a'}},
\end{align}
where $\Delta{\bf r} = {\bf r}_{i,a}-{\bf r}_{j,a'}$ with $a$ and $a'$ label the sublattices. Below we focus on the case ${\bf k}_0=(0,k_0)$ while other cases can be derived following the same procedure. Including up to the second (first) order for the quadratic (cubic and quartic) term(s), the Hamiltonian density, defined by $\frac{2}{3\sqrt{3}}\int_{\bf r}\mathcal{H}' = H'_0+H'_1$, is given by 
\begin{align}\label{eq:H_phi_low_k}
\frac{\mathcal{H}'}{n_0} \approx&\ \Delta\left[c(\partial_\parallel\phi)+c_\perp(\partial_\perp\phi)^2\right] + B(\partial_\parallel\phi)^2 + K_{20}(\partial^2_\parallel\phi)^2 + K_{02}(\partial^2_\perp\phi)^2 + K_{11}(\partial^2_\parallel\phi)(\partial^2_\perp\phi)
\nonumber\\
&\  + B_{30}(\partial_\parallel\phi)^3 + B_{12}(\partial_\parallel\phi)(\partial_\perp\phi)^2 + B_{40}(\partial_\parallel\phi)^4 + B_{04}(\partial_\perp\phi)^4 + B_{22}(\partial_\parallel\phi)^2(\partial_\perp\phi)^2 + \frac{U}{2n_0}\pi^2,
\end{align}
where $\Delta_{{\bf k}_0} = 1-|\Gamma_{\bar{{\bf k}}_0}|/|\Gamma_{{\bf k}_0}|$ with $\bar{\bf k}_0$ satisfying Eq.~(\ref{eq:k0}),
\begin{align}\label{eq:parameters_L_phi_low_k}
c = 12t_2\sin(3k_0/2) &,\quad c_\perp = 3t_2[2+\cos(3k_0/2)],\quad B = 4t_2[1-\cos(3k_0/2)],\quad B_{30} = B_{12} = 3t_2\sin(3k_0/2),
\nonumber\\
K_{20} =&\ \frac{t_2}{24}[-5+32\cos(3k_0/2)],\quad
K_{02} = \frac{9t_2}{8},\quad
K_{11} = \frac{3t_2}{4}[-1+4\cos(3k_0/2)],
\nonumber\\
B_{40} =&\ \frac{t_2}{24}[-5+32\cos(3k_0/2)],\quad B_{04} = \frac{9t_2}{8},\quad B_{22} = \frac{3t_2}{4}[-1+4\cos(3k_0/2)],
\end{align}
and the constant and linear in $\pi$ terms are dropped. In the above, the $\rho$ dependence is hidden in $k_0(\rho)$ and $t_2=t_1\rho$. By integrating over $\pi$, we obtain the zeroth-order Lagrangian density, (\ref{eq:L_phi_low_k}), with the parameters (\ref{eq:parameters_L_phi_low_k}), however, not satisfying the relation $B_{lm}^{(0)} = b_{lm}^{(0)}$. This inconsistency is due to the $\theta_{{\bf k}_0}$ factor in Eq.~(\ref{eq:H_0_pi_phi}), or more generally the two-band nature of the model, which invalids the simple continuum limit above that neglects the spatial variation between the two sublattice sites within a unit cell, as discussed in more details below.

\subsection{Continuous effective theory in the momentum space \label{app:c6_momentum_space}}

Alternatively, We can Fourier transform the Hamiltonian ${H}'_0$ and ${H}'_1$ followed by a change of basis defined as
\begin{align}\label{eq:trans_site_band}
    \left(\begin{array}{cccc}
    \pi_{{\bf q},1} \\ \pi_{{\bf q},2} \\ \phi_{{\bf q},1} \\ \phi_{{\bf q},2}
    \end{array}\right) =&\ \frac{1}{2}\left(\begin{array}{cc}
    U_{{\bf q}}+U_{-{\bf q}}^* & in_0(U_{{\bf q}}-U_{-{\bf q}}^*) \\ -i\frac{1}{n_0}(U_{{\bf q}}-U_{-{\bf q}}^*) & U_{{\bf q}}+U_{-{\bf q}}^*
    \end{array}\right)\left(\begin{array}{cccc}
    \pi_{{\bf q},+} \\ \pi_{{\bf q},-} \\ \phi_{{\bf q},+} \\ \phi_{{\bf q},-}
    \end{array}\right),
\end{align}
where the subscripts $1,2$ label the two sublattices and $\pm$ denote the two bands as in the main text. The unitary matrix above is given by
\begin{align}
    U_{\bf q} =&\ \frac{1}{\sqrt{2}}\biggl( \begin{array}{cc}
\exp(i\frac{\theta_{{\bf k}_0+{\bf q}} - \theta_{{\bf k}_0}}{2}) & \exp(i\frac{\theta_{{\bf k}_0+{\bf q}} - \theta_{{\bf k}_0}}{2})  \\
-\exp(-i\frac{\theta_{{\bf k}_0+{\bf q}} - \theta_{{\bf k}_0}}{2})  & \exp(-i\frac{\theta_{{\bf k}_0+{\bf q}} - \theta_{{\bf k}_0}}{2}) 
\end{array}\biggr).
\end{align}

At quadratic order, the action for the lower band is given by 
\begin{align}
S_0 = \frac{1}{2}\sum_{q} \left( \begin{array}{cc}
\pi_{-q,-} & \phi_{-q,-}
\end{array}\right)G_0^{-1}(q,\omega_n)
\left( \begin{array}{cc}
\pi_{q,-} \\ \phi_{q,-}
\end{array}\right),
\label{eq:quadratic_action_pi_phi}
\end{align}
where 
\begin{eqnarray}
G_0^{-1} (q,\omega_n)= \left( \begin{array}{cc}
\mathcal{E}_{\bf q}/(2n_0) + U\frac{1+\cos\Theta_{\bf q} }{2}& \omega_n + i\mathcal{E}_{{\bf q},2} \\
-\omega_n - i\mathcal{E}_{{\bf q},2} & 2n_0\mathcal{E}_{\bf q} + 4n_0^2 U\frac{1-\cos \Theta_{\bf q}}{2}
\end{array}\right).
\label{eq:G0}
\end{eqnarray}

In the above, we consider the weakly-interacting limit and can thereby drop the upper band contributions which only give $O(Un_0/t_1)$ corrections to the low energy physics. Before computing higher-order terms, we have two comments in order. 

Firstly, we note the quadratic action
reproduces the same Bogoliubov quasiparticle spectrum  in Eq.~(\ref{eq:disp}). Technically, both the density-phase representation (\ref{eq:pi_phi_rep_lattice}) and the Bogoliubov approximation (\ref{eq:d_decompose}) share the same classical background field, which in real space is given by $\sqrt{n_0}e^{\pm i\frac{\theta_{{\bf k}_0}}{2}+i{\bf k}_0\cdot {\bf r}_{i,1(2)}}$ with the $+$ and $-$ factors for the sites in the $1$ and $2$ sublattices respectively. The fluctuations around the condensate are given by $\pi$ and $\phi$ in
Eq.~(\ref{eq:pi_phi_rep_lattice}) or $d_-$ in Eq.~(\ref{eq:d_decompose}). An expansion of the former representation gives linear terms of $\pi$ and $\phi$ that takes the same form as the real and imaginary parts of $d_-$ in the latter representation, and thus both give the same form of the quadratic action. While the higher-order terms which modifies the $S_0$ in Eq.~(\ref{eq:quadratic_action_pi_phi}) do not have a simple relation compared with higher-order terms which corrects Eq.~(\ref{eq:hammatrix}), they should describe the same physics. 

Secondly, as shown above, an accurate calculation gives $B$ and $K$ identical with the ones from Bogoliubov theory in Eq.~(\ref{eq:disp_small_q}). In Appendix~\ref{app:c6_real_space}, fields on the two sites within a unit cell are considered to be the same, i.e., $\phi_{i,1} \approx \phi_{i,2} \approx \phi_{i}$. This is equivalent to approximate the transformation in Eq.~(\ref{eq:trans_site_band}) with $U_{{\bf q}} \approx U_{0}$, which gives $\phi_{{\bf q},1} \approx \phi_{{\bf q},2} \approx \frac{1}{\sqrt{2}}\phi_{{\bf q},-}$. Therefore, the discrepancy between $B$ and $K$ in (\ref{eq:parameters_L_phi_low_k}) and the accurate ones in Eq.~\eqref{eq:b_lm_Vertex} comes from the definition of the fields, which by inspection gives a factor of $1/2$ via the relation $\phi_{{\bf q}}=\frac{1}{\sqrt{2}}\phi_{{\bf q},-}$, and the difference between $U_{{\bf q}}$ and $U_0$. The latter can be neglected in the limit $k_0\to 0$, where the difference is of higher-order in ${\bf q}$.

The calculations for the nonlinear terms are quite tedious. Especially, it is hard to simplify the expressions after the transformation in Eq.~(\ref{eq:trans_site_band}). Here, we employ the same approximation $\phi_{{\bf q},1} \approx \phi_{{\bf q},2} \approx \phi_{{\bf q}}$ to derive below the cubic and quartic terms in $\phi$.
\begin{align}
    S_{1} \approx &\ -\frac{in_0}{\sqrt{V}}\sum_{{\bf q},{\bf q}'}\Gamma_3({\bf q})\phi_{{\bf q}}\phi_{{\bf q}'}\phi_{-{\bf q}-{\bf q}'} + \frac{n_0}{12V}\sum_{\{{\bf q}\}}\left[-4\Gamma_4({\bf q}_1)+3\Gamma_4({\bf q}_1+{\bf q}_2)\right]\times\phi_{{\bf q}_1}\phi_{{\bf q}_2}\phi_{{\bf q}_3}\phi_{-{\bf q}_1-{\bf q}_2-{\bf q}_3},
\end{align}
where
\begin{align}
    \Gamma_3({\bf q}) =&\ -t_1\text{Re}\biggl[(\Gamma_{{\bf k}_0+{\bf q}}-\Gamma_{{\bf k}_0-{\bf q}}) e^{-i\phi_{{\bf k}_0}} + 4\rho\sum_\alpha\sin\left({\bf k}_0 \cdot {\bf v}_\alpha\right)\sin({\bf q} \cdot {\bf v}_\alpha)\biggr]
    \nonumber\\
    \Gamma_4({\bf q}) =&\ -2t_1\text{Re}\biggl[\biggl(\frac{\Gamma_{{\bf k}_0+{\bf q}}+\Gamma_{{\bf k}_0-{\bf q}}}{2} - \Gamma_{{\bf k}_0}\biggr)e^{-i\phi_{{\bf k}_0}} - 2\rho\sum_\alpha\cos\left({\bf k}_0 \cdot {\bf v}_\alpha\right)\left[1-\cos\left({\bf q} \cdot {\bf v}_\alpha\right)\right]\biggr],
\end{align}
where an expansion of $\Gamma_3$ and $\Gamma_4$ in small ${\bf q}$ together with a Fourier transform back to the real space reproduce the nonlinear terms in the field theory (\ref{eq:H_phi_low_k}) with the same coefficients obtained earlier (\ref{eq:parameters_L_phi_low_k}).

\section{One-loop calculation of $b_\perp$}\label{app:b_perp}

Here we calculate the perpendicular curvature $b_{\perp} = \frac{1}{2N_0}\partial_{k_\perp}^2\Omega({\bf k}_0)$ at one-loop order. By expanding the condensate momentum ${\bf k}_0={\bf k}_0^{(0)}+{\bf k}_0^{(1)}$ around its mean-field value ${\bf k}_0^{(0)}$ and keeping the leading-order terms, we get
\begin{align}
    b_{\perp}^{(1)} = \frac{1}{2N_0}\partial_{k_\parallel}\partial_{k_\perp}^2\Omega^{(0)}\left({\bf k}_0^{(0)}\right)k_0^{(1)} + \frac{1}{2N_0}\partial_{k_\perp}^2\Omega^{(1)}\left({\bf k}_0^{(0)}\right),\quad    k_0^{(1)} = -\frac{\partial_{k_\parallel}\Omega^{(1)}\left({\bf k}_0^{(0)}\right)}{\partial_{k_\parallel}^2\Omega^{(0)}\left({\bf k}_0^{(0)}\right)},
\end{align}
where $k_0^{(1)}$ can be determined by $\nabla_{{\bf k}_0}\Omega({\bf k}_0)=0$ and
\begin{align}\label{eq:del_Omega^(1)}
     \partial_{k_\parallel} \Omega^{(1)}\left({\bf k}_0^{(0)}\right) =&\ \frac{1}{2}\sum_{\bf q}\coth\left(\beta E_{\bf q}/2\right)\partial_{k_\parallel}E_{\bf q}
    \nonumber\\
     \partial_{k_\perp}^2 \Omega^{(1)}\left({\bf k}_0^{(0)}\right) =&\ \sum_{\bf q}\left[\coth\left(\beta E_{\bf q}/2\right)\partial_{k_\perp}^2 E_{\bf q} - \frac{\beta}{2\sinh^2\left(\beta E_{\bf q}/2\right)}(\partial_{k_\perp}E_{\bf q})^2\right].
\end{align}
In the absence of interaction, $\partial_{k_\parallel}E_{\bf q}=\partial_{q_\parallel}E_{\bf q}$ and $\partial_{k_\perp}^2 E_{\bf q}=\partial_{q_\perp}^2 E_{\bf q}$. Accordingly, the integrands in $\partial_{k_\parallel} \Omega^{(1)}\left({\bf k}_0^{(0)}\right)$ and $\partial_{k_\perp}^2 \Omega^{(1)}\left({\bf k}_0^{(0)}\right)$ are total derivatives, which makes $\partial_{k_\parallel} \Omega^{(1)}\left({\bf k}_0^{(0)}\right)=\partial_{k_\perp}^2 \Omega^{(1)}\left({\bf k}_0^{(0)}\right)=0$ and subsequently $b_\perp^{(1)}=0$. In the presence of weak interaction, the dispersion deviates from the noninteracting form within a crossover scale ${\bf q}_c=(q_\parallel^c,q_\perp^c)=(\xi^{-1},(\lambda\xi)^{-1/2})$. Therefore, the integrals (\ref{eq:del_Omega^(1)}) are dominated by the region ${\bf q}<{\bf q}_c$. In this small $U$ limit, we obtain $b_\perp^{(1)} = b_{0\perp}f_b(Un_0/T)$ by using change of variables $q'_\parallel=\xi q_\parallel$ and $q'_\perp=(\lambda\xi)^{1/2}q_\perp$ in Eq.~(\ref{eq:del_Omega^(1)}), where $b_{0\perp}=(Un_0)^{5/4}t_1^{-1/4}g_b(\rho)$ with $g_b(\rho)$ a dimensionless $\mathcal{O}(1)$ constant and the scaling function $f_b$ is defined in Eq.~(\ref{eq:scaling_fn_f}).

\section{One-loop calculation of $B_\perp$}\label{app:B_perp}

A complete one-loop calculation should give the perpendicular stiffness of $\phi$ equals to the corresponding curvature in the thermodynamic potential, i.e., $B_\perp^{(1)}=b_\perp^{(1)}$, with $b_\perp^{(1)}$ being calculated in Appendix~\ref{app:b_perp}. Here, we instead perform an approximate calculation of $B_\perp^{(1)}$ in the long-wavelength limit, where the Lagrangian (\ref{eq:L_phi_low_k}) is valid. Specifically, we consider the leading derivative terms and thereby only keep $U\pi^2$ for the $\pi$ field, which is enabled by an $U$-dependent UV cutoff $\Lambda_U$ defined by $\mathcal{E}_{\bf q} < \mathcal{O}(1)\times Un_0$, which in the weakly-interacting limit reduces to ${\bf q}<\mathcal{O}(1)\times{\bf q}_c$. Although such calculation neglects the range of integration ${\bf q}>{\bf q}_c$, the contribution to $B_\perp$ in this range is actually very small as seen in the calculation of $b_\perp^{(1)}$ in Appendix~\ref{app:b_perp} that takes into account all momentum. Therefore, the integral of $B_\perp^{(1)}$ or $b_\perp^{(1)}$ actually self regulates at ${\bf q}_c$ in the weakly-interacting limit.

\subsection{Long-wavelength limit}

We consider the low-energy $\phi$-only theory (\ref{eq:L_phi_low_k}), where the bare Green function is given by ($k_B=1$)
\begin{align}
    G_{0,\phi}(q) =&\ \langle \phi_{-q}\phi_{q} \rangle_0 = \frac{1}{2n_0}\frac{1}{B_\tau\omega_n^2 + \mathcal{E}_{\bf q}},
\end{align}
where $\mathcal{E}_{\bf q} = Bq_\parallel^2 + K_{20} q_\parallel^4 + K_{02} q_\perp^4 + K_{11}q_\parallel^2 q_\perp^2$ and $B_\tau=1/2Un_0$.

We first calculate the coefficient of $\partial_\parallel\phi$, which corresponds to the slope of the thermodynamic potential at ${\bf k}_0$ along the $\parallel$ direction. At one-loop order, it is given by
\begin{align}
\begin{tikzpicture}[baseline={(X.base)}]
\node[] (X) at (0,0.5) {};
\node[fill,circle,inner sep=1.5pt] at (0,0.6) {};
\end{tikzpicture}
+
\begin{tikzpicture}[baseline={(X.base)}]
\node[] (X) at (0,0.8) {};
\node[fill,circle,inner sep=1.5pt] at (0,0.6) {};
\draw plot [smooth cycle, tension=1] coordinates { (0,0.6) (-0.3,0.9) (0,1.2) (0.3,0.9)};
\end{tikzpicture} =&\ n_0C\Delta_{{\bf k}_0}  + \frac{B_{12}}{2}\int_{q}^{\Lambda_U}\frac{3q_\parallel^2+q_\perp^2}{B_\tau\omega_n^2 + \mathcal{E}_{\bf q}},
\end{align}
where the first term is the zeroth-order contribution that vanishes at $k_0=\bar{k}_0$ and $\int_q = \frac{1}{\beta V}\sum_{{\bf q},\omega_n}$. With the one-loop correction, ${\bf k}_0$ needs to be shifted such that the coefficient of $\partial_\parallel\phi$ is zero again, which assures ${\bf k}_0$ is located at the local minimum of the loop-corrected thermodynamic potential with vanished first derivative. In the above, the tadpole diagrams are dropped as they cancel out up to all orders when choosing the correct $k_0$.

Next we calculate $B_\perp$, which at one-loop order is given by
\begin{align}\label{eq:B_perp^(1)_app}
B_\perp^{(1)} =&\ \begin{tikzpicture}[baseline={(X.base)}]
\node[] (X) at (0,0.5) {};
\node[fill,circle,inner sep=1.5pt] at (0,0.6) {};
\end{tikzpicture}
+
\begin{tikzpicture}[baseline={(X.base)}]
\node[] (X) at (0,0.8) {};
\node[fill,circle,inner sep=1.5pt] at (0,0.6) {};
\draw plot [smooth cycle, tension=1] coordinates { (0,0.6) (-0.3,0.9) (0,1.2) (0.3,0.9)};
\end{tikzpicture}
+
\begin{tikzpicture}[baseline={(X.base)}]
\node[] (X) at (0,-0.1) {};
\node[fill,circle,inner sep=1.5pt] at (0,0) {};
\node[fill,circle,inner sep=1.5pt] at (1.5,0) {};
\draw plot [smooth, tension=1.2] coordinates { (0,0) (0.75,0.3) (1.5,0)};
\draw plot [smooth, tension=1.2] coordinates { (0,0) (0.75,-0.3) (1.5,0)};
\end{tikzpicture} 
\nonumber\\
=&\ c_\perp n_0\Delta_{{\bf k}_0} + \frac{1}{8}\int_{q}^{\Lambda_U}\frac{B_{22}q_\parallel^2+6B_{04} q_\perp^2}{B_\tau\omega_n^2 + \mathcal{E}_{\bf q}} - B_{12}^2\int_{q}^{\Lambda_U}\frac{q_\parallel^2q_\perp^2}{(B_\tau\omega_n^2 + \mathcal{E}_{\bf q})^2}
\nonumber\\
=&\ -\frac{c_\perp B_{12}}{2c}\int_{q}^{\Lambda_U}\frac{3q_\parallel^2+q_\perp^2}{B_\tau\omega_n^2 + \mathcal{E}_{\bf q}} + \frac{1}{8}\int_{q}^{\Lambda_U}\frac{B_{22}q_\parallel^2+6B_{04} q_\perp^2}{B_\tau\omega_n^2 + \mathcal{E}_{\bf q}} - B_{12}^2\int_{q}^{\Lambda_U}\frac{q_\parallel^2q_\perp^2}{(B_\tau\omega_n^2 + \mathcal{E}_{\bf q})^2},
\end{align}
where in the last line we choose ${\bf k}_0$ at which the linear term $\partial_\parallel\phi$ vanished. In the above, the arguments of the coefficients are taken to be their mean-field value $k_0=\bar{k}_0$ as the difference is of higher order. With a change of variables $q'_\parallel=\xi q_\parallel$ and $q'_\perp=(\lambda\xi)^{1/2}q_\perp$, we can rewrite the expression above that in the weakly-interacting limit is given by
\begin{align}
    B_\perp^{(1)} \approx &\ -\frac{c_\perp B_{12} B_\tau}{4c\xi(\lambda\xi)^{3/2}}\int_{q'}^{\Lambda}\frac{q'_\perp{}^2}{\omega'_n{}^2 + q'_\parallel{}^2 + q'_\perp{}^4} + \frac{3B_{04} B_\tau}{4\xi(\lambda\xi)^{3/2}}\int_{q'}^{\Lambda}\frac{ q'_\perp{}^2}{\omega'_n{}^2 + q'_\parallel{}^2 + q'_\perp{}^4} - \frac{B_{12}^2 B_\tau^2}{\xi^{3}(\lambda\xi)^{3/2}}\int_{q'}^{\Lambda}\frac{q'_\parallel{}^2q'_\perp{}^2}{(\omega'_n{}^2 + q'_\parallel{}^2 + q'_\perp{}^4)^2}
    \nonumber\\
    =&\ \frac{3B_{04}-c_\perp B_{12}/c}{8\xi(\lambda\xi)^{3/2}}\int_{{\bf q}'}^{\Lambda}\frac{q'_\perp{}^2}{E'_{{\bf q}'}}\coth\left(\frac{Un_0 E'_{{\bf q}'}}{T}\right) - \frac{B_\tau B_{12}^2}{2\xi^{3}(\lambda\xi)^{3/2}}\int_{{\bf q}'}^{\Lambda}\frac{q'_\parallel{}^2q'_\perp{}^2}{2E'_{{\bf q}'}{}^3}\left[\coth\left(\frac{Un_0 E'_{{\bf q}'}}{T}\right)+\frac{Un_0 E'_{{\bf q}'}}{T}\text{csch}^2\left(\frac{Un_0 E'_{{\bf q}'}}{T}\right)\right]
    ,
\end{align}
where $\omega'_n=B_\tau\omega_n=\pi n T/Un_0$ (here $n$ is an integer, not confused with the particle density), $E'_{{\bf q}'}=q'_\parallel{}^2 + q'_\perp{}^4$ and the UV cutoff $\Lambda$ indicates ${\bf q}'=(q'_\parallel,q'_\perp)<(\mathcal{O}(1),\mathcal{O}(1))$. In the last line above, we can read $B_\perp^{(1)} = B_{0\perp}f_B(Un_0/T)$, where the dimensionless scaling function $f_B$ exhibits the same limits as $f_b$ in Eq.~(\ref{eq:scaling_fn_f}) and $B_{0\perp} = (Un_0)^{5/4}t_1^{-1/4}g_B(\rho)$ with $g_B(\rho)$ a dimensionless $\mathcal{O}(1)$ constant.

\subsection{Isotropic limit}

In the isotropic limit $\rho\to 1/6$ or $k_0\to 0$, the coefficients $K\equiv K_{20}=K_{02}=K_{11}/2$, $B/k_0^2 = B_{30}/k_0 = B_{12}/k_0 = B_{40}=B_{04}=B_{22}/2$ and $c=2k_0 c_\perp$. Consequently, $B_\perp^{(1)}$ in the last line of Eq.~(\ref{eq:B_perp^(1)_app}) vanishes
\begin{align}
B_\perp^{(1)} =&\ - \frac{B}{4k_0^2}\int_{q}\frac{3q_\parallel^2+q_\perp^2}{B_\tau\omega_n^2 +  Bq_\parallel^2 + Kq^4} + \frac{B}{4k_0^2}\int_{q}\frac{q_\parallel^2+3q_\perp^2}{B_\tau\omega_n^2 + Bq_\parallel^2+Kq^4} - \frac{B^2}{k_0^2}\int_{q}\frac{q_\parallel^2q_\perp^2}{(B_\tau\omega_n^2 + Bq_\parallel^2+Kq^4)^2} = 0,
\end{align}
where we used the following integration by part for the last term
\begin{align}
    &\int_{\bf q}\frac{q_\parallel^2q_\perp^2}{(B_\tau\omega_n^2 + Bq_\parallel^2+Kq^4)^2} = \int_{0}^\infty dq q\int_{0}^{2\pi} d\theta \frac{q^4\cos(\theta)^2\sin(\theta)^2}{[B_\tau\omega_n^2+Bq^2\cos(\theta)^2+Kq^4]^2}
    \nonumber\\
     =&\ \frac{1}{2B}\int_{0}^\infty dq q\int_{0}^{2\pi} d\theta \frac{q^2[\sin(\theta)^2-\cos(\theta)^2]}{B_\tau\omega_n^2 + Bq^2\cos(\theta)^2+Kq^4} = \frac{1}{2B}\int_{\bf q} \frac{q_\perp^2-q_\parallel^2}{B_\tau\omega_n^2 + Bq_\parallel^2+Kq^4}.
\end{align}

\bibliography{HoneycombSF}

\end{document}